\documentclass{llncs}

\usepackage[T1]{fontenc}

\usepackage{amsmath,amsfonts,amssymb,amstext}
\usepackage{graphicx}
\usepackage[usenames,dvipsnames]{xcolor}
\usepackage[colorlinks=true,linkcolor=MidnightBlue,citecolor=DarkGreen,urlcolor=Maroon]{hyperref}

\usepackage{fullpage}

\definecolor{DarkGreen}{rgb}{0.0, 0.5, 0.0}

\newcommand{\eels}[1]{\textsc{EELS}$_{#1}$}
\newcommand{\eelsfi}[1]{\ensuremath{\textsc{EELS}_{#1}^{\mathrm{fi}}}}
\newcommand{\eelsea}[1]{\ensuremath{\textsc{EELS}_{#1}^{\mathrm{ea}}}}

\title{
Evacuating from Eventually Emerging Targets on the Line\thanks{
Computations were performed on the Nibi supercomputer at the SciNet HPC Consortium. SciNet is funded by Innovation, Science and Economic Development Canada; the Digital Research Alliance of Canada; the Ontario Research Fund: Research Excellence; and the University of Toronto.}
\thanks{
The authors used ChatGPT as an auxiliary tool for exploratory discussion, code generation and debugging for preliminary experiments, and consistency checks of intermediate arguments. All mathematical claims, proofs, computational results, and final verification are the sole responsibility of the authors.
}
\\
{\small (Full Version)
\thanks{This is the full version of the paper with the same title
accepted to WALCOM 2027~\cite{GeorgiouJonesKrupsky2027}}
}
}

\author{
Konstantinos Georgiou\inst{1}\thanks{
ORCID: \url{https://orcid.org/0009-0007-9677-341X}. Research supported in part by an NSERC Discovery Grant and by the Toronto Metropolitan University Faculty of Science Dean's Research Fund.
}
\and
Caleb Jones\inst{1}
\and
Liza Krupsky\inst{1}
}

\institute{
Department of Mathematics, Toronto Metropolitan University,
Toronto, Ontario, Canada
}

\begin{document}
\maketitle
\thispagestyle{empty}

\begin{abstract}
We introduce and study Eventually Emerging Linear Search (\eels{n}), an
online evacuation problem on the line that, to the best of our knowledge,
has not been studied before.
In this problem, $n$ unit-speed agents with wireless communication start at the origin and, following a deterministic online algorithm, search for a hidden target at location $x$, where $|x|\geq1$, that becomes detectable at an unknown emergence time $T$. Visiting $x$ before time $T$ reveals nothing, and the search terminates only when all agents have reached the target.

We study two competitive measures. 
In the emergence-aware model, the offline algorithm learns the target location $x$ only when the target emerges at time $T$, giving cost $T+|x|$. In the fully-informed model, the offline algorithm knows $x$ from the beginning, giving cost $\max\{|x|,T\}$ (since emergence happens only at time $T$).
We denote the corresponding optimal competitive ratios by \eelsea{n} and \eelsfi{n}.
For one agent, we show that $\eelsea{1}=\eelsfi{1}=9$, the optimal value for classical
cow-path search. For two agents, we give an algorithm with competitive
ratio $3+\sqrt6$ under both measures, and prove it optimal within the
natural class of algorithms, in which each agent performs
strictly expanding origin-return excursions on a fixed half-line.
We also prove the unconditional bounds
$\eelsea{2}\geq4.35$ and $\eelsfi{2}\geq4.38$. These bounds are
computer-assisted; we model the movements of arbitrary algorithms by finite linear
programming relaxations, and after imposing necessary linear conditions for such algorithms to have competitive ratio at most $C$, we 
certify infeasibility using exact rational Farkas certificates, thereby showing that their competitive ratio is more than $C$. 
Thus, even with wireless communication, eventual emergence makes the two-agent
problem strictly harder than classical line evacuation, whose optimal
competitive ratio is $3$.

For three agents, we prove analytically the unconditional lower bounds
$\eelsea{3}\geq3.1014$ and $\eelsfi{3}\geq3.2823$, and introduce a
geometric three-peak family.
Different parameter choices give $\eelsea{3}\leq4.010599$ and $\eelsfi{3}\leq4.340445$. 
These upper bounds lie below the corresponding two-agent lower bounds, so a third
agent provably improves both measures. In contrast to the two-agent
algorithm, which achieves the same guarantee under both measures, the
stated three-agent bounds use different parameter choices.
\keywords{Linear Search \and Competitive Analysis \and Mobile Agents \and Evacuation}
\end{abstract}

\newpage
\thispagestyle{empty}
\tableofcontents

\newpage

\section{Introduction}
Searching for a hidden object with mobile agents is one of the basic
problems of search theory, with early formulations dating back to the work
of Bellman and Beck in the 1960s
\cite{beck1964linear,bellman1963optimal}. Among the simplest geometric
settings is search on the infinite line. In the classical linear-search
problem, also known as the cow-path problem, a mobile agent starts at the
origin and searches for a target at an unknown location. The problem was
later revisited in the theoretical computer science literature by
Baeza-Yates et al.~\cite{baezayates1993searching}, and has since served as a
basic model for studying online search under uncertainty. Its many
variations include changes to the information available to the searchers,
their motion capabilities, the number of participating agents, their
communication model, and the objective used to evaluate the search.

In the classical linear-search problem the target is present and detectable
from the beginning. Hence the uncertainty is spatial: once an agent visits
a point without finding the target, that point can be excluded from further
consideration. We introduce an additional source of uncertainty by allowing
the target to become detectable only after an unknown emergence time. The
target has an unknown location $x$ and an unknown emergence time $T$, and
a visit to $x$ before time $T$ reveals nothing. Consequently, visiting a
location no longer eliminates it permanently. A part of the line that has
already been searched may have to be searched again, and an algorithm must
decide not only how far to explore but also when previously visited regions
should be revisited.

We call the resulting problem \emph{Eventually Emerging Linear Search}. 
We study $n$ unit-speed agents starting at the origin and communicating wirelessly, for $n=1,2,3$.
Once the target is detected, its location is communicated
instantaneously, and the search is completed only when all agents have
reached it. 
This termination requirement is called \emph{evacuation}.
Wireless communication removes the difficulty of communicating
a discovery, but it does not remove the coordination problem before
detection: the agents must distribute their effort between exploring new
locations and revisiting old ones without knowing either where the target is
or when it has become detectable. Eventually Emerging Linear Search is
therefore naturally connected both to classical linear search and to
multi-agent group search and evacuation. For this problem, we consider two offline benchmarks. In the emergence-aware model, the target location becomes known at time $T$ and is associated with an offline cost of $T+|x|$. In the fully-informed model, the target location is known from the beginning, and is associated with offline cost of $\max\{T,|x|\}$.
We review the relevant developments in these areas next.

\subsection{Related Work}

The classical linear-search problem goes back to Bellman and Beck
\cite{bellman1963optimal,beck1964linear}, and was later revisited from an
algorithmic perspective by Baeza-Yates et al.~\cite{baezayates1993searching}.
For a unit-speed searcher on the line, the optimal competitive ratio is $9$,
achieved by the familiar exponentially expanding alternating (zig-zag) search.
Numerous variants have since been considered. Among those most relevant here
are models with partial information about the target distance
\cite{Bose13}, costs associated with changing direction
\cite{demaine2006online}, and movement speeds depending on the direction or
the terrain \cite{CzyzowiczKKNOS17}. These works preserve the basic
assumption that whenever the searcher reaches the target location, the target
is detected.

A closer connection to our work arises when this last assumption is relaxed.
In $p$-Faulty Search on a Half-Line, each visit to the target independently
detects it with a fixed probability $p$, and hence already searched locations need to be visited repeatedly. 
For the same problem, the natural family of monotone search strategies was shown to be suboptimal, and a refinement of zig-zag trajectories was introduced, in which an increasing number of local reversals improves the competitive ratio~\cite{BGMP2022pfaulty}.
This phenomenon is relevant
to the multiple-peak trajectories studied in the present paper: when visiting
a location does not permanently rule it out, local revisitations can be
beneficial. More recently, Coleman and Morales-Ponce considered linear search
in which detection is probabilistic at one speed and guaranteed at another
\cite{ColemanMoralesPonce2025}. In our model nondetection has a different
source: every visit before the adversarially chosen emergence time fails, whereas every visit
after emergence succeeds.

For multiple agents, the same termination requirement is also studied under the name \emph{group search}. 
On the line, Chrobak et al.~\cite{ChrobakGGM15} proved that under face-to-face
communication the optimal competitive ratio remains $9$, independently of
the number of unit-speed agents. In contrast, two unit-speed agents with
wireless communication admit the classical competitive ratio $3$
\cite{BS95,GLweightedLine2023}. The effect of heterogeneous speeds was studied
in \cite{BampasCGIKKP19}, while intermediate communication capabilities were
considered in \cite{CzyzowiczKKKNOPS21}. Other extensions on the line include
time-energy tradeoffs under both face-to-face and wireless communication
\cite{CzyzowiczGKKKLNOS21,CzyzowiczGKKKLN19}. Of particular relevance is
weighted group search on a line \cite{GLweightedLine2023}, where the
termination cost is changed from the last arrival time to a weighted average
of the agents' arrival times. That change alone leads to substantially
different optimal search trajectories and competitive ratios.

Search and evacuation by several communicating agents have also been studied
on other geometric domains. Representative examples include evacuation from
a disk \cite{CzyzowiczGGKMP14} and from multiple rays
\cite{BrandtFRW20}; a broader account of group search and evacuation and the
wireless and face-to-face communication models appears in
\cite{CGK19search}. These problems illustrate the role played by
communication in coordinating several searchers. In the present work we
assume the stronger wireless model, so that the additional difficulty comes
from the target's emergence rather than from communicating its location once
it has been detected.

A different source of uncertainty, and one that is particularly relevant
to our work, arises when the state of the search environment changes with
time. A closely related recent problem is the eventually-emerging
black-hole search problem of Bonnet, Bramas, and Lamani~\cite{BonnetBL26},
where a node of a ring may become a black hole at an unknown time during
the execution. Thus, a node visited safely in the past cannot be regarded
as permanently safe. Bonnet et al. relate this model to earlier gray- and
Byzantine-hole variants, in which visiting a dangerous node need not always
destroy the agent. Eventually Emerging Linear Search has a similar loss of
information from earlier visits, since visiting a point before the target emerges
does not eliminate that point as a possible target location later. However, the two
problems have different objectives and domains.

Finally, linear programming has previously been used to derive search lower
bounds, for example for linear search with turn costs
\cite{demaine2006online}, and more recently for weighted group search and
priority evacuation on the disk \cite{GeorgiouWang2025}. The lower-bound
arguments developed in this paper require a different finite formulation
that accounts for the possible visitation orders together with the unknown
emergence time.

\subsection{Results, Significance \& Paper Organization}

Eventually Emerging Linear Search introduces a temporal component into
classical linear search while otherwise keeping the underlying geometry
deliberately simple. An input is no longer determined only by the unknown
target location, but also by the unknown time at which that location
becomes detectable. Moreover, the emergence itself is not announced to the
agents. Consequently, information obtained through search is no longer
permanent: visiting a point rules it out only up to the time of that visit,
and previously searched parts of the line may have to be searched again.
The two competitive measures studied here separate two different forms of offline information, according to whether the target location is revealed only when the target emerges or is known from the beginning.
In this sense, the model adds a genuinely temporal source of online uncertainty to one of the most basic search problems.

Our results show that this additional uncertainty changes the problem in a
nontrivial way. 
For one agent, it does not change the optimal competitive
ratio (the classical cow-path strategy remains optimal under the
emergence-aware and fully-informed measures).
The situation changes already with two agents. Although we
work in the wireless model, where communication after detection is
instantaneous, Eventually Emerging Linear Search is provably harder than
classical two-agent wireless evacuation on the line. Thus the main
difficulty is not communicating the discovery of the target, but
coordinating the search before detection, when agents must balance the
exploration of new locations against renewed visits to locations searched
earlier. At the same time, two agents improve strictly over one, and our
analysis completely determines the optimal performance within the natural
class of origin-separated algorithms.

Adding a third agent produces another change. We prove that three agents
strictly outperform two under both competitive measures. A natural
geometric extension of the two-agent strategy, in which each side agent
uses one outer turning point before returning inward, is not sufficient to
establish this separation from our two-agent lower bounds, which apply to arbitrary algorithms.
We therefore refine the trajectories by introducing additional outer
turning points between successive scales. The resulting three-peak
algorithm gives the required improvement. This phenomenon is reminiscent
of $p$-faulty search~\cite{BGMP2022pfaulty}, where local non-monotone
excursions improve upon the natural monotone search strategy. 
In both problems, a visit does not permanently resolve a location, and
it can be advantageous to trade some expansion of the searched region
for carefully scheduled revisitations. 
Interestingly, the two competitive measures admit
the same optimal value within the two-agent origin-separated family,
whereas our three-agent upper bounds require different parameter choices
and give different guarantees.

The analysis also develops tools for handling the temporal component of
the problem. 
For the geometrically scaled trajectories used in our upper bounds,
scaling and reflection reduce the continuum of possible target
positions to a bounded interval.
More importantly, for a fixed target location the evacuation time remains
unchanged while the emergence time varies between two consecutive visits
to that location. This allows the continuum of possible emergence times to
be replaced by a merged visitation timeline. The resulting reduction is
used throughout our upper-bound analyses and is particularly useful for the
more complicated three-agent trajectories.

Our lower bounds require different ideas. For two agents we develop a
computer-assisted framework that applies to arbitrary trajectories. We
select finitely many target locations, describe necessary conditions on
their first and later visits by linear inequalities, and systematically
consider the relevant discrete visitation choices. The resulting linear
programs are used only as necessary conditions: if every possible algorithmic choice
leads to an empty feasible region, no search algorithm with the proposed
competitive ratio can exist. Numerical computation is used to locate useful
target placements and candidate infeasible systems, but the final proof is
exact. All exclusions are certified by rational Farkas certificates and
checked by an independent verifier. Thus the computational component
reduces a lower bound over arbitrary continuous trajectories to a finite,
reproducible certificate rather than relying on floating-point solver
output.

For three agents we obtain unconditional lower bounds analytically. The
argument first shows that a sufficiently competitive algorithm must force
different agents far into opposite directions over suitable time
intervals. We then compare two target placements having the same delayed
emergence time. Since two agents are committed on opposite sides, 
the remaining agent cannot detect both targets sufficiently quickly.
This gives a direct adversarial use of the emergence time and leads to
different lower-bound conditions for the two competitive measures. Together
with the computer-assisted two-agent argument, these results provide complementary approaches for lower bounds in a search problem where both
space and time are unknown.

Section~\ref{sec:preliminaries} formally defines the problem, the two
competitive measures, and the normalization tools used in the analysis.
Sections~\ref{sec: 1-agent evacuation} and~\ref{sec: 2-agent evacuation}
treat one and two agents, respectively, while
Section~\ref{sec: 3-agent evacuation} develops the single- and three-peak
three-agent algorithms together with the corresponding lower bounds. The
final section discusses the remaining questions suggested by these results.

\section{Preliminaries}
\label{sec:preliminaries}

In this section we formally define Eventually Emerging Linear Search and the two competitive measures studied in this paper, quantify our main results, and establish the basic observations and normalizations that will be used in the subsequent algorithmic and lower-bound arguments.

\subsection{Problem Definition \& Results}
\label{sec:ProblemDefinition-results}

We study the \emph{$n$-agent Eventually Emerging Linear Search} problem (\eels{n}). 
We consider deterministic algorithms throughout.
In this online problem, $n \geq 1$ mobile agents start at the origin of the real line and move with maximum speed~$1$, while all turns are performed instantaneously. 
A target is located at an unknown position $x\in\mathbb{R}$, with $|x|\geq 1$ and emerges at an unknown time $T\geq 0$. 
Before time $T$, the target cannot be detected, even if an agent passes through its location. The target is detected at the first time $t\geq T$ at which some agent is located at~$x$. 
Until the target is detected, the trajectories of the agents are completely independent of both $x$ and $T$ (justifying the classification of the problem as online). For $n\geq 2$, we consider the wireless communication model. Thus, immediately upon detection, the location of the target becomes known to all agents. The emergence time $T$ is never revealed to them. From the detection time onward, without loss of generality, every agent moves directly to the target at maximum speed. 

An algorithm for \eels{n} is \emph{feasible} if, for every $|x|\geq1$ and $T\geq0$, at least one agent visits $x$ at some time $t\geq T$. 
The algorithm terminates at the first time when every agent is located at the target after its emergence. For a fixed algorithm, we denote this time by $E(x,T)$, and call it the \emph{$(x,T)$-evacuation time}. 
Observe that when $n=1$, this reduces to the search problem, and $E(x,T)$ is simply the detection time of the target.
We next define two notions of competitive ratio, corresponding to two offline benchmarks that differ in when the target location $x$ becomes known. 

Suppose that an (optimal) offline algorithm learns the target location $x$ only when the target emerges at time $T$. Before then, an optimal strategy keeps all agents at the origin. Indeed, if some agent is away from the origin, the target may lie on the opposite side. Once $x$ is revealed, all agents move directly to it. Therefore, the optimal completion time is $T+|x|$. We call such an offline algorithm emergence-aware. 
\begin{definition}
The \emph{emergence-aware competitive ratio} of an online algorithm $\mathcal A$ is
\begin{equation*}
\mathrm{CR}_{\mathrm{ea}}(\mathcal A)
=
\sup_{|x|\geq 1,\,T\geq 0}
\frac{E(x,T)}{T+|x|}
\end{equation*}
\end{definition}
Suppose now that an (optimal) offline algorithm knows the target location $x$ from the beginning, and can move directly there. Since the target cannot be detected before it emerges, the earliest possible completion time is $\max\{|x|,T\}$. We call such an offline algorithm fully-informed. 
\begin{definition}
The \emph{fully-informed competitive ratio} of an online algorithm $\mathcal A$ is
\begin{equation*}
\mathrm{CR}_{\mathrm{fi}}(\mathcal A)
=
\sup_{|x|\geq 1,\,T\geq 0}
\frac{E(x,T)}
{\max\{|x|,T\}}.
\end{equation*}
\end{definition}

Consequently, taking the infimum over all feasible algorithms for \eels{n}, we define
$$
\begin{aligned}
\eelsfi{n} &= \inf_{\mathcal A}\mathrm{CR}_{\mathrm{fi}}(\mathcal A),\\
\eelsea{n} &= \inf_{\mathcal A}\mathrm{CR}_{\mathrm{ea}}(\mathcal A).
\end{aligned}
$$
With that notation in mind, our main results are summarized in the following table.
\begin{center}
\begin{tabular}{c|cc|cc|c}
& \multicolumn{2}{c|}{Fully-informed \eelsfi{n}}
& \multicolumn{2}{c|}{Emergence-aware \eelsea{n}}
& \\
$n$
& Lower bound & Upper bound
& Lower bound & Upper bound
& Reference\\
\hline
$1$
& $9$ & $9$
& $9$ & $9$
& Theorem~\ref{thm: opt 1 agent search}, Section~\ref{sec: 1-agent evacuation}\\
$2$
& $4.38$ & $3+\sqrt6$
& $4.35$ & $3+\sqrt6$
& Theorem~\ref{thm: all eels2 results},  Section~\ref{sec: 2-agent evacuation}\\
$3$
& $3.2823$ & $4.340445$
& $3.10142$ & $4.010599$
& Theorem~\ref{thm: 3-agent results in 3-agent section}, Section~\ref{sec: 3-agent evacuation}
\end{tabular}
\end{center}
For any number of agents, both optimal competitive ratios are at least $3$, as follows by considering an instance with emergence time $T=0$.
For $n=2$, the upper bound $3+\sqrt6 \approx 5.44949$ is also optimal, under both measures, within the origin-separated family defined below.
In particular, the results imply
$
\eelsfi{3}<\eelsfi{2}<\eelsfi{1}=9
$
and
$
\eelsea{3}<\eelsea{2}<\eelsea{1}=9.
$

\subsection{Notation \& Useful Remarks}
\label{sec:notation}

For ease of reference, we collect here notation that will be used throughout the paper.
The $n$ agents of problem \eels{n} are denoted by $A_0,\ldots,A_{n-1}$. For a fixed online algorithm $\mathcal A$, we write $p_i(t)$ for the position of agent $A_i$ at time $t$ in the execution in which no target is detected. Since the agents receive no information before detection, every actual execution agrees with these trajectories up to its detection time.
An instance of the problem is denoted by $(x,T)$, where $x$ is the target position and $T$ is its emergence time. We write $E_{\mathcal A} (x,T)$ for the evacuation time of algorithm $\mathcal A$, or simply $E(x,T)$ when the underlying algorithm is clear from the context.

We use $\mathrm{CR}_{\mathrm{fi}}(\mathcal A)$ and $\mathrm{CR}_{\mathrm{ea}}(\mathcal A)$ for the fully-informed and emergence-aware competitive ratios of algorithm $\mathcal A$, respectively, and $\eelsfi{n}$ and $\eelsea{n}$ for the corresponding optimal competitive ratios for \eels{n}.
Finally, for a fixed target position $x$, we denote by $v_j(x)$ the time of the $j$-th visitation of $x$ by any of the agents, and set $v_0(x)=0$. If more than one agent visits $x$ at the same time, this counts as a single visitation.

For every instance $(x,T)$ we have $\max\{|x|,T\}\leq |x|+T$. Consequently, for every algorithm $\mathcal A$,
$
\mathrm{CR}_{\mathrm{fi}}(\mathcal A)\geq
\mathrm{CR}_{\mathrm{ea}}(\mathcal A),
$
and hence
$
\eelsfi{n}\geq\eelsea{n}.
$

We describe the agents' trajectories only until the target is detected, since thereafter all agents move directly to the target at maximum speed. The trajectories used in our upper bounds are specified by sequences of turning points. For convenience, we index these sequences by $\mathbb Z$ rather than by $\mathbb N$. The negative-index turning points converge to the origin, while the total time required to traverse them is finite. Thus they define a continuous unit-speed trajectory starting at the origin, although infinitely many turns may accumulate at time $0$.

This convention is only notational. If one requires every bounded time interval to contain only finitely many turns, fix a sufficiently small time $t_*>0$ such that all trajectories before $t_*$ remain in $(-1,1)$. Since $|p_i(t_*)|\leq t_*$, each agent can instead wait at the origin and then move at unit speed to $p_i(t_*)$, arriving there at time $t_*$. From that time onward it follows the original trajectory. Since every valid target satisfies $|x|\geq1$, no target can be detected before $t_*$, and therefore this modification does not change the detection or evacuation time of any instance.

\subsection{Normalizing Detection and Emergence Times}
The upper-bound algorithms considered in this paper have scaling and reflection symmetries that allow us to simplify the range of instances that needs to be analyzed. In this section we formalize two reductions. First, we restrict the target position to a bounded interval, and second, for every fixed target position, we reduce the possible emergence times to the times between consecutive visits to the target.
We begin with the reduction of the target position, for which we use the following property.

\begin{definition}[$\lambda$-normalizable algorithm]
\label{def:lambda-normalizable}
For $\lambda>1$, we say that an algorithm $\mathcal A$ for \eels{n} is \emph{$\lambda$-normalizable} if the following properties hold: \\
- The trajectories are jointly invariant under scaling by $\lambda$, that is,
$
p_j(\lambda t)=\lambda p_j(t)
$
for every $t\geq0$ and every $j\in\{0,\ldots,n-1\}$. \\
- There exist a constant $\mu>0$ and a permutation $\pi$ of $\{0,\ldots,n-1\}$ such that
$
p_{\pi(j)}(\mu t)=-\mu p_j(t)
$
for every $t\geq0$ and every $j\in\{0,\ldots,n-1\}$. 
\end{definition}

The following lemma shows that for a $\lambda$-normalizable algorithm it is enough to consider positive target positions in the interval $(1,\lambda]$.
\begin{lemma}
\label{lem:normalization-reduction}
Let $\mathcal A$ be a $\lambda$-normalizable algorithm for \eels{n}. Then
\begin{equation*}
\begin{aligned}
\mathrm{CR}_{\mathrm{ea}}(\mathcal A)
&=
\sup_{\substack{x\in(1,\lambda]\\T\geq0}}
\frac{E(x,T)}{T+|x|},\\
\mathrm{CR}_{\mathrm{fi}}(\mathcal A)
&=
\sup_{\substack{x\in(1,\lambda]\\T\geq0}}
\frac{E(x,T)}
{\max\{|x|,T\}}.
\end{aligned}
\end{equation*}
\end{lemma}

\begin{proof}
We first show that both competitive ratios are preserved under scaling by $\lambda$.
Indeed, for every integer $m$, every $t\geq0$, and every agent $j$, we have
$
p_j(\lambda^m t)=\lambda^m p_j(t)
$.
Now, 
fix a target position $x$ and an emergence time $T$, and suppose that the target is detected at time $\tau$. The target at $\lambda^m x$ with emergence time $\lambda^mT$ is detected at time $\lambda^m\tau$. Moreover, at that time agent $j$ is at position
$
p_j(\lambda^m\tau)
=
\lambda^m p_j(\tau)
$.
Hence its remaining distance from the target is
$
\left|
p_j(\lambda^m\tau)-\lambda^m x
\right|
=
\lambda^m
\left|
p_j(\tau)-x
\right|
$.
Since every agent moves directly to the target after detection,
$
E(\lambda^m x,\lambda^mT)
=
\lambda^mE(x,T)
$.
Therefore,
$$
\begin{aligned}
\frac{E(\lambda^m x,\lambda^mT)}
{\lambda^mT+|\lambda^m x|}
&=
\frac{E(x,T)}
{T+|x|},\\
\frac{E(\lambda^m x,\lambda^mT)}
{\max\{|\lambda^m x|,\lambda^mT\}}
&=
\frac{E(x,T)}
{\max\{|x|,T\}}.
\end{aligned}
$$
For every fixed integer $m$, the map
$
T\mapsto\lambda^mT
$
is a bijection from $[0,\infty)$ onto itself. Hence
$$
\begin{aligned}
\sup_{T\geq0}
\frac{E(\lambda^m x,T)}
{T+|\lambda^m x|}
&=
\sup_{T\geq0}
\frac{E(x,T)}
{T+|x|},\\
\sup_{T\geq0}
\frac{E(\lambda^m x,T)}
{\max\{|\lambda^m x|,T\}}
&=
\sup_{T\geq0}
\frac{E(x,T)}
{\max\{|x|,T\}}.
\end{aligned}
$$We next show that negative targets can be replaced by positive targets. By the second property in Definition~\ref{def:lambda-normalizable},
$
p_{\pi(j)}(\mu t)=-\mu p_j(t)
$
for every agent $j$ and every $t\geq0$. Therefore, the execution for the target $-\mu x$ with emergence time $\mu T$ is obtained from the execution for the target $x$ with emergence time $T$ by reflecting the line, multiplying positions and times by $\mu$, and permuting the agents according to $\pi$. Since the evacuation time is the maximum over all agents, the permutation does not affect it. Thus
$E(-\mu x,\mu T)
=
\mu E(x,T)
$.
Consequently,
$$
\begin{aligned}
\frac{E(-\mu x,\mu T)}
{\mu T+|-\mu x|}
&=
\frac{E(x,T)}
{T+|x|},\\
\frac{E(-\mu x,\mu T)}
{\max\{|-\mu x|,\mu T\}}
&=
\frac{E(x,T)}
{\max\{|x|,T\}}.
\end{aligned}
$$
Consider now an arbitrary target position $x$ with $|x|\geq1$ and an arbitrary emergence time $T\geq0$. If $x>0$, let $m$ be the unique integer such that
$
1<\lambda^{-m}x\leq\lambda,
$
and set
$
y=\lambda^{-m}x,
\qquad
S=\lambda^{-m}T.
$
By scaling, the instances $(x,T)$ and $(y,S)$ have the same competitive ratios.

If $x<0$, let $m$ be the unique integer such that
$
1<-\frac{\lambda^{-m}x}{\mu}\leq\lambda,
$
and set
$
y=-\frac{\lambda^{-m}x}{\mu},
\qquad
S=\frac{\lambda^{-m}T}{\mu}.
$
Combining scaling and reflection, the instances $(x,T)$ and $(y,S)$ again have the same competitive ratios.
In both cases, $y\in(1,\lambda]$.
Thus every instance is equivalent, for both competitive ratios, to an instance with
$
x\in(1,\lambda]
$
and
$
T\geq0
$, as wanted. 
This proves that each global supremum is at most the corresponding supremum over the normalized instances. The reverse inequalities are immediate because every instance with
$
x\in(1,\lambda]
$
and
$
T\geq0
$
is possible.
\end{proof}

We next reduce the range of emergence times that needs to be considered. 
Fix a target position $x>0$, and suppose that the distinct times at which some agent visits $x$ form an unbounded discrete sequence
$
v_1(x)<v_2(x)<\cdots.
$
We refer to this sequence as the \emph{merged visitation timeline} of $x$, and set $v_0(x)=0$ for convenience.
For every $k\geq1$, we introduce notation
\begin{equation*}
M_k(x)
:=
v_k(x)
+
\max_{j\in\{0,1,\ldots,n-1\}}
\left|p_j(v_k(x))-x\right|.
\end{equation*}

\begin{lemma}
\label{lem:merged-visitation-timeline}
For a target position $x>0$ with an unbounded discrete merged visitation timeline, we have
$$
\begin{aligned}
\sup_{T\geq0}
\frac{E(x,T)}{\max\{T,x\}}
&=
\sup_{k\geq1}
\frac{M_k(x)}
{\max\{v_{k-1}(x),x\}},\\
\sup_{T\geq0}
\frac{E(x,T)}{T+x}
&=
\sup_{k\geq1}
\frac{M_k(x)}
{v_{k-1}(x)+x}.
\end{aligned}
$$
\end{lemma}

\begin{proof}
Fix $x$. If $T=0$, the target is detected at time $v_1(x)$. If $T>0$, there exists a unique $k\geq1$ such that
$
v_{k-1}(x)<T\leq v_k(x),
$
and the target is detected at time $v_k(x)$.
After detection, every agent moves directly to the target at maximum speed. 
Hence,
\begin{equation*}
E(x,T)
=
v_k(x)
+
\max_{j\in\{0,1,\ldots,n-1\}}
|p_j(v_k(x))-x|.
\end{equation*}
For every fixed $k$, the evacuation time is independent of $T$ between two consecutive visits, while both $\max\{T,x\}$ and $T+x$ are nondecreasing in $T$. For $k=1$, the supremum is obtained at $T=0$, while for $k\geq2$ it is obtained as
$
T\rightarrow v_{k-1}(x)^+.
$
The claim follows.
\end{proof}

Combining the two preceding reductions gives the form that we will use throughout our upper-bound analyses.
\begin{corollary}
\label{cor:normalized-visitation-reduction}
Let $\mathcal A$ be a $\lambda$-normalizable algorithm for \eels{n}, 
and suppose that every $x\in(1,\lambda]$ has an unbounded discrete merged visitation timeline. 
Then
\begin{equation*}
\begin{aligned}
\mathrm{CR}_{\mathrm{fi}}(\mathcal A)
&=
\sup_{\substack{x\in(1,\lambda]\\k\geq1}}
\frac{M_k(x)}
{\max\{v_{k-1}(x),x\}},\\
\mathrm{CR}_{\mathrm{ea}}(\mathcal A)
&=
\sup_{\substack{x\in(1,\lambda]\\k\geq1}}
\frac{M_k(x)}
{v_{k-1}(x)+x}.
\end{aligned}
\end{equation*}
\end{corollary}

\section{1-Agent Search - \eels1}
\label{sec: 1-agent evacuation}
In this section we prove the following theorem.

\begin{theorem}
\label{thm: opt 1 agent search}
$\eelsfi{1}=\eelsea{1}=9$.
\end{theorem}

For the lower bound argument, note that the adversary can be restricted to choose emergence time $T=0$. In that case \eels{1} reduces to the well-known linear-search problem (cow path), whose competitive ratio is at least $9$~\cite{baezayates1993searching}.
Hence,
$
\eelsfi{1}\geq \eelsea{1}\geq 9.
$
Therefore, Theorem~\ref{thm: opt 1 agent search} follows if we show that $\eelsfi{1}\leq9$.
To describe the algorithm that achieves this bound, we introduce the following trajectory, which will also be used later in the paper.

\begin{definition}[Geometric zig-zag trajectory]
\label{def:geometric-zig-zag}
For $b>1$, let $\mathcal Z_b$ denote the unit-speed trajectory that successively visits the turning points
$(-1)^i b^i, i\in\mathbb Z$
in increasing order of $i$.
\end{definition}

We prove the upper bound more generally for $\mathcal Z_b$. The proof also demonstrates the analysis framework implied by Corollary~\ref{cor:normalized-visitation-reduction}. We note that the proof of Theorem~\ref{thm: opt 1 agent search} follows by Lemma~\ref{lem: upper bound 1-agent fi} below, using $b=2$. 

\begin{lemma}
\label{lem: upper bound 1-agent fi}
For every $b>1$, the algorithm $\mathcal A$ for problem \eels{1} in which $A_0$ follows $\mathcal Z_b$ satisfies
$
\mathrm{CR}_{\mathrm{fi}}(\mathcal A)
\leq
1+\frac{2b^2}{b-1}.
$
\end{lemma}

\begin{proof}
Let $p(t)$ denote the position of the agent before detection. Since
$
p(b^2t)=b^2p(t)
$
and
$
p(bt)=-bp(t)
$
for every $t\geq0$, the algorithm is $b^2$-normalizable.
Therefore, by Lemma~\ref{lem:normalization-reduction}, it is enough to consider target placements
$
1<x\leq b^2.
$

First assume $1<x<b^2$. For every $j\geq0$, the odd visitation of $x$ during the excursion to $b^{2j+2}$ occurs at time
\begin{equation*}
v_{2j+1}(x)
=
2\sum_{i=-\infty}^{2j+1}b^i+x
=
\frac{2b^{2j+2}}{b-1}+x,
\end{equation*}
while the following even visitation occurs on the return from $b^{2j+2}$, at time
\begin{equation*}
v_{2j+2}(x)
=
\frac{2b^{2j+3}}{b-1}-x.
\end{equation*}
Since there is only one agent, it is at the target at every visitation. Therefore, using the notation of Corollary~\ref{cor:normalized-visitation-reduction}, we have 
$
M_k(x)=v_k(x)
$,
for every $k\geq1$.
Next, we distinguish three visitation families.
\begin{description}
\item[Competitive ratio for $M_1(x)$:]
Since $v_0(x)=0$,
\begin{equation*}
\frac{M_1(x)}
{\max\{v_0(x),x\}}
=
\frac{v_1(x)}{x}
=
1+\frac{2b^2}{(b-1)x}
<
1+\frac{2b^2}{b-1}.
\end{equation*}

\item[Competitive ratios for $M_{2j+2}(x)$:]
For every $j\geq0$, we have
$
v_{2j+1}(x)>x
$.
Hence,
\begin{equation*}
\frac{M_{2j+2}(x)}
{\max\{v_{2j+1}(x),x\}}
=
\frac{v_{2j+2}(x)}
{v_{2j+1}(x)}
=
\frac{
\frac{2b^{2j+3}}{b-1}-x
}{
\frac{2b^{2j+2}}{b-1}+x
}
<
b.
\end{equation*}

\item[Competitive ratios for $M_{2j+3}(x)$:]
For every $j\geq0$, we have
$
v_{2j+2}(x)>x
$.
Hence,
\begin{equation*}
\frac{M_{2j+3}(x)}
{\max\{v_{2j+2}(x),x\}}
=
\frac{v_{2j+3}(x)}
{v_{2j+2}(x)}
=
\frac{
\frac{2b^{2j+4}}{b-1}+x
}{
\frac{2b^{2j+3}}{b-1}-x
}.
\end{equation*}
The last expression is increasing in $x$, and therefore
\begin{equation*}
\frac{M_{2j+3}(x)}
{\max\{v_{2j+2}(x),x\}}
\leq
\frac{
\frac{2b^{2j+4}}{b-1}+b^2
}{
\frac{2b^{2j+3}}{b-1}-b^2
}
=
\frac{2b^{2j+2}+b-1}
{2b^{2j+1}-b+1}
\leq
\frac{2b^2+b-1}{b+1}
=
2b-1, 
\end{equation*}
where the last inequality follows because the preceding expression is decreasing in $j$.
\end{description}
For $x=b^2$, the first outward and return visits coincide at the turning point $b^2$. Hence they count as one visitation in the merged visitation timeline. Removing this duplicate only removes the corresponding ratio equal to $1$, while the denominator for the next visitation remains unchanged. Therefore, all the bounds above also hold for $x=b^2$.
Therefore, by Corollary~\ref{cor:normalized-visitation-reduction} we have 
\begin{equation*}
\mathrm{CR}_{\mathrm{fi}}(\mathcal A)
=
\sup_{\substack{1<x\leq b^2\\k\geq1}}
\frac{M_k(x)}
{\max\{v_{k-1}(x),x\}}
\leq
\max\left\{
1+\frac{2b^2}{b-1},
b,
2b-1
\right\}
=
1+\frac{2b^2}{b-1}.
\end{equation*}

\end{proof}


\section{2-Agent Evacuation - \eels2}
\label{sec: 2-agent evacuation}

In this section we prove upper and lower bounds for \eels{2}, for both competitive ratio measures. We first introduce a family of trajectories that will also be used for \eels{3}.

\begin{definition}[One-sided zig-zag trajectory]
\label{def:one-sided-zig-zag}
Let $\ell=(\ell_i)_{i\in\mathbb Z}$ and $u=(u_i)_{i\in\mathbb Z}$ be sequences of nonnegative real numbers such that $0\leq \ell_i<u_i$ and $\ell_{i+1}<u_i$ for every $i\in\mathbb Z$. We further assume that $\ell_i,u_i\rightarrow0$ as $i\rightarrow-\infty$ and that
\[
\sum_{i=-\infty}^{0}\bigl((u_i-\ell_i)+(u_i-\ell_{i+1})\bigr)<\infty.
\]
We denote by $\mathcal S_{\ell,u}$ the unit-speed trajectory on the positive half-line whose successive turning points are
$\ldots,\ell_i,u_i,\ell_{i+1},u_{i+1},\ldots$.
We call $\ell_i$ and $u_i$ the inner and outer turning points, respectively. If $\ell_i=0$ for every $i\in\mathbb Z$, we simply write $\mathcal S_u$.
\end{definition}

We call an algorithm for \eels{2} \emph{origin-separated} if agents $A_0$ and $A_1$ follow trajectories $-\mathcal S_x$ and $\mathcal S_y$, respectively, where $x=(x_i)_{i\in\mathbb Z}$ and $y=(y_i)_{i\in\mathbb Z}$ are strictly increasing sequences of positive real numbers satisfying $\sum_{i=-\infty}^{0}(x_i+y_i)<\infty$ and $x_i,y_i\rightarrow\infty$ as $i\rightarrow\infty$. Thus, each agent remains on a fixed half-line and returns to the origin after every excursion. In particular, the trajectories of the two agents intersect only at the origin.
We are now ready to summarize the results of this section.
\begin{theorem}
\label{thm: all eels2 results}
There exists an origin-separated algorithm $\mathcal A$ for \eels{2} such that $\mathrm{CR}_{\mathrm{fi}}(\mathcal A),\mathrm{CR}_{\mathrm{ea}}(\mathcal A)\leq3+\sqrt6$, and this bound is optimal, for both measures, among origin-separated algorithms. Moreover, unconditionally, $\eelsfi{2}\geq4.38$ and $\eelsea{2}\geq4.35$.
\end{theorem}
The upper bound in Theorem~\ref{thm: all eels2 results} is proved in Section~\ref{sec: upper bounds for 2 agents}, its optimality within the origin-separated family is proved in Section~\ref{sec: lower bounds for 2 agents}, and the unconditional lower bounds are proved in Section~\ref{sec:eels2-finite-lower-bound}.

\subsection{Upper bounds to \eelsfi{2}, \eelsea{2}}
\label{sec: upper bounds for 2 agents}

In this section we prove the upper bound statement of Theorem~\ref{thm: all eels2 results}, which we restate for completeness.
\begin{theorem}
\label{thm: 2 agent search upper}
There exists an origin-separated algorithm $\mathcal A$ for \eels{2} such that
$\mathrm{CR}_{\mathrm{fi}}(\mathcal A),\mathrm{CR}_{\mathrm{ea}}(\mathcal A)\leq 3+\sqrt6$.
\end{theorem}

To prove Theorem~\ref{thm: 2 agent search upper}, we consider origin-separated algorithms whose one-sided zig-zag trajectories have geometrically increasing outer turning points.
Theorem~\ref{thm: 2 agent search upper} follows by the lemma below, using $b=3+\sqrt6$, a value identified as the unique root more than $1$ to equation $3+\frac{2b}{b-1}=b$. 
\begin{lemma}
\label{lem: upper bound 2-agent fi}
For every $b>1$, let $u=(u_i)_{i\in\mathbb Z}$, where $u_i=b^i$. The origin-separated algorithm $\mathcal A$ in which $A_0$ follows $-\mathcal S_u$ and $A_1$ follows $\mathcal S_u$ satisfies
$
\mathrm{CR}_{\mathrm{fi}}(\mathcal A)
\leq
\max\left\{
3+\frac{2b}{b-1},b
\right\}.
$
\end{lemma}

\begin{proof}
Since
$
p_0(bt)=bp_0(t)
$
and
$
p_1(bt)=bp_1(t)
$
for every $t\geq0$, and
$
p_1(t)=-p_0(t),
$
the algorithm is $b$-normalizable.
Therefore, by Lemma~\ref{lem:normalization-reduction}, it is enough to consider target placements
$
1<x\leq b.
$
For every $j\geq0$, set
\begin{equation*}
S_j
=
2\sum_{i=-\infty}^{j}b^i
=
\frac{2b^{j+1}}{b-1}.
\end{equation*}
First assume $1<x<b$. The odd visitation of $x$ during the excursion to $b^{j+1}$ occurs at time
$
v_{2j+1}(x)
=
S_j+x.
$
The following even visitation occurs on the return from $b^{j+1}$, at time
$
v_{2j+2}(x)
=
S_j+x+2(b^{j+1}-x)
=
bS_j-x.
$

Since $x>0$, only agent $A_1$ visits the target. Moreover, whenever $A_1$ is at $x$, agent $A_0$ is at $-x$. Therefore, using the notation of Corollary~\ref{cor:normalized-visitation-reduction},
$
M_k(x)
=
v_k(x)+2x,
$
for every $k\geq1$. In particular,
\begin{equation*}
M_{2j+1}(x)
=
S_j+3x,
\qquad
M_{2j+2}(x)
=
bS_j+x.
\end{equation*}

Next, we distinguish three visitation families.

\begin{description}

\item[Competitive ratio for $M_1(x)$:]
Since $v_0(x)=0$,
\begin{equation*}
\frac{M_1(x)}
{\max\{v_0(x),x\}}
=
\frac{S_0+3x}{x}
=
3+\frac{2b}{(b-1)x}
<
3+\frac{2b}{b-1}.
\end{equation*}

\item[Competitive ratios for $M_{2j+2}(x)$:]
For every $j\geq0$, we have
$
v_{2j+1}(x)>x.
$
Hence,
\begin{equation*}
\frac{M_{2j+2}(x)}
{\max\{v_{2j+1}(x),x\}}
=
\frac{bS_j+x}{S_j+x}
<
b.
\end{equation*}

\item[Competitive ratios for $M_{2j+3}(x)$:]
For every $j\geq0$, we have
$
v_{2j+2}(x)>x.
$
Hence,
$$
\frac{M_{2j+3}(x)}
{\max\{v_{2j+2}(x),x\}}
=
\frac{bS_j+3x}{bS_j-x}.
$$
The last expression is increasing in $x$, and therefore
\begin{equation*}
\frac{M_{2j+3}(x)}
{\max\{v_{2j+2}(x),x\}}
\leq
\frac{bS_j+3b}{bS_j-b}
=
\frac{2b^{j+1}+3b-3}
{2b^{j+1}-b+1}
\leq
\frac{5b-3}{b+1}
<
\frac{5b-3}{b-1}
=
3+\frac{2b}{b-1}
,
\end{equation*}
where the second to last inequality follows because the preceding expression is decreasing in $j$. 
\end{description}
For $x=b$, the first outward and return visits coincide at the turning point $b$, and hence count as one visitation. Removing this duplicate creates no new case: the next visitation has the same preceding visitation time as in the analysis above. Therefore, all the preceding bounds also hold for $x=b$.
Therefore, by Corollary~\ref{cor:normalized-visitation-reduction},
\begin{equation*}
\mathrm{CR}_{\mathrm{fi}}(\mathcal A)
=
\sup_{\substack{1<x\leq b\\k\geq1}}
\frac{M_k(x)}
{\max\{v_{k-1}(x),x\}}
\leq
\max\left\{
3+\frac{2b}{b-1},
b
\right\}.
\end{equation*}
\end{proof}

\subsection{Origin-Separated Lower Bound for \eels{2}}
\label{sec: lower bounds for 2 agents}

In this section we prove the lower bound within the origin-separated family claimed in Theorem~\ref{thm: all eels2 results}.
\begin{theorem}
\label{thm: 2 agent search lower bound restricted ea}
Any origin-separated algorithm $\mathcal A$ for \eels{2} satisfies
$$
\mathrm{CR}_{\mathrm{fi}}(\mathcal A)
\geq
\mathrm{CR}_{\mathrm{ea}}(\mathcal A)
\geq
3+\sqrt6.
$$
\end{theorem}

In what follows we fix an origin-separated algorithm in which $A_0$ follows $-\mathcal S_x$ and $A_1$ follows $\mathcal S_y$, where
$x=(x_i)_{i\in\mathbb Z}$ and $y=(y_i)_{i\in\mathbb Z}$ are as in Definition~\ref{def:one-sided-zig-zag}.
For every $i\in\mathbb Z$, define
\begin{equation*}
\begin{aligned}
R_i&:=2\sum_{j=-\infty}^{i-1}y_j,\\
L_i&:=2\sum_{j=-\infty}^{i-1}x_j.
\end{aligned}
\end{equation*}
Thus, $R_i$ and $R_{i+1}$ are the times at which the right agent starts and finishes its excursion to $y_i$, respectively. Similarly, $L_i$ and $L_{i+1}$ are the times at which the left agent starts and finishes its excursion to $-x_i$, respectively.

\begin{lemma}
\label{lem:opposite-origin-visit}
Consider an origin-separated search algorithm $\mathcal A$ to \eels{2} for which 
$
\mathrm{CR}_{\mathrm{ea}}(\mathcal A)<3+\sqrt6
$.
Then, for every sufficiently large $i$, during the time interval $(R_i,R_{i+1})$, the left agent is at the origin at some time. Similarly, for every sufficiently large $j$, during the time interval $(L_j,L_{j+1})$, the right agent is at the origin at some time.
\end{lemma}

\begin{proof}
 We focus on the right agent, as the claim for the left agent is identical. 
Let
$
C=\mathrm{CR}_{\mathrm{ea}}(\mathcal A).
$
First we bound $C$ with respect to excursion times $R_i$, and show that 
$$
C \geq
\limsup_{i\rightarrow\infty}\frac{R_{i+1}}{R_i}
$$
For this, consider any arbitrarily large index $i$ for which $y_i>1$. For that index $i$, we choose target placement $x=1$ and
\[
\begin{aligned}
0<\epsilon&<2(y_i-1),\\
T&=R_i+1+\epsilon.
\end{aligned}
\]
Thus, the target emerges after the right agent visits $1$ while moving from the origin toward $y_i$, and before its return visit to $1$. Therefore, the target is  detected when the right agent returns from $y_i$ and visits $1$ again, which is in time $R_i+2y_i-1=R_{i+1}-1$, while the left agent is to the left of the origin, and hence at least $1$ away from the target. 
Overall, that gives $E(x,T)\geq R_{i+1}-1+1=R_{i+1}$, while the optimal offline algorithm has termination time $x+T=R_i+2+\epsilon$. 
Letting $\epsilon \rightarrow 0$, we obtain the emergence-aware competitive ratio
$
C\geq \frac{R_{i+1}}{R_i+2}.
$
Since $R_i\rightarrow\infty$, it follows that
$
C\geq
\limsup_{i\rightarrow\infty}\frac{R_{i+1}}{R_i},
$ as wanted. 

Next, for every sufficiently large $i$, set
$
t_i=R_i+y_{i-1},
$
and let $d_i$ be the distance of the left agent from the origin at time $t_i$.
For $0<\epsilon<y_i-y_{i-1}$, consider target placement
$
x=y_{i-1}+\epsilon
$
and emergence time $T=0$. The right agent detects the target at time
$t_i+\epsilon$. Let $d_i(\epsilon)$ be the distance of the left agent from
the origin at that time. Therefore,
$$
C\geq
2+\frac{R_i+d_i(\epsilon)}{y_{i-1}+\epsilon}.
$$
By continuity of the left-agent trajectory,
$
d_i(\epsilon)\rightarrow d_i
$
as $\epsilon\rightarrow0^+$. Hence,
$$
C\geq
2+\frac{R_i+d_i}{y_{i-1}}.
$$

During the left-agent excursion that contains time $t_i$, one of the two
adjacent visits to the origin occurs at a time $H_i$ satisfying
$
|H_i-t_i|=d_i.
$
We show that $R_i<H_i<R_{i+1}$. It is enough to prove
$
R_i<t_i-d_i
$
and
$
t_i+d_i<R_{i+1},
$
since
$
t_i-d_i\leq H_i\leq t_i+d_i.
$
\emph{The justification of $R_i<t_i-d_i$:} 
From the discussion above we have that
$
C\geq
\limsup_{i\rightarrow\infty}\frac{R_{i+1}}{R_i}.
$
Therefore, together with the premise $C<3+\sqrt6$, we see that 
$
\frac{R_{i+1}}{R_i}<3+\sqrt6
$
for all sufficiently large $i$. 
Hence, we obtain the following bound on $R_i$
$$
y_{i-1}
=
\frac{R_i-R_{i-1}}{2}
<
\frac{(2+\sqrt6)R_i}{2(3+\sqrt6)}
= \frac1{\sqrt6}R_i.
$$
We also derived $
C\geq
2+\frac{R_i+d_i}{y_{i-1}},
$
which together with the premise $C<3+\sqrt6$ imply that $2+\frac{R_i+d_i}{y_{i-1}}<3+\sqrt6$. Solving for $d_i$, and using the aforementioned bound on $R_i$ we get 
$$
d_i<(1+\sqrt6)y_{i-1}-R_i <(1+\sqrt6)y_{i-1}- \sqrt6 y_{i-1} = y_{i-1}.
$$
But then, we have $t_i-d_i= R_i+y_{i-1}-d_i >R_i$, as wanted. 

\emph{The justification of $R_{i+1}>t_i+d_i$:}
From the previous derivation, we have $d_i<y_{i-1}$. Therefore,
$$
t_i+d_i
=
R_i+y_{i-1}+d_i
<
R_i+y_{i-1}+y_{i-1}
<
R_i+2y_i
=
R_{i+1},
$$
as wanted.
\end{proof}

Lemma~\ref{lem:opposite-origin-visit} implies that the two sequences of
origin times eventually interlace strictly. First, sufficiently late origin
times of the two agents cannot coincide. Indeed, if $L_j=R_i$, then the
lemma gives a left-agent origin time in $(R_i,R_{i+1})$, so
$L_{j+1}<R_{i+1}$. Hence $(L_j,L_{j+1})$ contains no right-agent origin
time, contradicting the symmetric part of the lemma.

Next, if two consecutive left-agent origin times $L_j,L_{j+1}$ belonged to
the same interval $(R_i,R_{i+1})$, then $(L_j,L_{j+1})$ would again contain
no right-agent origin time. The symmetric argument applies to two
consecutive right-agent origin times. Thus, after shifting the indices of one
sequence, we may write
$
R_i<L_i<R_{i+1}<L_{i+1}
$
for every sufficiently large $i$.
We utilize this conclusion in the proof of Theorem~\ref{thm: 2 agent search lower bound restricted ea} next.

\begin{proof}[Proof of Theorem~\ref{thm: 2 agent search lower bound restricted ea}]
Suppose, toward a contradiction, that
$
C=\mathrm{CR}_{\mathrm{ea}}(\mathcal A)<3+\sqrt6.
$
By Lemma~\ref{lem:opposite-origin-visit}, we may assume that
$
R_i<L_i<R_{i+1}<L_{i+1}
$
for every sufficiently large $i$.

As in the proof of Lemma~\ref{lem:opposite-origin-visit}, for every sufficiently large $i$, let
$
t_i=R_i+y_{i-1},
$
and let $d_i$ be the distance of the left agent from the origin at time $t_i$. By considering target placement
$
x=y_{i-1}+\epsilon
$
and emergence time $T=0$, and then letting $\epsilon\rightarrow0$, continuity of the left-agent trajectory gives
$
C\geq
2+\frac{R_i+d_i}{y_{i-1}}.
$
The proof of Lemma~\ref{lem:opposite-origin-visit} also gives a left-agent
origin time $H_i\in(R_i,R_{i+1})$ satisfying
$
|H_i-t_i|=d_i.
$
By the uniqueness established above, $H_i=L_i$. Therefore, 
$$
L_i
\leq
t_i+d_i
=
R_i+y_{i-1}+d_i
\leq
(C-1)y_{i-1}
=
\frac{C-1}{2}(R_i-R_{i-1}).
$$
Since the indices are chosen so that $R_i<L_i<R_{i+1}<L_{i+1}$, the symmetric placement of the target at $-x_{i-1}-\epsilon$ and with emergence time $T=0$ gives 
$$
R_{i+1}
\leq
\frac{C-1}{2}(L_i-L_{i-1}).
$$
Define now
$$
M=
\max\left\{
\limsup_{i\rightarrow\infty}\frac{L_i}{R_i},
\limsup_{i\rightarrow\infty}\frac{R_{i+1}}{L_i}
\right\}.
$$
The two inequalities and the interlacing imply that both ratios appearing above are eventually at most $(C-1)/2$. Hence, $M$ is finite, and in particular, due to the interlacing, we have $M\geq 1$. So, for every $\epsilon>0$, all the ratios
$
\tfrac{L_i}{R_i},
\tfrac{R_i}{L_{i-1}},
\tfrac{L_{i-1}}{R_{i-1}},
\tfrac{R_{i+1}}{L_i}
$
are eventually at most $M+\epsilon$. Therefore,
$$
\frac{L_i}{R_i}
\leq
\frac{C-1}{2}
\left(
1-\frac{R_{i-1}}{R_i}
\right)\\
=
\frac{C-1}{2}
\left(
1-
\frac{1}{
\displaystyle
\frac{R_i}{L_{i-1}}
\frac{L_{i-1}}{R_{i-1}}
}
\right)\\
\leq
\frac{C-1}{2}
\left(
1-\frac{1}{(M+\epsilon)^2}
\right),
$$
and similarly,
$$
\frac{R_{i+1}}{L_i}
\leq
\frac{C-1}{2}
\left(
1-\frac{L_{i-1}}{L_i}
\right)\\
=
\frac{C-1}{2}
\left(
1-
\frac{1}{
\displaystyle
\frac{L_i}{R_i}
\frac{R_i}{L_{i-1}}
}
\right)\\
\leq
\frac{C-1}{2}
\left(
1-\frac{1}{(M+\epsilon)^2}
\right).
$$
Taking the limit superior in both inequalities and then letting $\epsilon\rightarrow0$ gives
$
M
\leq
\frac{C-1}{2}
\left(
1-\frac{1}{M^2}
\right).
$
The case $M=1$ is impossible as otherwise $M\leq 0$. We conclude that $M>1$, and so we can write equivalently that 
$$
\frac{C-1}{2}
\geq
\frac{M^3}{M^2-1}
\geq \frac{3\sqrt3}{2},
$$
where the last inequality follows by minimizing function $f(x)= x^3/(x^2-1)$ over $x>1$. 
Then, solving for $C$ gives 
$
C\geq1+3\sqrt3 \approx 6.19615
$,
a contradiction to the assumption that $C<3+\sqrt6 \approx 5.44949$. 

\end{proof}

\subsection{\eels{2} Computer-Assisted Unconditional Lower Bounds}
\label{sec:eels2-finite-lower-bound}

In this section we prove the unconditional lower bounds of Theorem~\ref{thm: all eels2 results}, which we restate for completeness.
\begin{theorem}
\label{thm: eels2 unconditional lb}
We have $\eelsfi{2}\geq219/50=4.38$ and $\eelsea{2}\geq87/20=4.35$.
\end{theorem}

We prove these bounds by fixing a proposed competitive ratio $C$ and finitely many target positions. An arbitrary algorithm does not give rise to a single finite LP that can be written down in advance, since the constraints depend on the order and manner in which these target positions are visited. We therefore enumerate the relevant visitation patterns and, for each one, derive linear constraints that every $C$-competitive algorithm realizing that pattern must satisfy. The variables of the resulting LP are selected visitation times and the positions of the two agents at those times. The formulation is a relaxation: every $C$-competitive algorithm induces a feasible solution to at least one of the LPs that we consider, while a feasible LP solution need not correspond to a feasible algorithm. Consequently, proving all relevant LPs infeasible is sufficient for a lower bound.

The numerical values of the visitation times are never enumerated. What must be considered discretely is which target positions are visited first, which agent makes a selected later visit, and where that visit occurs relative to the first-visitation events. Emergence times are not taken from a time grid either. For delayed targets we choose the emergence time immediately after a visitation time supplied by the hypothetical algorithm, and then let the delay tend to zero. We first illustrate the construction on one complete LP, and then give the general formulation and the finite enumeration used in the computer-assisted proof.

Throughout this section, $p_0(t)$ and $p_1(t)$ denote the left and right positions, respectively. This does not restrict the algorithms under consideration, since replacing the two trajectories by their pointwise minimum and maximum preserves the speed bound and the set of positions occupied at every time. Thus, if detection occurs at time $t$, then
\begin{equation}
\label{eq:eels2-detection-spread}
E(x,T)\geq t+p_1(t)-p_0(t),
\end{equation}
because one of the two agents is at the target and the other must still reach it.

\subsubsection{A complete LP example}

Consider target positions $\{-3,-2,-1,1,2,3\}$, fix $C$, and suppose that their first-visitation order is
$$
1,-1,2,-2,3,-3.
$$
Write the corresponding first-visitation times as $t_1,\ldots,t_6$. We also consider delayed targets at $1$. Suppose that, around the time $H_1=t_4$ by which $-2$ and $2$ have both been visited, the consecutive selected visits to $1$ occur at times $P_1,Q_1$, with
$$
t_2\leq P_1\leq t_3\leq t_4\leq Q_1\leq t_5.
$$
Suppose further that these two visits no longer surround $H_2=t_6$, and that the next selected pair occurs at times $P_2,Q_2$, with
$$
Q_1\leq t_5\leq P_2\leq t_6\leq Q_2.
$$
For this example, let the right agent make all four selected visits to $1$.

For each of the ten selected times
$$
t_1,t_2,P_1,t_3,t_4,Q_1,t_5,P_2,t_6,Q_2,
$$
we introduce the two position variables of the agents at that time. To make the LP explicit, write $L(s)$ and $R(s)$ for these left and right position variables at time $s$. The prescribed visits give
\begin{equation}
\label{eq:eels2-first-example}
\begin{aligned}
R(t_1)&=1, & L(t_2)&=-1, & R(t_3)&=2,\\
L(t_4)&=-2, & R(t_5)&=3, & L(t_6)&=-3,\\
R(P_1)&=R(Q_1)=R(P_2)=R(Q_2)=1.
\end{aligned}
\end{equation}
The chronological order is the one displayed above. Between every two consecutive times $s\leq s'$ in this order we impose
$$
|L(s')-L(s)|\leq s'-s,
\qquad
|R(s')-R(s)|\leq s'-s,
$$
and at every selected time $s$ we impose
$$
-s\leq L(s)\leq R(s)\leq s.
$$
Here and below, an absolute-value inequality denotes its two linear inequalities.

A selected later visit cannot place an agent beyond a target whose first visit has not yet occurred. Thus
\begin{equation}
\label{eq:eels2-example-no-pass}
\begin{aligned}
L(P_1)&\geq-2, & R(P_1)&\leq2,\\
L(Q_1)&\geq-3, & R(Q_1)&\leq3,\\
L(P_2)&\geq-3.
\end{aligned}
\end{equation}
For example, $P_1$ lies between the first visits to $-1$ and $2$, so neither $-2$ nor $2$ can have been passed before that time.

The six inputs with emergence time $0$ give
\begin{equation}
\label{eq:eels2-example-first-budgets}
\begin{aligned}
t_1+R(t_1)-L(t_1)&\leq C, &
 t_2+R(t_2)-L(t_2)&\leq C,\\
t_3+R(t_3)-L(t_3)&\leq2C, &
 t_4+R(t_4)-L(t_4)&\leq2C,\\
t_5+R(t_5)-L(t_5)&\leq3C, &
 t_6+R(t_6)-L(t_6)&\leq3C.
\end{aligned}
\end{equation}
These inequalities are common to the two competitive ratios. To obtain the delayed-emergence constraints, first suppose that $P_r<Q_r$, for $r\in\{1,2\}$, and consider a target at $1$ emerging at time $P_r+\epsilon$, where
$
0<\epsilon<Q_r-P_r.
$
There is no visit to $1$ between $P_r$ and $Q_r$, so the target is first detected at $Q_r$. If $P_r=Q_r$, then $P_r=Q_r=H_r$, and we instead consider the target emerging at time $H_r$. 
In both cases, letting $\epsilon\to0^+$ in the first case gives
\begin{equation}
\label{eq:eels2-delayed-example}
Q_r+R(Q_r)-L(Q_r)\leq C f_\rho(P_r),
~~ r\in\{1,2\},
\end{equation}
where
$$
\begin{aligned}
f_{\mathrm{ea}}(P)&=P+1,\\
f_{\mathrm{fi}}(P)&=P.
\end{aligned}
$$
For the fully-informed measure, $P_r\geq1$ because an agent needs at least one time unit to reach $1$.

Equations~\eqref{eq:eels2-first-example}--\eqref{eq:eels2-delayed-example}, together with the chronological and speed constraints, define one complete LP corresponding to the discrete choices made above. The LP does not enforce that no unselected visit occurs between $P_r$ and $Q_r$. This is deliberate: we retain only necessary conditions. An actual algorithm with this visitation pattern gives a feasible solution, while an arbitrary feasible solution may describe more freedom than an algorithm has.

There is one further possibility already visible in the example. If the pair $P_1,Q_1$ also satisfies $P_1\leq H_2\leq Q_1$, then no new pair $P_2,Q_2$ is needed at $H_2$; the same two visitation times are used again. The general formulation below considers both alternatives.

\subsubsection{First-visitation constraints}

Fix $C>0$ and distances
$$
1=X_0<X_1<\cdots<X_m,
$$
and let $\mathcal X=\{\pm X_0,\ldots,\pm X_m\}$. These are adversarial target positions, not prescribed turning points of the agents. Put $N=2m+2$. On each half-line the first visits occur in increasing order of distance from the origin. Hence a possible first-visitation order $\pi=(\pi_1,\ldots,\pi_N)$ is an interleaving of
$$
X_0,X_1,\ldots,X_m
\qquad\text{and}\qquad
-X_0,-X_1,\ldots,-X_m.
$$
Simultaneous first visits on opposite sides may be listed in either order.

For a fixed $\pi$, let $t_k,p_{0,k},p_{1,k}$ be real variables intended to represent the first visitation time of $\pi_k$ and the two agent positions at that time. Set $t_0=p_{0,0}=p_{1,0}=0$.

\begin{definition}
\label{def:1stVisitationPolytope}
Let $V_1(C,\mathcal X,\pi)$ be the set of all $(t_k,p_{0,k},p_{1,k})_{k=1}^N$ satisfying
\begin{equation}
\label{eq:eels2-first-lp}
\begin{aligned}
0=t_0&\leq t_1\leq\cdots\leq t_N,\\
p_{0,k}&\leq p_{1,k},&&k=1,\ldots,N,\\
|p_{j,k}-p_{j,k-1}|&\leq t_k-t_{k-1},&&k=1,\ldots,N,\quad j\in\{0,1\},\\
p_{1,k}&=\pi_k,&&\pi_k>0,\\
p_{0,k}&=\pi_k,&&\pi_k<0,\\
t_k+p_{1,k}-p_{0,k}&\leq C|\pi_k|,&&k=1,\ldots,N.
\end{aligned}
\end{equation}
\end{definition}

The constraints have the same meaning as in the example: chronological order and unit speed, the prescribed first visits, and the competitive-ratio inequality for the input $(\pi_k,0)$.

\begin{lemma}
\label{lem:first-visitation}
If an algorithm for \eels{2} has competitive ratio at most $C$ under either measure, then it induces a first-visitation order $\pi$ for which $V_1(C,\mathcal X,\pi)$ is nonempty.
\end{lemma}

\begin{proof}
Set $t_k=v_1(\pi_k)$ and $p_{j,k}=p_j(t_k)$. Chronology, the ordering of the agents, and the speed bound give the first three constraint families. At the first visit to a positive position the right agent is there, since otherwise it would already have passed that position. The analogous statement holds for the left agent at a negative position. Finally, input $(\pi_k,0)$ and~\eqref{eq:eels2-detection-spread} give
$$
t_k+p_{1,k}-p_{0,k}\leq E(\pi_k,0)\leq C|\pi_k|.
$$
\end{proof}

Before solving any LP we can discard many first-visitation orders analytically.
Suppose that $-X_a$ is visited no later than the first visit to $X_b$.
The agent that visits $-X_a$ must eventually reach $X_b$: either it is the detecting agent for input $(X_b,0)$, or it must reach the detected target during evacuation. Its path therefore has length at least
$$
X_a+(X_a+X_b)=X_b+2X_a,
$$
which gives the necessary condition
\begin{equation}
\label{eq:eels2-opposite-first-visits}
C\geq1+\frac{2X_a}{X_b}.
\end{equation}
The symmetric condition holds when $X_a$ is visited no later than the first visit to $-X_b$.
These inequalities depend only on the first-visitation order and the fixed target positions.

\subsubsection{Delayed emergence and one-side visitation histories}

The first-visitation LP does not use inputs whose target emerges after its position has already been visited. We now add such inputs. Fix $1\leq d\leq m$, a first-visitation order $\pi$, and one of the two positions $z\in\{-1,1\}$. For $i=1,\ldots,d$, let
$$
H_i=\max\{v_1(-X_i),v_1(X_i)\}.
$$
Once $\pi$ is fixed, $H_i=t_{h_i}$, where
$$
h_i=\max\{\pi^{-1}(-X_i),\pi^{-1}(X_i)\}.
$$
Thus $H_i$ is one of the first-visitation time variables.

For an actual algorithm, define
\begin{equation}
\label{eq:eels2-selected-visits}
\begin{aligned}
\widehat P_i^z
&=\max\{t\leq H_i:p_0(t)=z\text{ or }p_1(t)=z\},\\
\widehat Q_i^z
&=\inf\{t>H_i:p_0(t)=z\text{ or }p_1(t)=z\}.
\end{aligned}
\end{equation}
The first set is a nonempty closed subset of $[0,H_i]$, and hence its maximum exists. Feasibility makes the second set nonempty. If $\widehat Q_i^z>H_i$, continuity implies a visit to $z$ at time $\widehat Q_i^z$. If $\widehat Q_i^z=H_i$, continuity gives a visit at $H_i$, and then $\widehat P_i^z=\widehat Q_i^z=H_i$.

If $\widehat P_i^z<\widehat Q_i^z$, choose
$
0<\epsilon<\widehat Q_i^z-\widehat P_i^z,
$
consider input $(z,\widehat P_i^z+\epsilon)$, and let $\epsilon\to0^+$.
If the two times are equal, use input $(z,H_i)$ instead. In both cases~\eqref{eq:eels2-detection-spread} yields the necessary inequality
\begin{equation}
\label{eq:eels2-delayed-budget}
\widehat Q_i^z+p_1(\widehat Q_i^z)-p_0(\widehat Q_i^z)
\leq C f_\rho(\widehat P_i^z),
\end{equation}
where $f_{\mathrm{ea}}(P)=P+1$ and $f_{\mathrm{fi}}(P)=P$. The latter formula uses $\widehat P_i^z\geq1$.

As $i$ increases, the same two consecutive visits to $z$ may surround several successive completion times $H_i$. If the selected pair changes between $H_{i-1}$ and $H_i$, then
$$
\widehat Q_{i-1}^z\leq\widehat P_i^z.
$$
Indeed, if $H_i<\widehat Q_{i-1}^z$, then no visit to $z$ occurs between $H_{i-1}$ and $H_i$, so the selected pair cannot change. Otherwise the visit at $\widehat Q_{i-1}^z$ occurs no later than $H_i$ and hence no later than the last visit preceding $H_i$.

This observation leaves only finitely many discrete choices. A
\emph{one-side visitation history} $\sigma$ for $z$ records the indices at
which the pair of consecutive visits surrounding $H_i$ changes. Let
$$
\mathcal R=\{r_1=1<r_2<\cdots<r_q\}\subseteq\{1,\ldots,d\}
$$
be these indices. For every $r\in\mathcal R$, we introduce real variables
$P_r^z,Q_r^z$ and the two positions at these times. For each of the two
selected times, we also record the interval between consecutive
first-visitation times that contains it, and whether $p_0$ or $p_1$ is
equal to $z$ at that time. If $i\notin\mathcal R$, we use the most recently
introduced pair. The values of $P_r^z,Q_r^z$ remain LP variables and are
not enumerated.

For convenience, define
$$
\begin{aligned}
I_0&=[0,t_1],\\
I_s&=[t_s,t_{s+1}]\quad(1\leq s<N),\\
I_N&=[t_N,\infty).
\end{aligned}
$$
For every $r\in\mathcal R$, let $s_r^P$ and $s_r^Q$ denote the indices of
the intervals assigned to $P_r^z$ and $Q_r^z$, respectively, and let
$a_r,b_r\in\{0,1\}$ satisfy
$$
p_{a_r,P_r^z}=z,
\qquad
p_{b_r,Q_r^z}=z.
$$
If a selected time is equal to some $t_k$, either of the two adjacent
intervals may be used; this merely represents the same boundary case twice.
Finally, for every $i$, let
$$
r(i)=\max\{r\in\mathcal R:r\leq i\}.
$$

\begin{definition}
\label{def:2ndVisitationPolytope}
For a fixed one-side visitation history $\sigma$, let
$V_{2,z}^\rho(C,\mathcal X,\pi,\sigma)$ be the set of assignments to the first-visitation variables, the selected times, and their position variables satisfying the following constraints.
First, the pair currently in force surrounds every $H_i$, and a new pair begins after the preceding pair has ended:
\begin{equation}
\label{eq:eels2-pair-consistency}
P_{r(i)}^z\leq H_i\leq Q_{r(i)}^z,
\qquad
Q_{r_{k-1}}^z\leq P_{r_k}^z\quad(k=2,\ldots,q).
\end{equation}
For every $r\in\mathcal R$, the selected positions equal $z$, the first visit to $z$ occurs no later than $P_r^z$, and the delayed-emergence constraint holds:
\begin{equation}
\label{eq:eels2-event-visits}
\begin{aligned}
p_{a_r,P_r^z}&=z, & p_{b_r,Q_r^z}&=z,\\
t_{\pi^{-1}(z)}&\leq P_r^z, &
Q_r^z+p_{1,Q_r^z}-p_{0,Q_r^z}&\leq C f_\rho(P_r^z),
\end{aligned}
\qquad r\in\mathcal R.
\end{equation}
The position variables at every selected time $S\in\{P_r^z,Q_r^z:r\in\mathcal R\}$ satisfy
\begin{equation}
\label{eq:eels2-event-motion}
-S\leq p_{0,S}\leq p_{1,S}\leq S.
\end{equation}
If $S$ is assigned to an interval $I_s$ with $s<N$, then
\begin{equation}
\label{eq:eels2-slot-motion}
t_s\leq S\leq t_{s+1},
\qquad
|p_{j,S}-p_{j,s}|\leq S-t_s,
\qquad
|p_{j,s+1}-p_{j,S}|\leq t_{s+1}-S,
\end{equation}
for $j\in\{0,1\}$, where $t_0=p_{0,0}=p_{1,0}=0$. If $S$ is assigned to the last interval $I_N$, then
\begin{equation}
\label{eq:eels2-last-slot-motion}
t_N\leq S,
\qquad
|p_{j,S}-p_{j,N}|\leq S-t_N,
\qquad j\in\{0,1\}.
\end{equation}
We also impose
\begin{equation}
\label{eq:eels2-unvisited-position}
p_{1,S}\leq X_k\quad\text{whenever }\pi^{-1}(X_k)>s,
\qquad
p_{0,S}\geq-X_k\quad\text{whenever }\pi^{-1}(-X_k)>s.
\end{equation}
Finally, for every $r\in\mathcal R$ we impose
$$
|p_{j,Q_r^z}-p_{j,P_r^z}|
\leq Q_r^z-P_r^z,
\qquad j\in\{0,1\}.
$$
If two selected times $S,S'$ are assigned to distinct intervals $I_s$ and $I_{s'}$, with $s<s'$, we also impose
$$
|p_{j,S'}-p_{j,S}|\leq S'-S,
\qquad j\in\{0,1\}.
$$
If two selected times are assigned to the same interval and their order has not otherwise been fixed, we impose no additional speed constraint between them. This only enlarges the feasible region.
\end{definition}

All constraints in Definition~\ref{def:2ndVisitationPolytope} are linear once $C$, $\mathcal X$, $\pi$, and the discrete history $\sigma$ are fixed. The same first-visitation variables are used in $V_1$ and $V_{2,z}^\rho$.

\begin{lemma}
\label{lem:second-visitation}
Suppose that an algorithm for \eels{2} has $\rho$-competitive ratio at most $C$, and let $\pi$ be its first-visitation order on $\mathcal X$. For each $z\in\{-1,1\}$, there is a one-side visitation history $\sigma$ such that
$$
V_1(C,\mathcal X,\pi)\cap V_{2,z}^\rho(C,\mathcal X,\pi,\sigma)\neq\emptyset.
$$
\end{lemma}

\begin{proof}
Assign the first-visitation variables their actual values as in Lemma~\ref{lem:first-visitation}. 
For the fixed $z$, use the times in~\eqref{eq:eels2-selected-visits}. 
Include $1$ in $\mathcal R$, and for every $i>1$ include $i$ precisely when the selected pair differs from the one used for $H_{i-1}$. 
The inequality preceding the definition shows that successive pairs satisfy~\eqref{eq:eels2-pair-consistency}. 
Assign each selected time to an interval that contains it, and record whether $p_0$ or $p_1$ equals $z$ there. 
The actual trajectories satisfy all chronological, position, and speed constraints, as well as the no-passing inequalities~\eqref{eq:eels2-unvisited-position}. 
Finally, \eqref{eq:eels2-delayed-budget} gives the competitive constraints in~\eqref{eq:eels2-event-visits}.
\end{proof}

For completeness, the LP variables may be bounded without imposing an external time horizon. The first-visitation constraints give $t_k\leq C X_m$, while $P_r^z\leq H_r\leq C X_d$ and the delayed-emergence constraints give $Q_r^z\leq C(CX_d+1)$ for both measures. Hence every selected time is at most
\begin{equation}
\label{eq:eels2-event-bound}
B=\max\{CX_m,\ C(CX_d+1)\},
\end{equation}
and all corresponding position variables lie in $[-B,B]$.

The preceding lemmas give the lower-bound criterion used in the computation.

\begin{theorem}
\label{thm:eels2-finite-lower-bound}
Fix $C$, $\mathcal X$, $d$, and $\rho\in\{\mathrm{ea},\mathrm{fi}\}$. Consider one representative from each reflection pair of first-visitation orders that satisfy the necessary conditions~\eqref{eq:eels2-opposite-first-visits}. Suppose that for every such order $\pi$, either
$$
V_1(C,\mathcal X,\pi)=\emptyset,
$$
or there is a position $z\in\{-1,1\}$ such that
$$
V_1(C,\mathcal X,\pi)\cap
V_{2,z}^\rho(C,\mathcal X,\pi,\sigma)=\emptyset
$$
for every one-side visitation history $\sigma$. Then no feasible algorithm has $\rho$-competitive ratio at most $C$.
\end{theorem}

\begin{proof}
Suppose that such an algorithm exists. Reflecting the line if necessary, its first-visitation order has a representative $\pi$ among those considered in the theorem. Lemma~\ref{lem:first-visitation} gives a point in $V_1(C,\mathcal X,\pi)$. The order cannot have been eliminated by~\eqref{eq:eels2-opposite-first-visits}, since that inequality is necessary for every $C$-competitive algorithm. Therefore the second alternative of the theorem applies for some $z$. Lemma~\ref{lem:second-visitation} then supplies a one-side visitation history $\sigma$ for which the displayed intersection is nonempty, a contradiction.
\end{proof}

\subsubsection{Finite enumeration and exact certification}

The first-visitation orders are much fewer than $N!$. Since the order on each half-line is fixed, there are
$$
\binom{2m+2}{m+1}
$$
interleavings. Reflection pairs them, so we fix the first visit to be $X_0=1$ and retain half of these orders. Conditions~\eqref{eq:eels2-opposite-first-visits} eliminate further orders before any LP is solved.

For a fixed remaining first-visitation order, the computer-assisted proof does not construct every complete one-side visitation history in advance. It builds the history one completion time $H_i$ at a time. The procedure mirrors the proof above.

For $i>1$, it first asks whether the most recently selected pair can still satisfy $P\leq H_i\leq Q$ and the same delayed-emergence inequality. If so, that possibility is continued without introducing new selected times. The alternative is to introduce a new pair $P_i^z,Q_i^z$. Before fixing either visiting agent, the computation imposes only the necessary conditions
$$
p_{0,P_i^z}\leq z\leq p_{1,P_i^z},
\qquad
p_{0,Q_i^z}\leq z\leq p_{1,Q_i^z},
$$
together with $P_i^z\leq H_i\leq Q_i^z$, the speed constraint between the two selected times, the delayed-emergence constraint, and $Q\leq P_i^z$ when a previous pair is being replaced. It then fixes, in succession, the first-visitation interval containing $P_i^z$, the agent visiting $z$ at $P_i^z$, the interval containing $Q_i^z$, and the agent visiting $z$ at $Q_i^z$. Each choice adds the corresponding constraints of Definition~\ref{def:2ndVisitationPolytope}.

Every intermediate LP is a relaxation of all complete histories extending the choices already made. 
Therefore, once an intermediate LP has been proved infeasible, none of its extensions needs to be considered. 
Otherwise, the next discrete choice is fixed. 
The enumeration for $z$ eliminates a first-visitation order only if every possible history through $H_d$ is proved infeasible.

We certify every infeasibility statement using exact rational arithmetic. After all variables have been collected into a vector $\xi$, each LP is written in the canonical form
$$
G\xi\leq h,
$$
where $G$ and $h$ are rational. By Farkas' lemma, infeasibility is certified by a rational vector $y$ satisfying
\begin{equation}
\label{eq:eels2-farkas-certificate}
\begin{aligned}
y&\geq0,\\
y^{\mathsf T}G&=0,\\
y^{\mathsf T}h&<0.
\end{aligned}
\end{equation}
Indeed, if a feasible $\xi$ existed, then weak duality would give
$$
0=y^{\mathsf T}G\xi\leq y^{\mathsf T}h<0,
$$
a contradiction. A floating-point LP solver is used only to locate candidate infeasible systems and supports for certificates. No branch is pruned from numerical infeasibility alone: the corresponding rational certificate must first satisfy~\eqref{eq:eels2-farkas-certificate} exactly.

The final verification is independent of the numerical LP solver. From the fixed rational inputs, it reconstructs the first-visitation orders and every LP arising in the enumeration. It then checks each Farkas certificate using rational arithmetic and verifies that every retained first-visitation order is eliminated. In the certified instances below, each such order is eliminated either by the first-visitation LP itself or by considering all histories for one of the two positions $z=-1,1$. No combined enumeration for both delayed target positions is needed.

\subsubsection{The certified instances}

The target placements used in the proof were found before exact certification by numerical exploration over the placement vector and the proposed value of $C$. This search served only to identify promising inputs. The final vectors were made rational and then treated as fixed data. The theorem below depends only on the exact verification for these fixed vectors, not on the preliminary numerical search or on its exhaustiveness.

For the emergence-aware bound, take $C=87/20$, $m=8$, $d=7$, and
\begin{equation}
\label{eq:eels2-certified-placements-ea}
(X_0,\ldots,X_8)=\frac1{1000}
(1000,7719,15969,19145,24588,32707,48566,54394,136059).
\end{equation}
For the fully-informed bound, take $C=219/50$, again with $m=8$ and $d=7$, and
\begin{equation}
\label{eq:eels2-certified-placements-fi}
(X_0,\ldots,X_8)=\frac1{1000}
(1000,6205,10693,13525,15515,20455,28219,54968,63424).
\end{equation}
With $m=8$, there are
$\binom{18}{9}/2=24310$ possible first-visitation orders up to reflection.
Applying~\eqref{eq:eels2-opposite-first-visits} leaves $2400$ first-visitation orders for the emergence-aware instance and $2448$ for the fully-informed instance. 
The exact verification rules out every one of these remaining first-visitation orders. 
For each such order, either the first-visitation LP is already infeasible, or the delayed-emergence constraints for one of the two positions $z=-1,1$ suffice to make every
corresponding visitation history infeasible. 
Thus no simultaneous enumeration of delayed-visitation histories for both sides is needed.

\begin{proof}[Proof of Theorem~\ref{thm: eels2 unconditional lb}]
For each of the two parameter choices above, the exact verifier confirms the hypothesis of Theorem~\ref{thm:eels2-finite-lower-bound} for every first-visitation order that remains after the analytic restrictions. Hence no feasible algorithm has emergence-aware competitive ratio at most $87/20$, and no feasible algorithm has fully-informed competitive ratio at most $219/50$. Therefore
$$
\begin{aligned}
\eelsea{2}&\geq\frac{87}{20}=4.35,\\
\eelsfi{2}&\geq\frac{219}{50}=4.38.
\end{aligned}
$$
\end{proof}

\paragraph{Reproducibility.}
The publication package for these bounds contains the fixed rational target
placements, the code that constructs the LPs and enumerates the visitation
cases, and the rational verifier used to check the resulting Farkas
certificates. Thus, the complete certificates for both bounds can be
regenerated from the fixed inputs and then verified using exact arithmetic.
Neither step requires rerunning the preliminary search for target placements,
and the final verification does not rely on the floating-point LP solver.
The code and instructions for reproducing the computation are available at
\url{https://github.com/theory-computation/Eventually-Emerging-Linear-Search}.

\section{3-Agent Evacuation - \eels3}
\label{sec: 3-agent evacuation}

The main result proved in this section is the following
\begin{theorem}
\label{thm: 3-agent results in 3-agent section}
$ 3.2823 \leq \eelsfi{3} \leq 4.340445$
and
$ 3.1014 \leq \eelsea{3} \leq 4.010599$
\end{theorem}

We begin in Section~\ref{sec: geometric 3-agent algorithm} with a natural one-parameter extension of the geometric algorithms for one and two agents. Its analysis motivates the three-peak algorithm of Section~\ref{sec: three-peak algorithm}, where we establish the upper bounds of Theorem~\ref{thm: 3-agent results in 3-agent section}. Finally, Section~\ref{sec:eels3-lower-bounds} proves the corresponding lower bounds.

\subsection{A Single-Peak Geometric 3-Agent Algorithm}
\label{sec: geometric 3-agent algorithm}

In this section we analyze a natural one-parameter extension of the geometric algorithms for one and two agents. The resulting algorithm has one outer turning point between consecutive inner turning points of each side agent. The main result is the following.

\begin{theorem}
\label{thm:ells3-upper-simple}
$
\eelsfi{3}\leq 4\rho-3\approx4.770597,
$
where $\rho\in(3/2,2)$ is the unique root of
$2\rho^4-3\rho^3-2\rho^2-\rho+3=0$, and
$
\eelsea{3}\leq \frac{(\eta+1)(4\eta-3)}{2\eta}\approx4.475300,
$
where $\eta\in(2,3)$ is the unique root of
$4\eta^4-9\eta^3-2\eta^2+2\eta+3=0$.
\end{theorem}

For $b>1$, let $\ell=(\ell_i)_{i\in\mathbb Z}$ and $u=(u_i)_{i\in\mathbb Z}$ be given by
$\ell_i=b^{2i}$ and $u_i=(1+b+b^2)b^{2i}$.

\begin{definition}[Algorithm $\mathcal A_b$ for \eels{3}]
\label{def:three-agent-geometric}
The agents are denoted by $A_0,A_1,A_2$, where $A_1$ searches the negative half-line, $A_2$ searches the positive half-line, and $A_0$ is the middle agent. Before the target is detected, $A_0$ follows $\mathcal Z_b$, $A_2$ follows $\mathcal S_{\ell,u}$, and $A_1$ follows $-\mathcal S_{b\ell,bu}$.
\end{definition}
Since $A_0$ follows $\mathcal Z_b$, every point of the line is visited infinitely often. Hence $\mathcal A_b$ is feasible.
Thus, for every $i\in\mathbb Z$, the successive turning points of $A_2$ are
$
b^{2i},
(1+b+b^2)b^{2i},
b^{2i+2},
$
while those of $A_1$ are their reflected copies at the next odd powers of $b$. In particular, $A_0$ and $A_2$ meet at $b^{2i}$, and $A_0$ and $A_1$ meet at $-b^{2i+1}$.
The analysis below determines the competitive ratios of this family and identifies the limitation of using a single outer turning point. This motivates the three-peak construction of Section~\ref{sec: three-peak algorithm}.

To analyze Algorithm~\ref{def:three-agent-geometric}, we first record the main events and the positions of all three agents between two consecutive meetings of $A_0$ and $A_2$.
\begin{lemma}
\label{lem:three-agent-turning-events}
Fix an even $i\in\mathbb Z$. Between the consecutive meetings of $A_0$ and $A_2$ at $b^i$ and $b^{i+2}$, the events and the positions of the agents are as follows:
\begin{equation*}
\begin{array}{c|c|c}
\text{Time}
&
\text{Event}
&
(p_0,p_1,p_2)
\\
\hline
\frac{(b+1)b^i}{b-1}
&
A_0,A_2\text{ meet at }b^i
&
\left(
b^i,
-(b+2)b^{i-1},
b^i
\right)
\\
\frac{(b+1)b^{i+1}}{b-1}
&
A_0,A_1\text{ meet at }-b^{i+1}
&
\left(
-b^{i+1},
-b^{i+1},
(b+2)b^i
\right)
\\
\frac{(b^3+1)b^i}{b-1}
&
A_2\text{ turns at }b^i+b^{i+1}+b^{i+2}
&
\left(
(b^2-b-1)b^i,
-(b^2+b-1)b^i,
(b^2+b+1)b^i
\right)
\\
\frac{(b+1)b^{i+2}}{b-1}
&
A_0,A_2\text{ meet at }b^{i+2}
&
\left(
b^{i+2},
-(b+2)b^{i+1},
b^{i+2}
\right).
\end{array}
\end{equation*}
\end{lemma}

\begin{proof}
The times of the turning points of $A_0$ are
\begin{equation*}
\sum_{j=-\infty}^{m-1}(b^j+b^{j+1})
=
\frac{(b+1)b^m}{b-1}.
\end{equation*}
For even $i=2m$, the time at which $A_2$ reaches $b^i$ is
$$
\sum_{j=-\infty}^{m-1}
\left((u_j-\ell_j)+(u_j-\ell_{j+1})\right)
=
\sum_{j=-\infty}^{m-1}(b+1)^2b^{2j}\\
=
\frac{(b+1)b^i}{b-1}.
$$
Thus $A_0$ and $A_2$ meet at $b^i$ at this time. 
By reflection and scaling, $A_0$ and $A_1$ meet at $-b^{i+1}$ at time
$
\frac{(b+1)b^{i+1}}{b-1}.
$
The first, second, and fourth event times follow.

At time
$
\frac{(b+1)b^i}{b-1}
$,
agents $A_0$ and $A_2$ are at $b^i$. Agent $A_1$ started its excursion from $-b^{i-1}$ at time
$
\frac{(b+1)b^{i-1}}{b-1}
$
and has moved to the left for time
$
(b+1)b^{i-1}.
$
Hence,
\begin{equation*}
p_1
=
-b^{i-1}-(b+1)b^{i-1}
=
-(b+2)b^{i-1}.
\end{equation*}

At time
$
\frac{(b+1)b^{i+1}}{b-1}
$,
agents $A_0$ and $A_1$ meet at $-b^{i+1}$. During the time
$
(b+1)b^i
$
since the preceding meeting, $A_2$ has moved to the right from $b^i$. Therefore,
\begin{equation*}
p_2
=
b^i+(b+1)b^i
=
(b+2)b^i.
\end{equation*}

Agent $A_2$ then reaches its turning point after travelling distance
$
b^{i+1}+b^{i+2}
$.
Thus, its turning time is
\begin{equation*}
\frac{(b+1)b^i}{b-1}
+
b^{i+1}+b^{i+2}
=
\frac{(b^3+1)b^i}{b-1}.
\end{equation*}
The time elapsed since the meeting of $A_0$ and $A_1$ at $-b^{i+1}$ is
\begin{equation*}
\frac{(b^3+1)b^i}{b-1}
-
\frac{(b+1)b^{i+1}}{b-1}
=
(b^2-1)b^i.
\end{equation*}
Hence,
\begin{equation*}
p_0
=
-b^{i+1}+(b^2-1)b^i
=
(b^2-b-1)b^i,
\end{equation*}
and
\begin{equation*}
p_1
=
-b^{i+1}-(b^2-1)b^i
=
-(b^2+b-1)b^i.
\end{equation*}

Finally, at time
$
\frac{(b+1)b^{i+2}}{b-1}
$,
agents $A_0$ and $A_2$ meet at $b^{i+2}$. Agent $A_1$ has moved to the left from $-b^{i+1}$ for time
$
(b+1)b^{i+1},
$
and therefore
\begin{equation*}
p_1
=
-b^{i+1}-(b+1)b^{i+1}
=
-(b+2)b^{i+1}.
\end{equation*}
\end{proof}

The preceding description also makes the self-similarity of the algorithm explicit. We use it next to reduce the range of target positions that needs to be considered.
\begin{lemma}
\label{lem: Ab is b^2noemarlizable}
Algorithm $\mathcal A_b$ to \eels{3} is $b^2$-normalizable. 
\end{lemma}

\begin{proof}
Let $p_0,p_1,p_2$ denote the trajectories of $A_0,A_1,A_2$ before detection. By construction,
$
p_j(b^2t)=b^2p_j(t)
$
for every $t\geq0$ and every $j\in\{0,1,2\}$. Moreover,
\begin{equation*}
\begin{aligned}
p_0(bt)&=-bp_0(t),\\
p_1(bt)&=-bp_2(t),\\
p_2(bt)&=-bp_1(t).
\end{aligned}
\end{equation*}
Therefore, $\mathcal A_b$ is $b^2$-normalizable.
\end{proof}

By Lemma~\ref{lem: Ab is b^2noemarlizable}, $\mathcal A_b$ is $b^2$-normalizable, and so, by Lemma~\ref{lem:normalization-reduction}, it is enough to consider targets 
$
1<x\leq b^2
$.
The next lemma establishes the merged visitation timelines for all
$
x\in(1,y)\cup(y,b^2).
$

\begin{lemma}
\label{lem:merged-visitation-Ab-eels3}
Fix $b>1.94$, and let $\mathcal A_b$ be the three-agent geometric algorithm of Definition~\ref{def:three-agent-geometric}. Define
$y
:=
1+\frac1b+\frac1{b^2},
$
for which $1<y<b^2$. 
For every $1<x<b^2$, the visitation timeline of $A_0$ is given, for every $i\geq1$, by
\begin{equation*}
f_0(i,x)
=
\begin{cases}
\frac{2b^{i+1}}{b-1}+x,
&
\text{if $i$ is odd},
\\
\frac{2b^{i+1}}{b-1}-x,
&
\text{if $i$ is even}.
\end{cases}
\end{equation*}
The visitation timeline by $A_2$ is
\begin{equation*}
f_2(i,x)
=
\begin{cases}
x+\frac{2}{b^2(b-1)},
&
\text{if $i=1$ and $1<x<y$},
\\
\frac{2b}{b-1}-x,
&
\text{if $i=2$ and $1<x<y$},
\\
x+\frac{2}{b-1},
&
\text{if $i=3$ and $1<x<y$, or $i=1$ and $y<x<b^2$}.
\end{cases}
\end{equation*}
Moreover, the merged visitation timelines are as follows. 
If $x\in (1,y)$, we have 
\begin{equation*}
f_2(1,x)
<
f_2(2,x)
<
f_2(3,x)
<
f_0(1,x)
<
f_0(2,x)
<
f_0(3,x)
<
\cdots.
\end{equation*}
that is 
$
v_1(x)=f_2(1,x),
v_2(x)=f_2(2,x),
v_3(x)=f_2(3,x),
$
and
$
v_{k+3}(x)=f_0(k,x)
$
for every $k\geq1$.

If $x\in (y,b^2)$, we have 
$$
f_2(1,x)
<
f_0(1,x)
<
f_0(2,x)
<
f_0(3,x)
<
\cdots,
$$
or in other words,
$
v_1(x)=f_2(1,x),
v_{k+1}(x)=f_0(k,x)
$,
for every $k\geq1$.
\end{lemma}

\begin{proof}
We note that point 
$
y=
1+\frac1b+\frac1{b^2}
$
is the turning point of $A_2$ during the excursion
$
\frac1{b^2}
\rightarrow
\frac1{b^2}+\frac1b+1
\rightarrow
1
$.
Moreover,
\begin{equation*}
b^2-y
=
\frac{(b+1)(b^3-b^2-1)}{b^2}
>0,
\end{equation*}
since $b>1.94$. Therefore,
$
1<y<b^2
$, that is between the consecutive meetings of $A_0$ and $A_2$ at
$
1
$
and
$
b^2
$,
agent $A_2$ has exactly one rightmost turning point, namely $y$, which lies strictly between these two meeting points.
Next we compute the visitation times of $x\in (1,b^2]$ by either agent $A_0,A_2$. 

\emph{$A_0$-visitation timeline:}
For every $k\geq0$, the $(2k+1)$-st visit of $A_0$ to $x$ occurs while moving from $-b^{2k+1}$ to $b^{2k+2}$. Therefore,
\begin{equation*}
f_0(2k+1,x)
=
2\sum_{j=-\infty}^{2k+1}b^j+x
=
\frac{2b^{2k+2}}{b-1}+x.
\end{equation*}
The following visit occurs after $A_0$ reaches $b^{2k+2}$ and returns to $x$. Hence,
\begin{equation*}
f_0(2k+2,x)
=
f_0(2k+1,x)
+
2\left(b^{2k+2}-x\right)
=
\frac{2b^{2k+3}}{b-1}-x.
\end{equation*}

\emph{$A_2$-visitation timeline:}
We need to distinguish two cases depending on the value of $x$ relative to $y$. 

Suppose first that $1<x<y$. 
By Lemma~\ref{lem:three-agent-turning-events}, agents $A_0$ and $A_2$ meet at $b^{-2}$ at time
$$
\sum_{j=-\infty}^{-3}(b^j+b^{j+1})
=
\frac{b+1}{b^2(b-1)}.
$$
Therefore, the first visit of $A_2$ to $x$ occurs while moving from $b^{-2}$ to $y$, and
\begin{equation*}
f_2(1,x)
=
\frac{b+1}{b^2(b-1)}
+
x-\frac1{b^2}
=
x+\frac{2}{b^2(b-1)}.
\end{equation*}
The second visit occurs while moving from $y$ to $1$. Since $A_0$ and $A_2$ meet at $1$ at time $\frac{2}{b-1}+1$, 
we have
\begin{equation*}
f_2(2,x)
=
\frac{2}{b-1}+1
-(x-1)
=
\frac{2b}{b-1}
-x.
\end{equation*}
Finally, the third visit occurs while moving from $1$ to $b^2+b+1$, and
\begin{equation*}
f_2(3,x)
=
\frac{2}{b-1}+1
+(x-1)
=
\frac{2}{b-1}+x.
\end{equation*}

Second, suppose now that $y<x<b^2$. Since the excursion
$
b^{-2}
\rightarrow
y
\rightarrow
1
$
does not reach $x$, the first and only visit occurs while moving from $1$ to $b^2+b+1$, with
$
f_2(1,x)
=
\frac{2}{b-1}+1
+(x-1)
=
\frac{2}{b-1}+x.
$

We are now ready to proceed with the merged visitation timelines. Again, we distinguish two cases on $x$ and how it compares to $y$. 

Suppose first that
$
1<x<y
$.
We have the following calculations
\begin{align*}
f_2(2,x)-f_2(1,x)
=&
2(y-x)
>0,\\
f_2(3,x)-f_2(2,x)
=&
2(x-1)
>0,\\
f_0(1,x)-f_2(3,x)
=&
2(b+1)
>0.
\end{align*}
Therefore,
\begin{equation*}
f_2(1,x)
<
f_2(2,x)
<
f_2(3,x)
<
f_0(1,x).
\end{equation*}
Since
$
f_0(1,x)<f_0(2,x)<f_0(3,x)<\cdots
$,
the merged visitation timeline is
\begin{equation*}
f_2(1,x)
<
f_2(2,x)
<
f_2(3,x)
<
f_0(1,x)
<
f_0(2,x)
<
f_0(3,x)
<
\cdots.
\end{equation*}

Second, suppose now that
$
y<x<b^2
$.
In this case, $A_2$ visits $x$ only once. Since
$$
f_0(1,x)-f_2(1,x)
=
2(b+1)
>0,
$$
and
$
f_0(1,x)<f_0(2,x)<f_0(3,x)<\cdots,
$
the merged visitation timeline is
\begin{equation*}
f_2(1,x)
<
f_0(1,x)
<
f_0(2,x)
<
f_0(3,x)
<
\cdots.
\end{equation*}
\end{proof}

\begin{lemma}
\label{lem:three-agent-geometric}
Fix $b>1.94$, and let $\mathcal A_b$ be the three-agent geometric algorithm of Definition~\ref{def:three-agent-geometric}. Define
\begin{equation*}
D(b)
=
\frac{(b+1)(3b^2+b-3)}
{(b-1)(b^2+b+1)}.
\end{equation*}
Then
$$
\mathrm{CR}_{\mathrm{fi}}(\mathcal A_b)
=
\max\{D(b),4b-3\},
\qquad
\mathrm{CR}_{\mathrm{ea}}(\mathcal A_b)
=
\max\left\{
D(b),
\frac{(b+1)(4b-3)}{2b}
\right\}.
$$
\end{lemma}

\begin{proof}
By Lemmas~\ref{lem: Ab is b^2noemarlizable} and~\ref{lem:normalization-reduction}, it is enough to consider target placements
$
1<x\leq b^2
$.
Let also 
$
y
=
1+\frac1b+\frac1{b^2}
$
where $1<y<b^2$. 
The placements $x=y$ and $x=b^2$ do not require separate consideration, since the corresponding competitive-ratio values are limits of values obtained for targets in the interiors of the adjacent intervals. Therefore, it is enough to consider
$
x\in(1,y)\cup(y,b^2)
$.

We distinguish two cases.

\begin{description}
\item [Case $1<x<y$:]
By Lemma~\ref{lem:merged-visitation-Ab-eels3}, the merged visitation timeline is
\begin{equation*}
v_1(x)=f_2(1,x),
\quad
v_2(x)=f_2(2,x),
\quad
v_3(x)=f_2(3,x),
\quad
v_{k+3}(x)=f_0(k,x),
\quad
k\geq1.
\end{equation*}
By Lemma~\ref{lem:merged-visitation-timeline}, it remains to compute $M_k(x)$ for every visitation in this timeline.
\begin{description}
\item[Computation of $M_1(x)$ due to $A_2$:]
By Lemma~\ref{lem:merged-visitation-Ab-eels3},
$
v_1(x)
=
x+\frac{2}{b^2(b-1)}.
$
By Lemma~\ref{lem:three-agent-turning-events}, during this excursion the intermediate meeting of $A_0$ and $A_1$ occurs when
\begin{equation*}
(p_0,p_1,p_2)
=
\left(
-\frac1b,
-\frac1b,
\frac{b+2}{b^2}
\right).
\end{equation*}
If
$
1<x\leq\frac{b+2}{b^2},
$
then the first visitation occurs before this meeting. 
Hence,
$
p_0(v_1(x))
=
\frac2{b^2}-x
$, and 
$
p_1(v_1(x))
=
x-\frac{2(b+1)}{b^2}
$.
Therefore,
$$
\max_{j\in\{0,1,2\}}
|p_j(v_1(x))-x|
=
\frac{2(b+1)}{b^2},
$$
and
$
M_1(x)
=
\frac{2}{b-1}+x.
$

If
$
\frac{b+2}{b^2}\leq x<y,
$
then the first visitation occurs after this meeting. Hence,
$
p_0(v_1(x))
=
x-\frac{2(b+1)}{b^2}$,
and
$
p_1(v_1(x))
=
\frac2{b^2}-x
$.
Therefore,
$$
\max_{j\in\{0,1,2\}}
|p_j(v_1(x))-x|
=
2x-\frac2{b^2},
$$
and
$
M_1(x)
=
\frac{2}{b^2(b-1)}
-\frac2{b^2}
+3x
$. 
If $1.94<b<2$, then
\begin{equation*}
M_1(x)
=
\begin{cases}
\frac{2}{b-1}+x,
&
1<x\leq\frac{b+2}{b^2},
\\
\frac{2}{b^2(b-1)}-\frac2{b^2}+3x,
&
\frac{b+2}{b^2}\leq x<y.
\end{cases}
\end{equation*}
If $b\geq2$, then
\begin{equation*}
M_1(x)
=
\frac{2}{b^2(b-1)}-\frac2{b^2}+3x.
\end{equation*}

\item[Computation of $M_2(x)$ due to $A_2$:]
By Lemma~\ref{lem:merged-visitation-Ab-eels3},
$
v_2(x)
=
f_2(2,x)
=
\frac{2b}{b-1}-x.
$
By Lemma~\ref{lem:three-agent-turning-events}, when $A_2$ turns at $y$, we have
$$
p_0
=
1-\frac1b-\frac1{b^2},
\qquad
p_1
=
-1-\frac1b+\frac1{b^2},
\qquad
p_2
=
y.
$$
The second visitation occurs after time $y-x$. Therefore,
$
p_0(v_2(x))
=
2-x$,
and
$
p_1(v_2(x))
=
x-2-\frac2b
$
Hence,
\begin{equation*}
|p_0(v_2(x))-x|
=
2(x-1),
\qquad
|p_1(v_2(x))-x|
=
2+\frac2b.
\end{equation*}
Since $x<y$,
\begin{equation*}
2(x-1)
<
2(y-1)
=
\frac2b+\frac2{b^2}
<
2+\frac2b.
\end{equation*}
Thus, $A_1$ is farthest from the target, and
\begin{equation*}
M_2(x)
=
v_2(x)+2+\frac2b
=
\frac{2(2b^2-1)}{b(b-1)}-x.
\end{equation*}

\item[Computation of $M_3(x)$ due to $A_2$:]
By Lemma~\ref{lem:merged-visitation-Ab-eels3},
$
v_3(x)
=
f_2(3,x)
=
\frac{2}{b-1}+x
$.
Since $A_2$ is at the target, it remains to compare the distances of $A_0$ and $A_1$ from $x$.

By Lemma~\ref{lem:three-agent-turning-events}, when $A_0$ and $A_2$ meet at $1$, we have
\begin{equation*}
p_0=1,
\qquad
p_1=-\frac{b+2}{b},
\qquad
p_2=1.
\end{equation*}
Until $A_2$ reaches $x$, the elapsed time is $x-1$. Therefore,
\begin{equation*}
p_0(v_3(x))
=
2-x,
\qquad
p_1(v_3(x))
=
-x-\frac2b.
\end{equation*}
Hence,
$
|p_0(v_3(x))-x|
=
2(x-1)
$, and
$
|p_1(v_3(x))-x|
=
2x+\frac2b
$.
Since $A_2$ is at the target, its distance from $x$ is $0$. Moreover,
$
2x+\frac2b
-
2(x-1)
=
2+\frac2b
>
0
$
that is, 
$
\max_{j\in\{0,1,2\}}
|p_j(v_3(x))-x|
=
2x+\frac2b
$
and therefore, 
\begin{equation*}
M_3(x)
=
v_3(x)
+
2x+\frac2b
=
\frac{2}{b-1}
+\frac2b
+3x.
\end{equation*}

\item[Computation of $M_{2n+4}(x)$ due to $A_0$:]
For every $n\geq0$, the visitation
$
v_{2n+4}(x)=f_0(2n+1,x)
$
is made by $A_0$ while moving from $-b^{2n+1}$ to $b^{2n+2}$. By Lemma~\ref{lem:merged-visitation-Ab-eels3},
$
v_{2n+4}(x)
=
\frac{2b^{2n+2}}{b-1}+x.
$
Since $A_0$ is at the target, it remains to compare the distances of $A_1$ and $A_2$ from $x$.

By Lemma~\ref{lem:three-agent-turning-events}, applied with $i=2n$, when $A_0$ and $A_1$ meet at $-b^{2n+1}$, we have
\begin{equation*}
(p_0,p_1,p_2)
=
\left(
-b^{2n+1},
-b^{2n+1},
(b+2)b^{2n}
\right).
\end{equation*}
Until $A_0$ reaches $x$, agent $A_1$ moves to the left for time
$
b^{2n+1}+x
$.
Therefore,
\begin{equation*}
\left|p_1(v_{2n+4}(x))-x\right|
=
2b^{2n+1}+2x.
\end{equation*}

The same lemma shows that $A_2$ turns when $A_0$ reaches
$
(b^2-b-1)b^{2n}
$,
and that its turning point is
$
(b^2+b+1)b^{2n}.
$
Hence,
\begin{equation*}
\left|p_2(v_{2n+4}(x))-x\right|
=
\begin{cases}
2(b+1)b^{2n},
&
x\leq(b^2-b-1)b^{2n},
\\
2b^{2n+2}-2x,
&
x\geq(b^2-b-1)b^{2n}.
\end{cases}
\end{equation*}

For $n=0$, if $1.94<b<2$, then $A_2$ has already turned and
$
x>1>\frac{b(b-1)}2
$.
If $b\geq2$, then either $A_2$ has not yet turned, in which case its distance is smaller than that of $A_1$ since $x>1$, or $A_2$ has already turned, in which case its distance is again smaller than that of $A_1$. Hence, $A_1$ is farther from the target when $n=0$.

For every $n\geq1$, we have
\begin{equation*}
x
<
y
<
(b^2-b-1)b^2
\leq
(b^2-b-1)b^{2n}.
\end{equation*}
Thus, $A_2$ has not yet turned. Moreover,
$
x<b^2\leq b^{2n}
$,
and hence its distance exceeds that of $A_1$ by
$
2b^{2n}-2x>0
$.
Therefore,
\begin{equation*}
M_{2n+4}(x)
=
\begin{cases}
\frac{2b^2}{b-1}
+3x
+2b,
&
n=0,
\\
\frac{2b^{2n+2}}{b-1}
+x
+2(b+1)b^{2n},
&
n\geq1.
\end{cases}
\end{equation*}

\item[Computation of $M_{2n+5}(x)$ due to $A_0$:]
For every $n\geq0$, the visitation
$
v_{2n+5}(x)=f_0(2n+2,x)
$
is made by $A_0$ while moving from $b^{2n+2}$ to $-b^{2n+3}$. By Lemma~\ref{lem:merged-visitation-Ab-eels3},
$
v_{2n+5}(x)
=
\frac{2b^{2n+3}}{b-1}-x
$.
Since $A_0$ is at the target, it remains to compare the distances of $A_1$ and $A_2$ from $x$.

By Lemma~\ref{lem:three-agent-turning-events}, applied with $i=2n$, when $A_0$ and $A_2$ meet at $b^{2n+2}$, we have
\begin{equation*}
(p_0,p_1,p_2)
=
\left(
b^{2n+2},
-(b+2)b^{2n+1},
b^{2n+2}
\right).
\end{equation*}
Until $A_0$ reaches $x$, the elapsed time is
$
b^{2n+2}-x.
$
During this time, $A_1$ moves to the left and $A_2$ moves to the right. Therefore,
\begin{equation*}
\left|p_1(v_{2n+5}(x))-x\right|
=
2(b+1)b^{2n+1},
\end{equation*}
and
\begin{equation*}
\left|p_2(v_{2n+5}(x))-x\right|
=
2b^{2n+2}-2x.
\end{equation*}
The first distance exceeds the second by
$
2b^{2n+1}+2x>0.
$
Hence, $A_1$ is farthest from the target, and
\begin{equation*}
M_{2n+5}(x)
=
\frac{2b^{2n+3}}{b-1}-x
+
2(b+1)b^{2n+1}
=
\frac{2b(2b^2-1)}{b-1} b^{2n}-x.
\end{equation*}
\end{description}

\item[Case $y<x<b^2$:]
By Lemma~\ref{lem:merged-visitation-Ab-eels3}, the merged visitation timeline is
$$
v_1(x)=f_2(1,x),
\,
v_{k+1}(x)=f_0(k,x),
\,
k\geq1.
$$
Again, by Lemma~\ref{lem:merged-visitation-timeline}, it remains to compute $M_k(x)$ for every visitation in this timeline.

\begin{description}

\item[Computation of $M_1(x)$ due to $A_2$:]
By Lemma~\ref{lem:merged-visitation-Ab-eels3},
$
v_1(x)
=
f_2(1,x)
=
\frac{2}{b-1}+x
$.
Since $A_2$ is at the target, it remains to compare the distances of $A_0$ and $A_1$ from $x$.

By Lemma~\ref{lem:three-agent-turning-events}, when $A_0$ and $A_2$ meet at $1$, we have
$$
\begin{aligned}
p_0&=1,\\
p_1&=-\frac{b+2}{b},\\
p_2&=1.
\end{aligned}
$$
Agent $A_1$ turns when $A_2$ reaches
$
b+1-\frac1b
$,
while $A_0$ and $A_1$ meet at $-b$ when $A_2$ reaches $b+2$.

If
$
y<x\leq b+1-\frac1b,
$
then
$
p_0(v_1(x))=2-x
$
and
$
p_1(v_1(x))=-x-\frac2b
$.
Hence,
$
|p_0(v_1(x))-x|=2(x-1)
$
and
$
|p_1(v_1(x))-x|=2x+\frac2b
$.
Therefore,
$$
M_1(x)
=
\frac{2}{b-1}
+\frac2b
+3x.
$$

If
$
b+1-\frac1b\leq x\leq b+2,
$
then
$
p_0(v_1(x))=2-x
$
and
$
p_1(v_1(x))=x-2b-2
$.
Hence,
$
|p_0(v_1(x))-x|=2(x-1)
$
and
$
|p_1(v_1(x))-x|=2b+2
$.
Therefore,
$$
M_1(x)
=
\frac{2}{b-1}
+x+2b+2.
$$

If
$
b+2\leq x<b^2,
$
then
$
p_0(v_1(x))=x-2b-2
$
and
$
p_1(v_1(x))=2-x
$.
Hence,
$
|p_0(v_1(x))-x|=2b+2
$
and
$
|p_1(v_1(x))-x|=2x-2
$.
Therefore,
$$
M_1(x)
=
\frac{2}{b-1}
+3x-2.
$$
To summarize, what we showed is 
\begin{equation*}
M_1(x)
=
\begin{cases}
\frac{2}{b-1}
+\frac2b
+3x,
&
y<x\leq b+1-\frac1b,
\\
\frac{2}{b-1}
+x+2b+2,
&
b+1-\frac1b\leq x\leq b+2,
\\
\frac{2}{b-1}
+3x-2,
&
b+2\leq x<b^2.
\end{cases}
\end{equation*}

\item[Computation of $M_{2n+2}(x)$ due to $A_0$:]
For every $n\geq0$, the visitation
$
v_{2n+2}(x)=f_0(2n+1,x)
$
is made by $A_0$ while moving from $-b^{2n+1}$ to $b^{2n+2}$. By Lemma~\ref{lem:merged-visitation-Ab-eels3},
$
v_{2n+2}(x)
=
\frac{2b^{2n+2}}{b-1}+x
$.
Since $A_0$ is at the target, it remains to compare the distances of $A_1$ and $A_2$ from $x$.

By Lemma~\ref{lem:three-agent-turning-events}, applied with $i=2n$, when $A_0$ and $A_1$ meet at $-b^{2n+1}$, we have
$$
(p_0,p_1,p_2)
=
\left(
-b^{2n+1},
-b^{2n+1},
(b+2)b^{2n}
\right).
$$
Until $A_0$ reaches $x$, agent $A_1$ moves to the left for time
$
b^{2n+1}+x
$.
Therefore,
$$
|p_1(v_{2n+2}(x))-x|
=
2b^{2n+1}+2x.
$$

The same lemma shows that $A_2$ turns when $A_0$ reaches
$
(b^2-b-1)b^{2n}
$,
and that its turning point is
$
(b^2+b+1)b^{2n}
$.
Hence,
\begin{equation*}
|p_2(v_{2n+2}(x))-x|
=
\begin{cases}
2(b+1)b^{2n},
&
x\leq(b^2-b-1)b^{2n},
\\
2b^{2n+2}-2x,
&
x\geq(b^2-b-1)b^{2n}.
\end{cases}
\end{equation*}

For $n=0$, agent $A_1$ is farther from the target. Indeed, if $A_2$ has not yet turned, the difference between the distances of $A_2$ and $A_1$ is
$
2-2x<0
$.
If $A_2$ has already turned, then $x\geq b^2-b-1$. For $b\geq2$, this and $x>y$ imply $x>b(b-1)/2$, while for $1.94<b<2$ we have $x>1>b(b-1)/2$. Thus, the difference between the two distances is
$
2b(b-1)-4x<0.
$
Therefore,
\begin{equation*}
M_2(x)
=
\frac{2b^2}{b-1}
+
3x
+
2b.
\end{equation*}

For every $n\geq2$, we have
\begin{equation*}
x
<
b^2
<
(b^2-b-1)b^{2n}.
\end{equation*}
Thus, $A_2$ has not yet turned. Moreover,
$
x<b^2\leq b^{2n}
$,
and hence its distance exceeds that of $A_1$ by
$
2b^{2n}-2x>0
$.
Therefore,
\begin{equation*}
M_{2n+2}(x)
=
\frac{2b^{2n+2}}{b-1}
+
x
+
2(b+1)b^{2n},
\qquad
n\geq2.
\end{equation*}

It remains to consider $n=1$. If $b\geq2$, then
$
x<b^2\leq(b^2-b-1)b^2
$,
so $A_2$ has not yet turned and is farther from the target. Therefore,
\begin{equation*}
M_4(x)
=
\frac{2b^4}{b-1}
+
x
+
2(b+1)b^2.
\end{equation*}

Finally, suppose that $1.94<b<2$. We have
\begin{equation*}
y
<
(b^2-b-1)b^2
<
\frac{b^3(b-1)}2
<
b^2.
\end{equation*}
Before $A_2$ turns, its distance exceeds that of $A_1$ by
$
2b^2-2x>0
$.
After $A_2$ turns, its distance exceeds that of $A_1$ precisely when
$
x\leq\frac{b^3(b-1)}2
$.
Hence,
\begin{equation*}
M_4(x)
=
\begin{cases}
\frac{2b^4}{b-1}
+x
+2(b+1)b^2,
&
y<x\leq(b^2-b-1)b^2,
\\
\frac{2b^5}{b-1}-x,
&
(b^2-b-1)b^2
\leq x
\leq
\frac{b^3(b-1)}2,
\\
\frac{2b^4}{b-1}
+3x
+2b^3,
&
\frac{b^3(b-1)}2
\leq x<b^2.
\end{cases}
\end{equation*}

\item[Computation of $M_{2n+3}(x)$ due to $A_0$:]
For every $n\geq0$, the visitation
$
v_{2n+3}(x)=f_0(2n+2,x)
$
is made by $A_0$ while moving from $b^{2n+2}$ to $-b^{2n+3}$. By Lemma~\ref{lem:merged-visitation-Ab-eels3},
$
v_{2n+3}(x)
=
\frac{2b^{2n+3}}{b-1}-x
$.
Since $A_0$ is at the target, it remains to compare the distances of $A_1$ and $A_2$ from $x$.

By Lemma~\ref{lem:three-agent-turning-events}, applied with $i=2n$, when $A_0$ and $A_2$ meet at $b^{2n+2}$, we have
$$
(p_0,p_1,p_2)
=
\left(
b^{2n+2},
-(b+2)b^{2n+1},
b^{2n+2}
\right).
$$
Until $A_0$ reaches $x$, the elapsed time is
$
b^{2n+2}-x
$.
During this time, $A_1$ moves to the left and $A_2$ moves to the right. Therefore,
$
\left|p_1(v_{2n+3}(x))-x\right|
=
2(b+1)b^{2n+1}
$,
and
$
\left|p_2(v_{2n+3}(x))-x\right|
=
2b^{2n+2}-2x
$.
The first distance exceeds the second by
$
2b^{2n+1}+2x>0
$.
Therefore,
$$
\max_{j\in\{0,1,2\}}
\left|p_j(v_{2n+3}(x))-x\right|
=
2(b+1)b^{2n+1},
$$
and hence
$$
M_{2n+3}(x)
=
\frac{2b^{2n+3}}{b-1}
+
2(b+1)b^{2n+1}
-
x
=
\frac{2b^{2n+1}(2b^2-1)}{b-1}
-
x.
$$

\end{description}

\end{description}

We now use the preceding computations to determine the two competitive ratios. 
For each visitation family, we first identify the relevant preceding visitation and then bound the fully-informed and emergence-aware ratios in parallel.
In the analysis we use the following notation, for the expressions that appear in the statement of the lemma. 
\begin{equation*}
\begin{aligned}
Q(b)&=4b-3,\\
R(b)&=\frac{(b+1)(4b-3)}{2b}.
\end{aligned}
\end{equation*}
By Lemma~\ref{lem:merged-visitation-timeline}, for every $x\in(1,b^2)$,
\begin{equation*}
\sup_{T\geq0}
\frac{E(x,T)}{\max\{T,x\}}
=
\sup_{k\geq1}
\frac{M_k(x)}
{\max\{v_{k-1}(x),x\}},
\qquad
\sup_{T\geq0}
\frac{E(x,T)}{T+x}
=
\sup_{k\geq1}
\frac{M_k(x)}
{v_{k-1}(x)+x}.
\end{equation*}

We distinguish the two target-placement cases considered in the preceding analysis.

\begin{description}
\item[Case $1<x<y$:] ~~

\begin{description}
\item[Competitive ratios for $M_1(x)$:]
If $1.94<b<2$, the two expressions for $M_1(x)/x$ are decreasing in $x$ and coincide at
$
x=(b+2)/b^2
$.
Hence,
\begin{equation*}
\frac{M_1(x)}{x}
<
1+\frac{2}{b-1}
<
D(b),
\end{equation*}
where
\begin{equation*}
D(b)
-
\left(
1+\frac{2}{b-1}
\right)
=
\frac{2(b+1)(b^2-2)}
{(b-1)(b^2+b+1)}
>
0.
\end{equation*}

If $b\geq2$, then
\begin{equation*}
\frac{M_1(x)}{x}
=
3+
\frac{
\frac{2}{b^2(b-1)}-\frac2{b^2}
}{x}
\end{equation*}
is nondecreasing in $x$. Therefore,
\begin{equation*}
\frac{M_1(x)}{x}
\leq
3+
\frac{
\frac{2}{b^2(b-1)}-\frac2{b^2}
}{y}
<
D(b),
\end{equation*}
where
\begin{equation*}
D(b)
-
\left(
3+
\frac{
\frac{2}{b^2(b-1)}-\frac2{b^2}
}{y}
\right)
=
\frac{4(b+1)}{b^2+b+1}
>
0.
\end{equation*}

\item[Competitive ratios for $M_2(x)$:]
Since
$
M_2(x)
$
arises only when
$
1<x<y,
$
we have
$
v_1(x)>x
$.
Hence, the corresponding fully-informed and emergence-aware ratios are
\begin{equation*}
\frac{M_2(x)}{v_1(x)}
\qquad\text{and}\qquad
\frac{M_2(x)}{v_1(x)+x}.
\end{equation*}
Both are decreasing in $x$, and hence are maximized as $x\rightarrow1^+$. Direct calculation gives
\begin{equation*}
Q(b)
-
\lim_{x\rightarrow1^+}
\frac{M_2(x)}{v_1(x)}
=
\frac{2(b-1)^2(2b-3)}
{b^2-2b+2}
>
0,
\end{equation*}
and
\begin{equation*}
R(b)
-
\lim_{x\rightarrow1^+}
\frac{M_2(x)}{v_1(x)+x}
=
\frac{
(b-1)^2(b+1)(4b^2-2b-3)
}{
2b(b^3-b^2+1)
}
>
0.
\end{equation*}

\item[Competitive ratios for $M_3(x)$:]
We have
\begin{equation*}
v_2(x)-x
=
\frac{2b}{b-1}-2x
>
\frac{2b}{b-1}-2y
=
\frac{2}{b^2(b-1)}
>
0.
\end{equation*}
Thus, the fully-informed denominator is $v_2(x)$. The corresponding fully-informed and emergence-aware ratios are
\begin{equation*}
\frac{M_3(x)}{v_2(x)}
\qquad\text{and}\qquad
\frac{M_3(x)}{v_2(x)+x}.
\end{equation*}
Both are increasing in $x$, and hence are maximized as $x\rightarrow y^-$. Direct calculation gives
\begin{equation*}
Q(b)
-
\lim_{x\rightarrow y^-}
\frac{M_3(x)}{v_2(x)}
=
\frac{2b(b-1)(2b-3)}
{b^2-b+1}
>
0,
\end{equation*}
and
\begin{equation*}
R(b)
-
\lim_{x\rightarrow y^-}
\frac{M_3(x)}{v_2(x)+x}
=
\frac{
(b-1)(b+1)(4b^2-2b-3)
}{
2b^3
}
>
0.
\end{equation*}

\item[Competitive ratios for $M_{2n+4}(x)$:]
For $n=0$, this is the first right-moving visitation of $A_0$. We have
\begin{equation*}
M_4(x)
=
\frac{2b(2b-1)}{b-1}
+
3x,
\end{equation*}
while the preceding visitation time is
$
v_3(x)
=
\frac{2}{b-1}+x
$.
Hence, the corresponding fully-informed and emergence-aware ratios are
\begin{equation*}
\frac{
\frac{2b(2b-1)}{b-1}+3x
}{
\frac{2}{b-1}+x
},
\qquad
\frac{
\frac{2b(2b-1)}{b-1}+3x
}{
\frac{2}{b-1}+2x
},
\end{equation*}
respectively. Both are decreasing in $x$, and hence are maximized as
$
x\rightarrow1^+.
$
Moreover,
\begin{align*}
\lim_{x\rightarrow1^+}
\frac{
\frac{2b(2b-1)}{b-1}+3x
}{
\frac{2}{b-1}+x
}
&=
4b-3
=
Q(b),
\\
\lim_{x\rightarrow1^+}
\frac{
\frac{2b(2b-1)}{b-1}+3x
}{
\frac{2}{b-1}+2x
}
&=
\frac{(b+1)(4b-3)}{2b}
=
R(b).
\end{align*}

For every $n\geq1$, the preceding visitation is
$
v_{2n+3}(x)
=
\frac{2b^{2n+1}}{b-1}-x
$.
Moreover,
$
v_{2n+3}(x)>x
$,
since
\begin{equation*}
v_{2n+3}(x)-x
=
\frac{2b^{2n+1}}{b-1}-2x
>
\frac{2b^{2n+1}}{b-1}-2b^{2n}
=
\frac{2b^{2n}}{b-1}
>
0.
\end{equation*}
Hence, the corresponding fully-informed and emergence-aware ratios are
\begin{equation*}
\frac{
\frac{2b^{2n+2}}{b-1}
+x
+2(b+1)b^{2n}
}{
\frac{2b^{2n+1}}{b-1}-x
},
\qquad
\frac{
\frac{2b^{2n+2}}{b-1}
+x
+2(b+1)b^{2n}
}{
\frac{2b^{2n+1}}{b-1}
},
\end{equation*}
respectively. Both are increasing in $x$. Since
$
x<b^2\leq b^{2n},
$
we obtain
\begin{align*}
\frac{
\frac{2b^{2n+2}}{b-1}
+x
+2(b+1)b^{2n}
}{
\frac{2b^{2n+1}}{b-1}-x
}
&<
\frac{
\frac{2b^{2n+2}}{b-1}
+b^{2n}
+2(b+1)b^{2n}
}{
\frac{2b^{2n+1}}{b-1}-b^{2n}
}
=
Q(b),
\\
\frac{
\frac{2b^{2n+2}}{b-1}
+x
+2(b+1)b^{2n}
}{
\frac{2b^{2n+1}}{b-1}
}
&<
\frac{
\frac{2b^{2n+2}}{b-1}
+b^{2n}
+2(b+1)b^{2n}
}{
\frac{2b^{2n+1}}{b-1}
}
=
R(b).
\end{align*}

\item[Competitive ratios for $M_{2n+5}(x)$:]
For every $n\geq0$, the preceding visitation is
$
v_{2n+4}(x)
=
\frac{2b^{2n+2}}{b-1}+x
$.
Since
$
v_{2n+4}(x)>x
$,
the corresponding fully-informed and emergence-aware ratios are
\begin{equation*}
\frac{
\frac{2b(2b^2-1)}{b-1}b^{2n}-x
}{
\frac{2b^{2n+2}}{b-1}+x
},
\qquad
\frac{
\frac{2b(2b^2-1)}{b-1}b^{2n}-x
}{
\frac{2b^{2n+2}}{b-1}+2x
},
\end{equation*}
respectively. Both are decreasing in $x$. Therefore, both are smaller than
\begin{equation*}
\frac{
\frac{2b(2b^2-1)}{b-1}b^{2n}
}{
\frac{2b^{2n+2}}{b-1}
}
=
\frac{2b^2-1}{b}.
\end{equation*}
Moreover,
\begin{equation*}
Q(b)
-
\frac{2b^2-1}{b}
=
\frac{(b-1)(2b-1)}{b}
>
0,
\end{equation*}
and
\begin{equation*}
R(b)
-
\frac{2b^2-1}{b}
=
\frac{b-1}{2b}
>
0.
\end{equation*}
Hence, the fully-informed and emergence-aware ratios are smaller than $Q(b)$ and $R(b)$, respectively.

\end{description}

\item[Case $y<x<b^2$:]~~

\begin{description}

\item[Competitive ratios for $M_1(x)$:]
For
$
y<x\leq b+1-\frac1b,
$
we have
\begin{equation*}
\frac{M_1(x)}{x}
=
3+
\frac{
\frac{2}{b-1}+\frac2b
}{x},
\end{equation*}
which is decreasing in $x$. Hence,
\begin{equation*}
\sup_{y<x\leq b+1-\frac1b}
\frac{M_1(x)}{x}
=
\lim_{x\rightarrow y^+}
\frac{M_1(x)}{x}
=
D(b).
\end{equation*}

For
$
b+1-\frac1b
\leq
x
\leq
b+2,
$
the ratio
$
M_1(x)/x
$
is also decreasing in $x$, and coincides with the preceding expression at
$
x=b+1-\frac1b.
$
Hence,
$
M_1(x)/x<D(b)
$.

Finally, the interval
$
b+2\leq x<b^2
$
is nonempty only when $b>2$. In this interval,
\begin{equation*}
\frac{M_1(x)}{x}
=
3+
\frac{
\frac{2}{b-1}-2
}{x},
\end{equation*}
which is increasing in $x$. Moreover,
\begin{equation*}
D(b)
-
\left(
3+
\frac{
\frac{2}{b-1}-2
}{b^2}
\right)
=
\frac{
2(b+1)(2b^3-2b^2+b-2)
}{
b^2(b-1)(b^2+b+1)
}
>
0.
\end{equation*}
Together with the bound for $M_1(x)$ in the case $1<x<y$, this gives
\begin{equation}
\label{eq:first-visitation-bound-Ab}
\sup_{1<x<b^2}
\frac{M_1(x)}{x}
=
D(b).
\end{equation}

\item[Competitive ratios for $M_{2n+2}(x)$:]
For $n=0$, the expressions for $M_2(x)$ and its preceding visitation coincide with those considered for $M_4(x)$ when $1<x<y$. Hence, the fully-informed and emergence-aware ratios are smaller than $Q(b)$ and $R(b)$, respectively.

For every $n\geq2$, and also for $n=1$ when $b\geq2$, the expressions for $M_{2n+2}(x)$ and its preceding visitation coincide with those considered for $M_{2n+4}(x)$ when $1<x<y$. Hence, the same bounds hold.

It remains to consider $n=1$ when $1.94<b<2$. The preceding visitation is
$
v_3(x)
=
\frac{2b^3}{b-1}-x
$,
and
$
v_3(x)>x
$.

For
$
y<x\leq(b^2-b-1)b^2,
$
the expression for $M_4(x)$ is the same as in the preceding right-moving family. Hence, the fully-informed and emergence-aware ratios are smaller than $Q(b)$ and $R(b)$, respectively.

For
$
(b^2-b-1)b^2
\leq
x
\leq
\frac{b^3(b-1)}2,
$
we have
\begin{equation*}
M_4(x)
=
\frac{2b^5}{b-1}-x.
\end{equation*}
Thus, the corresponding ratios are
\begin{equation*}
\frac{
(2b^5)/(b-1)-x
}{
(2b^3)/(b-1)-x
},
\qquad
\frac{
(2b^5)/(b-1)-x
}{
(2b^3)/(b-1)
},
\end{equation*}
respectively. The first is increasing in $x$, while the second is decreasing in $x$. At the right endpoint, the first ratio agrees with the fully-informed ratio in the following interval, while at the left endpoint, the second ratio agrees with the emergence-aware ratio in the preceding interval.

Finally, for
$
\frac{b^3(b-1)}2
\leq
x<b^2,
$
we have
\begin{equation*}
M_4(x)
=
\frac{2b^4}{b-1}
+
3x
+
2b^3.
\end{equation*}
The corresponding ratios are increasing in $x$. Therefore,
\begin{align*}
\frac{
\frac{2b^4}{b-1}+3x+2b^3
}{
\frac{2b^3}{b-1}-x
}
&<
\frac{
\frac{2b^4}{b-1}+3b^2+2b^3
}{
\frac{2b^3}{b-1}-b^2
}
=
Q(b),
\\
\frac{
\frac{2b^4}{b-1}+3x+2b^3
}{
\frac{2b^3}{b-1}
}
&<
\frac{
\frac{2b^4}{b-1}+3b^2+2b^3
}{
\frac{2b^3}{b-1}
}
=
R(b).
\end{align*}
Hence, the fully-informed and emergence-aware ratios are smaller than $Q(b)$ and $R(b)$, respectively, also for
$
(b^2-b-1)b^2
\leq
x
\leq
\frac{b^3(b-1)}2.
$

\item[Competitive ratios for $M_{2n+3}(x)$:]
The expressions for $M_{2n+3}(x)$ and its preceding visitation coincide with those considered for $M_{2n+5}(x)$ when $1<x<y$. Hence, the corresponding fully-informed and emergence-aware ratios are smaller than $Q(b)$ and $R(b)$, respectively.
\end{description}
\end{description}

Combining the preceding bounds with Corollary~\ref{cor:normalized-visitation-reduction}, we obtain
\begin{equation*}
\begin{aligned}
\mathrm{CR}_{\mathrm{fi}}(\mathcal A)
&=
\sup_{\substack{1<x<b^2\\T\geq0}}
\frac{E(x,T)}{\max\{T,x\}}
\leq
\max\{D(b),Q(b)\},\\
\mathrm{CR}_{\mathrm{ea}}(\mathcal A)
&=
\sup_{\substack{1<x<b^2\\T\geq0}}
\frac{E(x,T)}{T+x}
\leq
\max\{D(b),R(b)\}.
\end{aligned}
\end{equation*}

The preceding bounds are tight. By~\eqref{eq:first-visitation-bound-Ab},
$
\lim_{x\rightarrow y^+}
\frac{M_1(x)}{x}
=
D(b).
$
Moreover, for the first right-moving visitation of $A_0$,
$
\lim_{x\rightarrow1^+}
\frac{M_4(x)}{v_3(x)}
=
Q(b),
\qquad
\lim_{x\rightarrow1^+}
\frac{M_4(x)}{v_3(x)+x}
=
R(b).
$
Therefore,
\begin{equation*}
\mathrm{CR}_{\mathrm{fi}}(\mathcal A_b)
=
\max\{D(b),Q(b)\},
\qquad
\mathrm{CR}_{\mathrm{ea}}(\mathcal A_b)
=
\max\{D(b),R(b)\}.
\end{equation*}
\end{proof}

We are now ready to prove Theorem~\ref{thm:ells3-upper-simple}.

\begin{proof}[Proof of Theorem~\ref{thm:ells3-upper-simple}]
By Lemma~\ref{lem:three-agent-geometric},
$
\eelsfi{3}
\leq
\inf_{b>1.94}
\max\{D(b),4b-3\}.
$
Moreover,
\begin{equation*}
D'(b)
=
-\frac{2(2b^4-2b^3+4b-1)}
{(b-1)^2(b^2+b+1)^2}
<0,
\end{equation*}
since
$
2b^4-2b^3+4b-1
=
2b^3(b-1)+4b-1>0
$
for every $b>1$.
Therefore, $D(b)$ is strictly decreasing, while $4b-3$ is strictly increasing. Moreover,
$
D\left(\frac32\right)>3,
$
$
D(1.94)>4(1.94)-3,
$
and
$
D(2)<5.
$
Hence the two expressions are equal at a unique point
$\rho\in(3/2,2)$. The second inequality shows that $\rho>1.94$, so
$\rho$ lies in the parameter range considered above, and their maximum is
minimized there.
The equation
$
D(b)=4b-3
$
is equivalent to
$
2b^4-3b^3-2b^2-b+3=0
$.
Then
\begin{equation*}
\eelsfi{3}
\leq
4\rho-3
\approx4.770596950171800.
\end{equation*}

Similarly, by Lemma~\ref{lem:three-agent-geometric},
$
\eelsea{3}
\leq
\inf_{b>1.94}
\max\left\{
D(b),
\frac{(b+1)(4b-3)}{2b}
\right\}.
$
Also,
$$
\frac{d}{db}
\left(
\frac{(b+1)(4b-3)}{2b}
\right)
=
\frac{4b^2+3}{2b^2}
>0.
$$
Therefore, $D(b)$ is strictly decreasing, while
$
\frac{(b+1)(4b-3)}{2b}
$
is strictly increasing. At $b=2$ the first expression is larger, while at $b=3$ the second is larger. Hence the two expressions are equal at a unique point $\eta\in(2,3)$, and their maximum is minimized there. The corresponding equation is
$
4b^4-9b^3-2b^2+2b+3=0.
$
Hence
\begin{equation*}
\eelsea{3}
\leq
\frac{(\eta+1)(4\eta-3)}{2\eta}
\approx4.475299799087579.
\end{equation*}
\end{proof}

\subsection{The Three-Peak Algorithm for \eels{3}}
\label{sec: three-peak algorithm}

In this section we prove the upper bounds of Theorem~\ref{thm: 3-agent results in 3-agent section}, which we restate for completeness.

\begin{theorem}
\label{thm:three-peak}
We have $\eelsfi{3}\leq\frac{868089}{200000}=4.340445$ and $\eelsea{3}\leq\frac{4010599}{1000000}=4.010599$. The two bounds use different choices of parameters.
\end{theorem}

We refine the one-sided trajectories of Algorithm $\mathcal A_b$, while keeping $A_0$ on $\mathcal Z_b$. Agent $A_2$ now visits three outer turning points and two intermediate inner turning points during its motion from $b^{2i}$ to $b^{2i+2}$.

Fix $b>1$ and parameters $(u_1,v_1,u_2,v_2,u_3)$ satisfying $1<v_1<u_1<v_2<u_2<b^2<u_3<b^4$. Let $\ell=(\ell_i)_{i\in\mathbb Z}$ and $w=(w_i)_{i\in\mathbb Z}$ be given, for every $i\in\mathbb Z$, by
\begin{align*}
(\ell_{3i},\ell_{3i+1},\ell_{3i+2})&=b^{2i}(1,v_1,v_2),\\
(w_{3i},w_{3i+1},w_{3i+2})&=b^{2i}(u_1,u_2,u_3).
\end{align*}

\begin{definition}[The three-peak algorithm]
\label{def:three-peak}
Before the target is detected, $A_0$ follows $\mathcal Z_b$, $A_2$ follows $\mathcal S_{\ell,w}$, and $A_1$ follows $-\mathcal S_{b\ell,bw}$. Upon detection, all three agents move directly to the target at maximum speed.
\end{definition}

Thus, for $s=b^{2i}$, agent $A_2$ follows $s\to u_1s\to v_1s\to u_2s\to v_2s\to u_3s\to b^2s$. The negative trajectory satisfies $p_1(t)=-b\,p_2(t/b)$. Unlike Algorithm $\mathcal A_b$, we do not require meetings with $A_0$ at the turning points $b^{2i}$ and $-b^{2i+1}$.

\paragraph{Turning times.}
Set
\begin{equation*}
h=\frac{2u_1-2v_1+2u_2-2v_2+2u_3-1-b^2}{b^2-1}.
\end{equation*}
The time spent on the motion from $s$ to $b^2s$ is $h(b^2-1)s$. Summing the durations of the preceding motions gives $\sum_{k<i}h(b^2-1)b^{2k}=hb^{2i}$, so this motion starts at time $hs$ and ends at time $hb^2s$.

For the calculations, let $A=2/(b-1)$, $c=(b+1)/(b-1)$, $K_1=h+2u_1-2v_1-1$, and $K_2=K_1+2u_2-2v_2$. Also set $H_1=K_1+2v_1=h+2u_1-1$ and $H_2=K_2+2v_2=K_1+2u_2$. Agent $A_0$ reaches $(-1)^i b^i$ at time $cb^i$. The successive turning times of $A_2$ in the motion above are $sr_0,\ldots,sr_6$, where
\begin{equation}
\label{eq:three-peak-times}
\begin{aligned}
r_0&=h, & r_1&=h+u_1-1, & r_2&=K_1+v_1, & r_3&=K_1+u_2,\\
r_4&=K_2+v_2, & r_5&=K_2+u_3, & r_6&=hb^2.
\end{aligned}
\end{equation}
Each successive time difference equals the corresponding distance travelled. In particular, the last equality follows from
\begin{equation}
\label{eq:three-peak-cycle}
K_2+2u_3=b^2(h+1).
\end{equation}

\paragraph{Conditions on the parameters.}
We analyze two sets of parameter choices, one for each competitive ratio.
The conditions below determine the order of visits, which agent is farthest
from the target at each visit, and where the resulting ratios are maximized.
They are sufficient conditions for our analysis, not additional requirements
in Definition~\ref{def:three-peak}.
For these inequalities and the calculations that follow, define
\begin{equation}
\label{eq:three-peak-cuts}
\begin{aligned}
a&=u_3/b^2, & \alpha&=r_1/b-K_2/b^2, & \beta&=h+1-r_2/b,\\
\delta_0&=r_3/b-h+1, & \delta_1&=r_4/b-h+1, & \delta_2&=H_1-r_5/b,\\
\delta_3&=bh-K_1, & z_0&=r_5-Ab^2, & z_1&=br_2-Ab^2,\\
z_2&=hb^2-Ab^2, & z_3&=br_3-Ab^2, & z_4&=br_4-Ab^2,\\
\theta&=Ab^3-br_4, & \omega&=Ab^3-b^2r_1.
\end{aligned}
\end{equation}
Here $a$ is the outer turning point $u_3/b^2$ visited during the preceding
motion. The constants $\alpha,\beta,\delta_0,\ldots,\delta_3$ identify target
positions at which $A_1$ changes direction during a visit by $A_2$.
The constants $z_0,\ldots,z_4,\theta,\omega$ serve the same purpose for the
two side agents during visits by $A_0$.

Both sets satisfy the common conditions
\begin{gather}
1<a<v_1<u_1<v_2<u_2<b^2,\qquad 1<h<b^4,
\label{eq:conditions-space}\\
\begin{gathered}
\frac cb<1+\frac{K_2}{b^2}<\frac{r_1}b<\frac{r_5}{b^2}<\frac{r_2}b<h,\\
h<\frac{r_3}b<\frac{r_4}b<r_1<\frac{r_5}b<r_2<bh,\\
bh<r_3<r_4<K_2+b^2<br_1,
\end{gathered}
\label{eq:conditions-phases}\\
\begin{gathered}
h<c<r_1,\qquad r_3<bc<K_2+b^2,\\
\max\{br_1,r_4\}<Ab^2,\qquad Ab^3<\min\{br_5,b^2r_2\},
\end{gathered}
\label{eq:conditions-times}\\
\delta_0<a,\qquad \delta_1<v_1,\qquad \delta_3<u_1,
\label{eq:conditions-early}\\
0<z_0<z_1<z_2<z_3<z_4,\qquad b^{-2}<z_0,\qquad 0<\omega<1,
\label{eq:conditions-central}\\
\begin{gathered}
\frac{2K_2}{b^2}>\frac{h-1}b,\qquad
2(h-1)>\frac{K_2}b,\qquad 2K_1>b(h-1),\\
\frac{H_1}b>\frac{K_2}{b^2},\qquad
bH_2>Ab^2,\qquad 2Ab^3>bK_2+b^2H_1.
\end{gathered}
\label{eq:conditions-signs}
\end{gather}
For the first set, let $\sigma=(b^2(h+1)-bH_1)/2$ and require
\begin{gather}
z_0<\sigma<1<z_1<a<z_2,\qquad
u_2<z_3<b^2<z_4,\qquad u_2<\theta<b^2,\nonumber\\
\delta_1<a,\qquad
3(h-1)<2Ab^2-bK_1<3K_1,\qquad b^4(h+1)>Ab^3,\nonumber\\
u_1<b^2z_0<v_2,\qquad u_2<b^2\sigma.
\label{eq:conditions-F}
\end{gather}
For the second set, let $\sigma=(b^2(h+1)-2Ab^2+bK_1)/4$ and require
\begin{gather}
z_1<1<\sigma<a<z_2,\qquad
u_2<z_3<z_4<b^2<\theta,\qquad a<\delta_1,\nonumber\\
3K_2<2\min\{2Ab^2-bK_1,\,2Ab^2-bK_2\},\qquad K_2<2bH_2,
\qquad u_2<b^2z_0.
\label{eq:conditions-E}
\end{gather}
All inequalities are strict. 
The analysis below assumes the common conditions together with one of the two additional sets. 
We substitute the rational parameters of~\eqref{eq:parameters-F} and~\eqref{eq:parameters-E} only in the proof of Theorem~\ref{thm:three-peak}, where the corresponding conditions are verified.
We call the common conditions together with~\eqref{eq:conditions-F} the fully-informed regime, and the common conditions together with~\eqref{eq:conditions-E} the emergence-aware regime.

For $A_2$, the total duration of all cycles preceding the one starting at
$b^{2n}$ is
$
\sum_{k<n}h(b^2-1)b^{2k}=hb^{2n},
$
which tends to $0$ as $n\to-\infty$, while all corresponding positions
tend to the origin. The same holds for $A_0$, and for $A_1$ by reflection.
Thus the doubly indexed descriptions above define unit-speed trajectories
starting at the origin, and the finite-prefix observation of
Section~\ref{sec:notation} applies.
Moreover, since $A_0$ follows $\mathcal Z_b$, every legal target position
is visited by $A_0$ at arbitrarily large times. Hence the three-peak
algorithm is feasible.

The next lemma shows that $A_0$ remains between the other two agents and
records the identities needed for the normalization reduction.
\begin{lemma}
\label{lem:three-peak-order}
For every $t\geq0$,
$
p_1(t)\leq p_0(t)\leq p_2(t).
$
Moreover, for every $j\in\{0,1,2\}$,
\[
\begin{aligned}
p_j(b^2t)&=b^2p_j(t),\\
p_0(bt)&=-bp_0(t),\\
p_1(bt)&=-bp_2(t),\\
p_2(bt)&=-bp_1(t).
\end{aligned}
\]
\end{lemma}
\begin{proof}
The scaling and reflection identities follow directly from the turning
points and the definition of $p_1$. It remains to prove the ordering. 
Since $h<c<r_1$, agent $A_2$ is on its
first outward segment at time $c$, and $p_2(c)=c-h+1>1$. Thus, at the
negative turning time $cb$ of $A_0$,
\[
p_1(cb)=-b(c-h+1)<-b=p_0(cb).
\]
The same holds at every negative turning time by scaling. Between a positive
and the next negative turn of $A_0$, the difference $p_0-p_1$ is
nonincreasing: $A_0$ moves left at unit speed and $p_1$ is $1$-Lipschitz.
Between a negative and the next positive turn, this difference is
nondecreasing. Consequently its minimum on either segment is attained at
the negative turn, where it is positive. This proves $p_1\leq p_0$.
Reflection and scaling then give $p_0\leq p_2$.
\end{proof}

The scaling identities of Lemma~\ref{lem:three-peak-order} show that the
three-peak algorithm is $b^2$-normalizable in the sense of
Definition~\ref{def:lambda-normalizable}, with $\mu=b$ and the permutation
that fixes $A_0$ and exchanges $A_1$ and $A_2$. Hence, by
Lemma~\ref{lem:normalization-reduction}, it is enough to consider target
positions $1<x\leq b^2$. We consider $x=1$ separately below, which is
equivalent to $x=b^2$.

By Corollary~\ref{cor:normalized-visitation-reduction}, it remains to
determine the successive visitation times and the corresponding evacuation
times for the normalized target positions. We first consider
$1<x<b^2$ and determine the merged visitation timelines. 
The endpoint targets are considered separately below.

\subsubsection{Merged visitation timelines}

Fix $1<x<b^2$. Agent $A_1$ stays negative. Of the cycles of $A_2$, only
$n=-1,0$ can visit $x$: all earlier peaks are at most $u_3/b^4<1$,
and all later minima are at least $b^2$.
For $n\in\mathbb Z$, set $s=b^{2n}$. We denote by $e_{n,j}(x)$ the
visitation time of $x$ on the $j$-th segment of this motion, whenever that
segment visits $x$.
Starting from the successive states in~\eqref{eq:three-peak-times} and
adding the distance to $x$ gives the following visits, on closed domains:

\begin{equation}\label{eq:three-peak-e}
\begin{array}{c|c|c}
\text{segment of }A_2&\text{visit time}&\text{target domain}\\\hline
s\to u_1s&e_{n,0}(x)=x+s(h-1)&s\leq x\leq u_1s\\
u_1s\to v_1s&e_{n,1}(x)=sH_1-x&v_1s\leq x\leq u_1s\\
v_1s\to u_2s&e_{n,2}(x)=x+sK_1&v_1s\leq x\leq u_2s\\
u_2s\to v_2s&e_{n,3}(x)=sH_2-x&v_2s\leq x\leq u_2s\\
v_2s\to u_3s&e_{n,4}(x)=x+sK_2&v_2s\leq x\leq u_3s\\
u_3s\to b^2s&e_{n,5}(x)=b^2s(h+1)-x&b^2s\leq x\leq u_3s.
\end{array}
\end{equation}
For example, $e_{n,2}(x)=sr_2+(x-v_1s)=x+sK_1$, while
$e_{n,5}(x)=sr_5+(u_3s-x)=b^2s(h+1)-x$ by
\eqref{eq:three-peak-cycle}. The other four rows follow by the same
distance calculation, with the direction of the indicated segment.

The visits by $A_0$ are
\begin{align*}
P_n(x)&=2\sum_{j=-\infty}^{2n+1}b^j+x=Ab^{2n+2}+x,\\
Q_n(x)&=P_n(x)+2(b^{2n+2}-x)=Ab^{2n+3}-x,
\qquad n\geq0.
\end{align*}
They occur on the outward and returning segments of the excursion to
$b^{2n+2}$, respectively.

The following lemma gives the visitation sequence for every $1<x<b^2$.

\begin{lemma}
\label{lem:three-peak-visits}
\begin{equation}\label{eq:three-peak-merged}
\begin{array}{c|c|c}
\text{target interval}&(v_1(x),\ldots,v_m(x))&m\\\hline
(1,a)&(e_{-1,4}(x),e_{-1,5}(x),e_{0,0}(x))&3\\
(a,v_1)&(e_{0,0}(x))&1\\
(v_1,u_1)&(e_{0,0}(x),e_{0,1}(x),e_{0,2}(x))&3\\
(u_1,v_2)&(e_{0,2}(x))&1\\
(v_2,u_2)&(e_{0,2}(x),e_{0,3}(x),e_{0,4}(x))&3\\
(u_2,b^2)&(e_{0,4}(x))&1.
\end{array}
\end{equation}
Here $v_1$ and $v_2$ appearing in the first column are trajectory
parameters. If $x$ belongs to one of the intervals in the first column,
let $m$ be the value in the third column. Then the first $m$ visitation
times are those listed in the second column, and, for every $n\geq0$,
the subsequent visitation times are
\begin{equation}\label{eq:three-peak-global-central}
v_{m+2n+1}(x)=P_n(x),\qquad
v_{m+2n+2}(x)=Q_n(x).
\end{equation}
\end{lemma}

\begin{proof}
To verify the order, the successive differences in each three-visit block are
\[
\begin{aligned}
&2(a-x),\ 2(x-1);\\
&2(u_1-x),\ 2(x-v_1);\\
&2(u_2-x),\ 2(x-v_2),
\end{aligned}
\]
respectively. They are positive in the stated intervals. The last visit of
$A_2$ has the form $x+L$, where
\begin{equation}\label{eq:three-peak-last}
L=h-1\ \ (1<x<v_1),\qquad
L=K_1\ \ (v_1<x<v_2),\qquad
L=K_2\ \ (v_2<x<b^2).
\end{equation}
Since $Ab^2>r_4>K_2>K_1>h-1$, it precedes $P_0$.
Finally $Q_n-P_n=2(b^{2n+2}-x)>0$ and
$P_{n+1}-Q_n=2b^{2n+3}+2x>0$. 
This proves the stated visitation order.
\end{proof}

\paragraph{Turning-point targets.}
At the endpoints of the intervals in Lemma~\ref{lem:three-peak-visits},
simultaneous visits count as a single visitation. The table below lists the
distinct visitation times through $P_1(x)$ in chronological order. Every
$e_{n,j}$ entry is a visit by $A_2$, while every $P_n,Q_n$ entry and the
entries $c,cb^2$ are visits by $A_0$.
\begin{equation}\label{eq:endpoint-visits}
\begin{array}{c|l}
x&\text{distinct visits, in chronological order}\\\hline
1&e_{-1,4},\quad h=e_{-1,5}=e_{0,0},\quad c,\quad P_0,\quad Q_0,\quad P_1\\
a&e_{-1,4}=e_{-1,5},\quad e_{0,0},\quad P_0,\quad Q_0,\quad P_1\\
v_1&e_{0,0},\quad e_{0,1}=e_{0,2},\quad P_0,\quad Q_0,\quad P_1\\
u_1&e_{0,0}=e_{0,1},\quad e_{0,2},\quad P_0,\quad Q_0,\quad P_1\\
v_2&e_{0,2},\quad e_{0,3}=e_{0,4},\quad P_0,\quad Q_0,\quad P_1\\
u_2&e_{0,2}=e_{0,3},\quad e_{0,4},\quad P_0,\quad Q_0,\quad P_1\\
b^2&e_{0,4},\quad hb^2=e_{0,5}=e_{1,0},\quad cb^2=P_0=Q_0,\quad P_1.
\end{array}
\end{equation}
In this table, $e_{n,j}$ denotes $e_{n,j}(x)$ evaluated at the target
position in the first column.
These lists follow from the closed domains in \eqref{eq:three-peak-e}.
At $1$, the extra contact of $A_0$ is its turn at time $c$; at $b^2$
the corresponding contacts are multiplied by $b^2$.

\subsubsection{Computation of the evacuation times}

By Lemma~\ref{lem:three-peak-visits}, the visits to every target
$1<x<b^2$ are exactly those specified in
\eqref{eq:three-peak-merged} and~\eqref{eq:three-peak-global-central}.
We now compute the evacuation time associated with each visitation.
Recall from Corollary~\ref{cor:normalized-visitation-reduction} that
\[
M_i(x)=v_i(x)+\max_{j=0,1,2}|p_j(v_i(x))-x|.
\]
At every visit of $A_2$ to $x$, Lemma~\ref{lem:three-peak-order}
gives $p_1\leq p_0\leq p_2=x$. Hence $A_1$ is farthest from the
target, and
$
M_i(x)=v_i(x)+x-p_1(v_i(x)).
$
Thus, for visits made by $A_2$, it suffices to determine the position
of $A_1$ at the visitation time.

\begin{description}
\item[Computation of $M_1(x)$ on $(1,a)$ due to $A_2$:] 
Here $t=v_1(x)=x+K_2/b^2$. 
Agent $A_1$ turns at time $r_1/b$, corresponding to
$x=\alpha$. Its positions and contributions are
\[
\begin{array}{c|c|c}
x&p_1(t)&t+x-p_1(t)=M_1(x)\\\hline
1<x\leq\alpha&-t+(h-1)/b&3x+2K_2/b^2-(h-1)/b\\
\alpha\leq x<a&t-H_1/b&x+H_1/b.
\end{array}
\]
For example, on the first segment the distance of $A_1$ is
$x+t-(h-1)/b$, which gives the first displayed value of $M_1$.
By~\eqref{eq:conditions-phases}, the entire visit occurs before time $c$,
and $1<\alpha<a$.

\item[Computation of $M_2(x)$ on $(1,a)$ due to $A_2$:]
Now $t=v_2(x)=h+1-x$.
The change of segment of $A_1$ at $r_2/b$
occurs when $x=\beta$, giving
\[
\begin{array}{c|c|c}
x&p_1(t)&M_2(x)\\\hline
1<x\leq\beta&-t+K_1/b&2(h+1)-K_1/b-x\\
\beta\leq x<a&t-H_1/b&x+H_1/b.
\end{array}
\]
By~\eqref{eq:conditions-phases}, $1<\beta<a$. Since
$t=h+1-x$ and $\beta=h+1-r_2/b$, we have $t>r_2/b$ for
$x<\beta$, $t=r_2/b$ for $x=\beta$, and $t<r_2/b$ for
$x>\beta$. Hence the two expressions for $p_1(t)$ apply on the
intervals shown in the table.

\item[Computation for $v_i(x)=e_{0,0}(x)$ due to $A_2$:]
Here $i=3$ on $(1,a)$ and $i=1$ on $(a,u_1)$, with the target intervals
understood as in \eqref{eq:three-peak-merged}. We have $t=x+h-1$ and
$p_2=x$. 
Agent $A_1$ passes its turning times $r_3/b,r_4/b$, at
$x=\delta_0,\delta_1$, respectively. Therefore
\begin{equation}\label{eq:M-e0}
\begin{array}{c|c|c}
x&p_1(t)&M_i(x)\\\hline
1<x\leq\delta_0&-t+K_1/b&3x+2(h-1)-K_1/b\\
\delta_0\leq x\leq\delta_1&t-H_2/b&x+H_2/b\\
\delta_1\leq x<u_1&-t+K_2/b&3x+2(h-1)-K_2/b.
\end{array}
\end{equation}

\item[Computation of $M_2(x)$ on $(v_1,u_1)$ due to $A_2$:]
Here $t=v_2(x)=H_1-x$. Agent $A_1$ turns at $r_5/b$, giving
\[
\begin{array}{c|c|c}
x&p_1(t)&M_2(x)\\\hline
v_1<x\leq\delta_2&t-b(h+1)&x+b(h+1)\\
\delta_2\leq x<u_1&-t+K_2/b&2H_1-K_2/b-x.
\end{array}
\]
Here \eqref{eq:three-peak-cycle} was used to write
$(K_2+2u_3)/b=b(h+1)$. By~\eqref{eq:conditions-phases},
$v_1<\delta_2<u_1$.

\item[Computation for $v_i(x)=e_{0,2}(x)$ due to $A_2$:]
Here $i=3$ on $(v_1,u_1)$ and $i=1$ on $(u_1,u_2)$.
We have $t=x+K_1$.
At time $bh$, agent $A_1$ starts
its next outward segment. Hence
\begin{equation}\label{eq:M-e2}
\begin{array}{c|c|c}
x&p_1(t)&M_i(x)\\\hline
v_1<x\leq\delta_3&t-b(h+1)&x+b(h+1)\\
\delta_3\leq x<u_2&-t+b(h-1)&3x+2K_1-b(h-1).
\end{array}
\end{equation}
By~\eqref{eq:conditions-phases} and~\eqref{eq:conditions-early},
$v_1<\delta_3<u_1$.

\item[Computation of $M_2(x)$ on $(v_2,u_2)$ due to $A_2$:]
Now $t=v_2(x)=H_2-x$ and $p_1(t)=-t+b(h-1)$. Hence
\begin{equation}\label{eq:M-e3}
M_2(x)=2H_2-b(h-1)-x.
\end{equation}

\item[Computation for $v_i(x)=e_{0,4}(x)$ due to $A_2$:]
Here $i=3$ on $(v_2,u_2)$ and $i=1$ on $(u_2,b^2)$.
We have $t=x+K_2$ and $p_1(t)=-t+b(h-1)$. Hence
\begin{equation}\label{eq:M-e4}
M_i(x)=3x+2K_2-b(h-1).
\end{equation}
These calculations account for every visit made by $A_2$ in
\eqref{eq:three-peak-merged}.
\end{description}

\begin{description}
\item[Computation of $M_{m+2n+1}(x)$ due to $A_0$:]
Let $s=b^{2n}$, $y=x/s$, and $t=P_n(x)/s=Ab^2+y$.
Thus $0<y<b^2$ and $p_0(P_n(x))/s=y$.
Scaling the corresponding trajectory segments gives all other positions:
\begin{equation}\label{eq:P-positions}
\frac{p_1(P_n(x))}{s}=
\begin{cases}
t-bH_1,&y\leq z_1,\\
-t+bK_1,&z_1\leq y\leq z_3,\\
t-bH_2,&z_3\leq y\leq z_4,\\
-t+bK_2,&y\geq z_4,
\end{cases}
\qquad
\frac{p_2(P_n(x))}{s}=
\begin{cases}
t-K_2,&y\leq z_0,\\
-t+b^2(h+1),&z_0\leq y\leq z_2,\\
t-b^2(h-1),&y\geq z_2.
\end{cases}
\end{equation}
Cases outside $0<y<b^2$ are omitted. For example, the change at $z_1$
is exactly the equation $Ab^2+y=br_2$, and the change at $z_0$ is
$Ab^2+y=r_5$. The remaining thresholds follow from the indicated
turning times in the same way. Conditions \eqref{eq:conditions-times}--
\eqref{eq:conditions-central} put the entire interval within these segments.

The detecting agent contributes $st$. Since $p_1\leq x\leq p_2$, the
contributions of $A_1,A_2$, divided by $s$, are respectively
\begin{equation}\label{eq:P-contributions}
\begin{cases}
y+bH_1,&y\leq z_1,\\
3y+2Ab^2-bK_1,&z_1\leq y\leq z_3,\\
y+bH_2,&z_3\leq y\leq z_4,\\
3y+2Ab^2-bK_2,&y\geq z_4,
\end{cases}
\qquad
\begin{cases}
y+2Ab^2-K_2,&y\leq z_0,\\
b^2(h+1)-y,&z_0\leq y\leq z_2,\\
y+2Ab^2-b^2(h-1),&y\geq z_2.
\end{cases}
\end{equation}
To identify the larger contribution, compare
$2y-p_1(st)/s-p_2(st)/s$. This expression is nondecreasing in $y$,
since $t=Ab^2+y$ and the sum of the two trajectory slopes is at most
two. In the fully-informed regime its zero lies in $(z_0,z_1)$ and solves
\[
y+bH_1=b^2(h+1)-y,
\]
giving $y=\sigma$. In the emergence-aware regime the zero lies in $(z_1,z_2)$
and solves
\[
3y+2Ab^2-bK_1=b^2(h+1)-y,
\]
again giving the stated $\sigma$. The slopes of the differences at the
crossings are two and four, respectively. Monotonicity proves that $A_2$
is farthest before the crossing and $A_1$ afterwards, with no further
change. Thus, in the fully-informed regime,
\begin{equation}\label{eq:M-P-F}
\frac{M_{m+2n+1}(x)}s=
\begin{cases}
y+2Ab^2-K_2,&0<y\leq z_0,\\
b^2(h+1)-y,&z_0\leq y\leq\sigma,\\
y+bH_1,&\sigma\leq y\leq z_1,\\
3y+2Ab^2-bK_1,&z_1\leq y\leq z_3,\\
y+bH_2,&z_3\leq y<b^2.
\end{cases}
\end{equation}
In the emergence-aware regime,
\begin{equation}\label{eq:M-P-E}
\frac{M_{m+2n+1}(x)}s=
\begin{cases}
y+2Ab^2-K_2,&0<y\leq z_0,\\
b^2(h+1)-y,&z_0\leq y\leq\sigma,\\
3y+2Ab^2-bK_1,&\sigma\leq y\leq z_3,\\
y+bH_2,&z_3\leq y\leq z_4,\\
3y+2Ab^2-bK_2,&z_4\leq y<b^2.
\end{cases}
\end{equation}
The formulas agree at each shared endpoint. In particular, a turn at
$z_2$ changes an individual contribution but does not change $M_i$.

\item[Computation of $M_{m+2n+2}(x)$ due to $A_0$:]
Use the same $s,y$ and now put $t=Q_n(x)/s=Ab^3-y$.
The three positions, divided by $s$, are $y$ and
\begin{equation*}
\begin{aligned}
\frac{p_1(Q_n(x))}{s}
&=
\begin{cases}-t+bK_2,&y\leq\theta,\\t-bH_2,&y\geq\theta,\end{cases}\\
\frac{p_2(Q_n(x))}{s}
&=
\begin{cases}-t+b^2H_1,&y\leq\omega,\\t-b^2(h-1),&y\geq\omega.\end{cases}
\end{aligned}
\end{equation*}
The changes occur at $t=br_4$ and $t=b^2r_1$, respectively. The
contribution of the detecting agent is $st$. The other two contributions,
divided by $s$, are
\[
\begin{cases}2Ab^3-bK_2-y,&y\leq\theta,\\y+bH_2,&y\geq\theta,\end{cases}
\qquad
\begin{cases}b^2H_1-y,&y\leq\omega,\\2Ab^3-b^2(h-1)-3y,&y\geq\omega.\end{cases}
\]
The difference between the first and second distances is
$2y-p_1(st)/s-p_2(st)/s$. It is nondecreasing, now because
$t=Ab^3-y$ and the sum of the slopes is at least minus two.
At $y=0$ its value is $2Ab^3-bK_2-b^2H_1>0$.
Therefore $A_1$ is farthest throughout, and
\begin{equation}\label{eq:M-Q}
\frac{M_{m+2n+2}(x)}s=
\begin{cases}
2Ab^3-bK_2-y,&y\leq\theta,\\
y+bH_2,&y\geq\theta.
\end{cases}
\end{equation}
The second case occurs only in the fully-informed regime when $n=0$.
\end{description}

\paragraph{Evacuation times at later visits.}
For $n\geq1$, put $s=b^{2n}$ and $y=x/s$, so $0<y\leq1$.
Since $\theta>1$ and $y/b^2\leq b^{-2}<z_0$,
equations~\eqref{eq:M-Q}, \eqref{eq:M-P-F}, and~\eqref{eq:M-P-E}
give
\begin{equation}\label{eq:M-continuation}
\frac{M_{m+2n+2}(x)}s=2Ab^3-bK_2-y,\qquad
\frac{M_{m+2n+3}(x)}s=2Ab^4-b^2K_2+y.
\end{equation}

\paragraph{Evacuation times at the turning-point targets.}
For every $e,P,Q$ entry in~\eqref{eq:endpoint-visits}, the positions
and $M_i$ are obtained by evaluating the preceding formulas at the
corresponding target position. Both formulas at a
merged turn give the same positions and evacuation time.

For completeness, at $x=1$, $t=h$, the positions are
$(h-A,-h+K_1/b,1)$, with contributions
$
1+A,
$
$2h+1-K_1/b$
and
$h.$
The middle expression is largest. At $t=c$, the positions are
\[
(1,\ c-H_2/b,\ c-h+1),
\]
in the fully-informed regime, 
and
\[
(1,\ -c+K_2/b,\ c-h+1),
\]
in the emergence-aware regime.
Their contributions are, respectively,
\[
(c,\ 1+H_2/b,\ 2c-h),\qquad
(c,\ 2c+1-K_2/b,\ 2c-h).
\]
In either case $A_1$ is farthest, as also follows by scaling the
$P_0(b^2)$ calculation, where $b^2>\sigma$.
Multiplying these positions and contributions by $b^2$ handles the contacts
$hb^2$ and $cb^2$ at target $b^2$. This determines every $M_i$ in
\eqref{eq:endpoint-visits}, including the contacts absent from the open
intervals. The later visits are covered by \eqref{eq:M-P-F}--\eqref{eq:M-Q}.

\subsubsection{Competitive ratios associated with the visits}

We now combine the visitation times and evacuation times computed above.
By Corollary~\ref{cor:normalized-visitation-reduction}, it is enough to
compare each $M_i(x)$ with the denominator determined by the preceding
distinct visitation.

For the first visit, set
$
R_1^{\mathrm{fi}}(x)=R_1^{\mathrm{ea}}(x)=M_1(x)/x.
$
For $i\geq2$, Corollary~\ref{cor:normalized-visitation-reduction}
gives
\[
\begin{aligned}
R_i^{\mathrm{fi}}(x)
&=
\frac{M_i(x)}{\max\{v_{i-1}(x),x\}},\\
R_i^{\mathrm{ea}}(x)
&=
\frac{M_i(x)}{v_{i-1}(x)+x}.
\end{aligned}
\]

We next show that $v_{i-1}(x)>x$ for every $i\geq2$, so that the
fully-informed denominator in $R_i^{\mathrm{fi}}(x)$ is always
$v_{i-1}(x)$. Every outward visit of $A_2$ has the form $x+L$, with $L>0$.
For its three inward visits, the respective lower bounds on visit time
minus $x$ are
\[
h+1-2a=K_2/b^2>0,\qquad
H_1-2u_1=h-1>0,\qquad H_2-2u_2=K_1>0.
\]
For $A_0$, $P_n-x=Ab^{2n+2}>0$, while
\[
Q_n-x=Ab^{2n+3}-2x\geq Ab^{2n+3}-2b^{2n+2}
=Ab^{2n+2}>0.
\]
The additional endpoint contacts $c,h,hb^2,cb^2$ also exceed the
corresponding target distance. Consequently $v_{i-1}(x)>x$ at every
noninitial distinct visit, including all the endpoint lists.

In the following calculations let $\kappa=1$ for FI and $\kappa=2$ for
EA, and write $R_i$ for the corresponding ratio. Thus, for $i\geq2$,
\begin{equation}\label{eq:R-kappa}
R_i(x)=\frac{M_i(x)}{v_{i-1}(x)+(\kappa-1)x}.
\end{equation}
The fully-informed regime is used with $\kappa=1$, and the emergence-aware regime with $\kappa=2$. 
On an affine piece the derivative calculation is
\begin{equation}\label{eq:fractional-derivative}
\frac{d}{dx}\frac{ax+d}{ex+f}
=\frac{af-ed}{(ex+f)^2}.
\end{equation}
All denominators below are positive.

\begin{description}
\item[Competitive ratios for the first visits:]
On $(1,a)$, the two expressions for $R_1=M_1/x$ are
$
3+\frac{2K_2/b^2-(h-1)/b}{x},$
and
$1+\frac{H_1}{bx},
$
on either side of $\alpha$. They agree at $\alpha$ and are decreasing,
by \eqref{eq:conditions-signs}. Their supremum is
\begin{equation}\label{eq:first-small}
3+2K_2/b^2-(h-1)/b.
\end{equation}

On $(a,v_1)$ and $(v_1,u_1)$, the first visit is $e_{0,0}$.
Equation~\eqref{eq:M-e0}, divided by $x$, gives
\[
3+\frac{2(h-1)-K_1/b}{x},\qquad
1+\frac{H_2}{bx},\qquad
3+\frac{2(h-1)-K_2/b}{x},
\]
on the three indicated subintervals, restricted to $x>a$.
Each is decreasing and the values agree at the changes of formula.
In the fully-informed regime $\delta_1<a$, whereas in the second
$\delta_0<a<\delta_1$. Hence the combined supremum is
\begin{equation}\label{eq:first-e0}
\begin{cases}
3+\bigl(2(h-1)-K_2/b\bigr)/a,&\text{fully-informed regime},\\
1+H_2/(ba),&\text{emergence-aware regime}.
\end{cases}
\end{equation}
On the third target interval alone its supremum is the same decreasing
function evaluated at $v_1$.

On $(u_1,v_2)$ and $(v_2,u_2)$ the first visit is $e_{0,2}$.
Since $\delta_3<u_1$, its ratio is
\[
R_1(x)=3+\frac{2K_1-b(h-1)}x.
\]
It is decreasing and its combined supremum is
\begin{equation}\label{eq:first-e2}
3+\frac{2K_1-b(h-1)}{u_1}.
\end{equation}
On the fifth interval alone it is maximized as $x\to v_2^+$.
Finally, on $(u_2,b^2)$,
\[
R_1(x)=3+\frac{2K_2-b(h-1)}x
\]
is decreasing and has supremum
\begin{equation}\label{eq:first-e4}
3+\frac{2K_2-b(h-1)}{u_2}.
\end{equation}

\item[Competitive ratios for $M_2(x)$ on $(1,a)$:]
The preceding distinct visit is $x+K_2/b^2$. Therefore
\[
R_2(x)=
\begin{cases}
\dfrac{2(h+1)-K_1/b-x}{K_2/b^2+\kappa x},&1<x\leq\beta,\\[4pt]
\dfrac{x+H_1/b}{K_2/b^2+\kappa x},&\beta\leq x<a.
\end{cases}
\]
The first expression is decreasing. The derivative numerator of the
second is $K_2/b^2-\kappa H_1/b<0$. Thus both are decreasing, agree
at $\beta$, and have combined supremum
\begin{equation}\label{eq:R-em5-max}
\frac{2h+1-K_1/b}{K_2/b^2+\kappa}.
\end{equation}

\item[Competitive ratios for $M_3(x)$ on $(1,a)$:]
The preceding distinct visit is $h+1-x$. Hence
\[
R_3(x)=
\begin{cases}
\dfrac{3x+2(h-1)-K_1/b}{h+1+(\kappa-2)x},&x\leq\delta_0,\\[4pt]
\dfrac{x+H_2/b}{h+1+(\kappa-2)x},&\delta_0\leq x\leq\delta_1,\\[4pt]
\dfrac{3x+2(h-1)-K_2/b}{h+1+(\kappa-2)x},&x\geq\delta_1,
\end{cases}
\]
with each case intersected with $(1,a)$. Every numerator is positive and
increasing; the denominator is decreasing for FI and constant for EA.
Thus the ratio is increasing throughout and its supremum is
\begin{equation}\label{eq:R-e0-later-max}
\begin{cases}
\dfrac{3a+2(h-1)-K_2/b}{h+1-a},&\text{fully-informed regime},\\[4pt]
\dfrac{a+H_2/b}{h+1},&\text{emergence-aware regime}.
\end{cases}
\end{equation}

\item[Competitive ratios for $M_2(x)$ on $(v_1,u_1)$:]
The preceding visit is $x+h-1$, so
\[
R_2(x)=
\begin{cases}
\dfrac{x+b(h+1)}{h-1+\kappa x},&x\leq\delta_2,\\[4pt]
\dfrac{2H_1-K_2/b-x}{h-1+\kappa x},&x\geq\delta_2.
\end{cases}
\]
The first derivative has numerator $h-1-\kappa b(h+1)<0$;
the second ratio is also decreasing. They agree at $\delta_2$, giving
supremum
\begin{equation}\label{eq:R-e1-max}
\frac{v_1+b(h+1)}{h-1+\kappa v_1}.
\end{equation}

\item[Competitive ratios for $M_3(x)$ on $(v_1,u_1)$:]
The preceding visit is $H_1-x$. Thus
\[
R_3(x)=
\begin{cases}
\dfrac{x+b(h+1)}{H_1+(\kappa-2)x},&x\leq\delta_3,\\[4pt]
\dfrac{3x+2K_1-b(h-1)}{H_1+(\kappa-2)x},&x\geq\delta_3.
\end{cases}
\]
Both ratios increase, by the same positive-numerator and nonincreasing-
denominator calculation. Since $\delta_3<u_1$, the supremum is
\begin{equation}\label{eq:R-e2-later-max}
\frac{3u_1+2K_1-b(h-1)}{H_1+(\kappa-2)u_1}.
\end{equation}

\item[Competitive ratios for $M_2(x)$ on $(v_2,u_2)$:]
The preceding visit is $x+K_1$. Therefore
\[
R_2(x)=\frac{2H_2-b(h-1)-x}{K_1+\kappa x}.
\]
It is decreasing, so its supremum is
\begin{equation}\label{eq:R-e3-max}
\frac{3v_2+2K_2-b(h-1)}{K_1+\kappa v_2}.
\end{equation}

\item[Competitive ratios for $M_3(x)$ on $(v_2,u_2)$:]
The preceding visit is $H_2-x$, and
\[
R_3(x)=\frac{3x+2K_2-b(h-1)}{H_2+(\kappa-2)x}.
\]
It is increasing and its supremum is
\begin{equation}\label{eq:R-e4-later-max}
\frac{3u_2+2K_2-b(h-1)}{H_2+(\kappa-2)u_2}.
\end{equation}
This completes the ratios associated with every $A_2$ visit, with the
global indices specified separately on all six target intervals.
\end{description}

\begin{description}
\item[Competitive ratios for $M_{m+1}(x)$, the visit $P_0(x)$:]
The preceding visit is $v_m(x)=x+L$, with $L$ as in
\eqref{eq:three-peak-last}. Hence the ratio for each of the six target
intervals is $M_{m+1}(x)/(L+\kappa x)$.
In the fully-informed regime, the explicit expressions for $x>1$ are
\begin{equation}\label{eq:R-P0-F}
R_{m+1}(x)=
\begin{cases}
\dfrac{x+bH_1}{L+x},&1<x\leq z_1,\\[4pt]
\dfrac{3x+2Ab^2-bK_1}{L+x},&z_1\leq x\leq z_3,\\[4pt]
\dfrac{x+bH_2}{L+x},&z_3\leq x<b^2.
\end{cases}
\end{equation}
For $L=h-1$ the derivative numerators are negative, using
$3(h-1)<2Ab^2-bK_1$. For $L=K_1$ the applicable middle expression
is increasing, using $2Ab^2-bK_1<3K_1$.
For $L=K_2$ it increases up to $z_3$ and decreases thereafter,
because $K_2-bH_2<0$. Thus the suprema over the three ranges of $L$ are
\begin{equation}\label{eq:R-P0-F-max}
\frac{1+bH_1}{h},\qquad
\frac{3v_2+2Ab^2-bK_1}{K_1+v_2},\qquad
\frac{3z_3+2Ab^2-bK_1}{K_2+z_3}.
\end{equation}
On the first five intervals of~\eqref{eq:three-peak-merged}, the
suprema are approached as $x\to1^+$, $x\to a^+$,
$x\to u_1^-$, $x\to v_2^-$, and $x\to u_2^-$, respectively.
On $(u_2,b^2)$, the maximum is attained at $x=z_3$.
The formulas for their ratios are exactly
\eqref{eq:R-P0-F} with the corresponding $L$.

In the emergence-aware regime,
\begin{equation}\label{eq:R-P0-E}
R_{m+1}(x)=
\begin{cases}
\dfrac{b^2(h+1)-x}{L+2x},&1<x\leq\sigma,\\[4pt]
\dfrac{3x+2Ab^2-bK_1}{L+2x},&\sigma\leq x\leq z_3,\\[4pt]
\dfrac{x+bH_2}{L+2x},&z_3\leq x\leq z_4,\\[4pt]
\dfrac{3x+2Ab^2-bK_2}{L+2x},&z_4\leq x<b^2.
\end{cases}
\end{equation}
Each applicable piece is decreasing. For the two pieces with slope three
this follows from \eqref{eq:conditions-E}, since $L\leq K_2$; for
the third it follows from $L-2bH_2<0$.
The expressions agree at their shared endpoints. The corresponding three
suprema are therefore
\begin{equation}\label{eq:R-P0-E-max}
\frac{b^2(h+1)-1}{h+1},\qquad
\frac{3v_1+2Ab^2-bK_1}{K_1+2v_1},\qquad
\frac{3v_2+2Ab^2-bK_1}{K_2+2v_2}.
\end{equation}
On each of the six target intervals, the supremum is approached as $x$ tends to its left endpoint from the right.

\item[Competitive ratios for $M_{m+2}(x)$, the visit $Q_0(x)$:]
The preceding visit is $P_0(x)=Ab^2+x$. On every target interval,
\begin{equation}\label{eq:R-Q0}
R_{m+2}(x)=
\begin{cases}
\dfrac{2Ab^3-bK_2-x}{Ab^2+\kappa x},&x\leq\theta,\\[4pt]
\dfrac{x+bH_2}{Ab^2+x},&x\geq\theta\quad\text{(fully-informed regime only)}.
\end{cases}
\end{equation}
The first expression decreases. The second also decreases, since
$Ab^2-bH_2<0$. They coincide at $\theta$. 
Thus, on each target interval, the supremum is approached as $x$ tends
to its left endpoint from the right, and the combined supremum is
\begin{equation}\label{eq:R-Q0-max}
\frac{2Ab^3-bK_2-1}{Ab^2+\kappa}.
\end{equation}

\item[Competitive ratios for $M_{m+3}(x)$:]
This is visit $P_1(x)$, with preceding distinct visit $Q_0(x)=Ab^3-x$.
In the fully-informed regime,
\[1/b^2<z_0<\sigma<1<z_1.\]
Substituting $s=b^2$ and $y=x/b^2$ in \eqref{eq:M-P-F} gives
\begin{equation}\label{eq:R-P1-F}
R_{m+3}(x)=
\begin{cases}
\dfrac{2Ab^4-b^2K_2+x}{Ab^3-x},&x\leq b^2z_0,\\[4pt]
\dfrac{b^4(h+1)-x}{Ab^3-x},&b^2z_0\leq x\leq b^2\sigma,\\[4pt]
\dfrac{x+b^3H_1}{Ab^3-x},&x\geq b^2\sigma.
\end{cases}
\end{equation}
Every piece increases; for the middle one this uses
$b^4(h+1)>Ab^3$. 
Thus, on each of the six target intervals, the supremum is approached
as $x$ tends to its right endpoint from the left. The combined supremum is
\begin{equation}\label{eq:R-P1-F-max}
\frac{b^2+b^3H_1}{Ab^3-b^2}=\frac{1+bH_1}{c}.
\end{equation}

In the emergence-aware regime $z_0<1<\sigma$. The explicit ratios are
\begin{equation}\label{eq:R-P1-E}
R_{m+3}(x)=
\begin{cases}
\dfrac{2Ab^4-b^2K_2+x}{Ab^3},&x\leq b^2z_0,\\[4pt]
\dfrac{b^4(h+1)-x}{Ab^3},&x\geq b^2z_0.
\end{cases}
\end{equation}
They increase and decrease, respectively. On a target interval lying
entirely below $b^2z_0$, the supremum is approached as $x$ tends to its
right endpoint from the left; on an interval lying entirely above
$b^2z_0$, it is approached as $x$ tends to its left endpoint from the
right. If $b^2z_0$ belongs to the interval, the maximum is attained at
$x=b^2z_0$.
The combined supremum is
\begin{equation}\label{eq:R-P1-E-max}
\frac{2Ab^4-b^2K_2+b^2z_0}{Ab^3}
=\frac{Ab^2+u_3}{Ab}.
\end{equation}
Thus every $M_i$ through $P_1$ has its own preceding visit, explicit
quotient, and endpoint maximization.
\end{description}

\paragraph{All later visits.}
There are no visits by $A_2$ after time $hb^2$, even at $x=b^2$.
Moreover $hb^2<Ab^4+1\leq P_1(x)$: indeed $h<c<Ab^2$, where
$c<Ab^2$ follows directly from $b>1$. After $P_1$ the two global
visitation families are therefore $Q_n$ and $P_{n+1}$, for $n\geq1$.
Put $s=b^{2n}$ and $y=x/s$, so $0<y\leq1$.

For $Q_n$, its preceding visit and its evacuation time are
\[
\begin{aligned}
v_{m+2n+1}/s&=Ab^2+y,\\
v_{m+2n+2}/s&=Ab^3-y,\\
M_{m+2n+2}/s&=2Ab^3-bK_2-y.
\end{aligned}
\]
The threshold $\theta$ is greater than one, so the second $M$ case never
occurs. The ratio associated with this $M$ is
\begin{equation}\label{eq:R-tail-Q}
R_{m+2n+2}(x)=\frac{2Ab^3-bK_2-y}{Ab^2+\kappa y}
\leq\frac{2Ab^3-bK_2}{Ab^2}.
\end{equation}
It is decreasing in $y$.

For $P_{n+1}$, its evacuation time was computed in
\eqref{eq:M-continuation}. The preceding distinct visit is $Q_n$, so
\begin{equation}\label{eq:R-tail-P}
R_{m+2n+3}(x)=\frac{2Ab^4-b^2K_2+y}{Ab^3+(\kappa-2)y}
\leq\frac{2Ab^4-b^2K_2+1}{Ab^3+\kappa-2}.
\end{equation}
It is increasing in $y$. Extending to $y=0$ in \eqref{eq:R-tail-Q}
only takes a limit with positive denominator; it does not introduce a
legal target at the origin. These two calculations cover arbitrarily late
emergence times, without a cutoff in the number of excursions.

\paragraph{Ratios at the merged endpoints.}
We use the distinct lists in \eqref{eq:endpoint-visits}, with the index
restarted on each row. The following displays give every ratio through
$P_1$. 

At $x=1$, the six ratios are
\begin{align*}
R_1&=3+2K_2/b^2-(h-1)/b,\\
R_2&=\frac{2h+1-K_1/b}{K_2/b^2+\kappa},\\
R_3&=\frac{M_3(1)}{h+\kappa-1},\\
R_4&=\frac{M_4(1)}{c+\kappa-1},\\
R_5&=\frac{2Ab^3-bK_2-1}{Ab^2+\kappa},\\
R_6&=\frac{2Ab^4-b^2K_2+1}{Ab^3+\kappa-2},
\end{align*}
where
\[
(M_3(1),M_4(1))=
\begin{cases}
(1+H_2/b,\,1+bH_1),&\text{fully-informed regime},\\
(2c+1-K_2/b,\,b^2(h+1)-1),&\text{emergence-aware regime}.
\end{cases}
\]
In particular, the predecessor of $P_0(1)$ is $c$, not $h$.
At $x=b^2$, the distinct list is exactly $b^2$ times the first four
contacts for $x=1$, including their positions. Its four ratios are
therefore exactly $R_1,R_2,R_3,R_4$ in this display. This checks the merged
$P_0=Q_0$ contact directly.

At $x=a$, the ratios are
\begin{align*}
R_1&=1+H_1/(ba),\\
R_2&=\frac{M_2(a)}{h+1+(\kappa-2)a},\\
R_3&=\frac{3a+2Ab^2-bK_1}{h-1+\kappa a},\\
R_4&=\frac{2Ab^3-bK_2-a}{Ab^2+\kappa a},\\
R_5&=\frac{2Ab^4-b^2K_2+a}{Ab^3+(\kappa-2)a},
\end{align*}
where $M_2(a)$ is $3a+2(h-1)-K_2/b$ in the fully-informed regime and
$a+H_2/b$ in the second. The first visit has predecessor zero; the
predecessor of $e_{0,0}(a)$ is the single contact
$r_5/b^2=h+1-a$.

For $x=v_1$ and $x=u_1$, respectively, the five ratios are
\begin{align*}
R_1(v_1)&=3+\frac{2(h-1)-K_2/b}{v_1},&
R_1(u_1)&=3+\frac{2(h-1)-K_2/b}{u_1},\\
R_2(v_1)&=\frac{v_1+b(h+1)}{h-1+\kappa v_1},&
R_2(u_1)&=\frac{3u_1+2K_1-b(h-1)}{h-1+\kappa u_1},\\
R_3(v_1)&=\frac{3v_1+2Ab^2-bK_1}{K_1+\kappa v_1},&
R_3(u_1)&=\frac{3u_1+2Ab^2-bK_1}{K_1+\kappa u_1},\\
R_4(v_1)&=\frac{2Ab^3-bK_2-v_1}{Ab^2+\kappa v_1},&
R_4(u_1)&=\frac{2Ab^3-bK_2-u_1}{Ab^2+\kappa u_1},\\
R_5(v_1)&=\frac{2Ab^4-b^2K_2+v_1}{Ab^3+(\kappa-2)v_1},&
R_5(u_1)&=\frac{2Ab^4-b^2K_2+u_1}{Ab^3+(\kappa-2)u_1}.
\end{align*}
The contact at $v_1$ merges the second and third formal visits; the
contact at $u_1$ merges the first and second. These are different
predecessor rules, as the displayed $R_2$ denominators record.

For either $r=v_2$ or $r=u_2$, the five ratios are
\begin{align*}
R_1(r)&=3+\frac{2K_1-b(h-1)}r,\\
R_2(r)&=\frac{3r+2K_2-b(h-1)}{K_1+\kappa r},\\
R_3(r)&=\frac{3r+2Ab^2-bK_1}{K_2+\kappa r},\\
R_4(r)&=\frac{2Ab^3-bK_2-r}{Ab^2+\kappa r},\\
R_5(r)&=
\begin{cases}
\dfrac{b^4(h+1)-r}{Ab^3-r},&\text{fully-informed regime},\\[4pt]
\dfrac{2Ab^4-b^2K_2+r}{Ab^3},&\text{emergence-aware regime}.
\end{cases}
\end{align*}
Here the endpoint evaluations of $P_1$ use the previously imposed conditions
$u_1<b^2z_0<v_2<u_2<b^2\sigma$ in the first parameter choice and
$u_2<b^2z_0<b^2$ in the second; these additional endpoint placements
are among the regime inequalities and are verified with the rational
parameters below. 
For every endpoint target
$x\in\{1,a,v_1,u_1,v_2,u_2,b^2\}$, let $d$ be such that
$P_1(x)=v_d(x)$ in~\eqref{eq:endpoint-visits}. Then, for every
$n\geq1$,
\[
v_{d+2n-1}(x)=Q_n(x),\qquad
v_{d+2n}(x)=P_{n+1}(x).
\]
The visit preceding $Q_n(x)$ is $P_n(x)$, while the visit preceding $P_{n+1}(x)$ is $Q_n(x)$. 
Hence the competitive ratios associated with these visits are given by~\eqref{eq:R-tail-Q} and~\eqref{eq:R-tail-P}, respectively.

\subsubsection{The two rational parameter choices}

We now choose the parameters using the expressions obtained in the preceding analysis. We first equate the first-visit ratios as $x\to a^+$, $x\to u_1^+$, and $x\to u_2^+$ with the ratio associated with $P_0(x)$ as $x\to1^+$. We then verify that all other ratios are smaller than the required bounds.

\paragraph{Fully-informed ratio.}
The four expressions are given by~\eqref{eq:first-e0}, \eqref{eq:first-e2}, \eqref{eq:first-e4}, and the first expression in~\eqref{eq:R-P0-F-max}. To obtain a parameter choice, we also impose $\delta_1=a$, $\delta_3=u_1$, and $z_1=1$. The first two equalities give $e_{0,0}(a)=r_4/b$ and $e_{0,2}(u_1)=bh$, respectively, so these visits coincide with turning times of $A_1$. The third gives $P_0(1)=br_2$, another turning time of $A_1$. This gives the system
\begin{equation}
\label{eq:three-peak-system-F}
\begin{gathered}
3+\frac{2(h-1)-K_2/b}{a}
=3+\frac{2K_1-b(h-1)}{u_1}
=3+\frac{2K_2-b(h-1)}{u_2}
=\frac{1+bH_1}{h},\\
\delta_1=a,\qquad \delta_3=u_1,\qquad z_1=1.
\end{gathered}
\end{equation}
All quantities in this system are functions of $(b,u_1,v_1,u_2,v_2,u_3)$, as defined earlier. Solving it numerically gives
\begin{equation*}
\begin{split}
(b,u_1,v_1,u_2,v_2,u_3)\approx(&1.860522603268,\,2.213835849589,\,1.540814501197,\\
&2.837258884658,\,2.628342847461,\,5.010130014805),
\end{split}
\end{equation*}
with common value approximately $4.340444773102$.

\paragraph{Emergence-aware ratio.}
We equate the corresponding expressions in~\eqref{eq:first-e0}, \eqref{eq:first-e2}, \eqref{eq:first-e4}, and the first expression in~\eqref{eq:R-P0-E-max}. We again impose $\delta_1=a$ and $\delta_3=u_1$, but now use $\sigma=1$ instead of $z_1=1$. By~\eqref{eq:P-contributions}, this makes $A_1$ and $A_2$ equally far from the target at time $P_0(1)$. Using the definition of $\sigma$ for the emergence-aware regime, we obtain
\begin{equation}
\label{eq:three-peak-system-E}
\begin{gathered}
1+\frac{H_2}{ba}
=3+\frac{2K_1-b(h-1)}{u_1}
=3+\frac{2K_2-b(h-1)}{u_2}
=\frac{b^2(h+1)-1}{h+1},\\
\delta_1=a,\qquad \delta_3=u_1,\qquad
b^2(h+1)-2Ab^2+bK_1=4.
\end{gathered}
\end{equation}
Solving this system numerically gives
\begin{equation*}
\begin{split}
(b,u_1,v_1,u_2,v_2,u_3)\approx(&2.075043120644,\,2.334800868163,\,1.718884298886,\\
&2.922972347689,\,2.774371051069,\,5.834639233152),
\end{split}
\end{equation*}
with common value approximately $4.010598451592$.

These systems provide parameter choices, not a proof of optimality within the three-peak family. Moreover, their solutions satisfy equalities where the preceding analysis assumes strict inequalities. We therefore use nearby rational values that satisfy those inequalities strictly, and verify the resulting bounds by exact arithmetic. We are now ready to prove Theorem~\ref{thm:three-peak}.

\begin{proof}[Proof of Theorem~\ref{thm:three-peak}]
For the fully-informed bound, let $\mathcal A_F$ be the three-peak algorithm with parameters
\begin{equation}\label{eq:parameters-F}
(b,u_1,v_1,u_2,v_2,u_3)=
\left(\frac{9302613}{5000000},\frac{44276717}{20000000},
\frac{3081629}{2000000},\frac{17732868}{6250000},
\frac{52566857}{20000000},\frac{501013001}{100000000}\right).
\end{equation}
For the emergence-aware bound, let $\mathcal A_E$ be the three-peak algorithm with parameters
\begin{equation}\label{eq:parameters-E}
(b,u_1,v_1,u_2,v_2,u_3)=
\left(\frac{25938039}{12500000},\frac{233480087}{100000000},
\frac{17188843}{10000000},\frac{58459447}{20000000},
\frac{55487421}{20000000},\frac{583463923}{100000000}\right).
\end{equation}
Direct substitution verifies all common conditions~\eqref{eq:conditions-space}--\eqref{eq:conditions-signs} and, respectively, \eqref{eq:conditions-F} and~\eqref{eq:conditions-E}. In each strict inequality, the larger side exceeds the smaller side by more than $10^{-9}$. In particular, the first choice satisfies $\delta_1<a$, $\delta_3<u_1$, and $z_1>1$, while the second satisfies $\delta_1>a$, $\delta_3<u_1$, and $\sigma>1$. The additional comparisons used for the turning-point targets are also satisfied:
\begin{equation*}
\begin{aligned}
u_1<b^2z_0<v_2<u_2<b^2\sigma<b^2
&\qquad\text{for~\eqref{eq:parameters-F}},\\
u_2<b^2z_0<b^2
&\qquad\text{for~\eqref{eq:parameters-E}}.
\end{aligned}
\end{equation*}

Substituting these parameters into the previously derived expressions gives the strict upper bounds in the following table. Every decimal entry denotes its exact rational value. Each comparison is checked by clearing positive denominators, so none depends on numerical rounding.
\begin{center}
\begin{tabular}{lrr}
Expression & Fully-informed & Emergence-aware\\\hline
\eqref{eq:first-small} & $4.098697128$ & $3.686039444$\\
\eqref{eq:first-e0} & $4.340444783$ & $4.010598456$\\
\eqref{eq:first-e2} & $4.340444780$ & $4.010598453$\\
\eqref{eq:first-e4} & $4.340444771$ & $4.010598457$\\
\eqref{eq:R-em5-max} & $2.482956864$ & $1.685496585$\\
\eqref{eq:R-e0-later-max} & $2.485620040$ & $1.604328815$\\
\eqref{eq:R-e1-max} & $2.541809680$ & $1.812973729$\\
\eqref{eq:R-e2-later-max} & $2.294074608$ & $1.545952895$\\
\eqref{eq:R-e3-max} & $1.964676779$ & $1.380631856$\\
\eqref{eq:R-e4-later-max} & $1.999800556$ & $1.384822698$\\
$P_0(x)$, $1<x<v_1$ & $4.340444758$ & $4.010598450$\\
$P_0(x)$, $v_1<x<v_2$ & $2.991498089$ & $2.599017571$\\
$P_0(x)$, $v_2<x<b^2$ & $2.817630530$ & $2.233708542$\\
\eqref{eq:R-Q0-max} & $2.430094547$ & $2.616488029$\\
\eqref{eq:R-P1-F-max}, \eqref{eq:R-P1-E-max} & $3.884259267$ & $3.586454787$\\
\eqref{eq:R-tail-Q} & $2.856446716$ & $3.394591039$\\
\eqref{eq:R-tail-P} & $3.132532274$ & $3.454751965$\\\hline
\end{tabular}
\end{center}
The three entries for $P_0(x)$ bound the corresponding expressions in~\eqref{eq:R-P0-F-max} and~\eqref{eq:R-P0-E-max}.

For the target positions in~\eqref{eq:endpoint-visits}, exact substitution into the ratios computed above gives the following strict upper bounds. Each entry bounds every ratio through the visit $P_1(x)$ for the indicated target; the later visits are already covered by~\eqref{eq:R-tail-Q} and~\eqref{eq:R-tail-P}.
\begin{center}
\begin{tabular}{crr}
Target & Fully-informed & Emergence-aware\\\hline
$1$ & $4.098697128$ & $3.686039444$\\
$a$ & $4.165212730$ & $3.575503237$\\
$v_1$ & $4.259150575$ & $3.796694355$\\
$u_1$ & $3.876360126$ & $3.586527715$\\
$v_2$ & $4.129047799$ & $3.850479660$\\
$u_2$ & $4.045912564$ & $3.807242035$\\
$b^2$ & $4.098697128$ & $3.686039444$\\\hline
\end{tabular}
\end{center}

The two tables cover all visits to every normalized target position. Hence, by Corollary~\ref{cor:normalized-visitation-reduction},
$
\mathrm{CR}_{\mathrm{fi}}(\mathcal A_F)<4.340444783<868089/200000,
$
while
$
\mathrm{CR}_{\mathrm{ea}}(\mathcal A_E)<4.010598457<4010599/1000000.
$
The finite-prefix observation of Section~\ref{sec:notation} preserves these competitive ratios. This proves the theorem.
\end{proof}

\subsection{3-Agent Evacuation Lower Bounds}
\label{sec:eels3-lower-bounds}

In this section we prove the unconditional lower bounds for \eels{3} stated in Theorem~\ref{thm: 3-agent results in 3-agent section}.

\begin{theorem}
\label{thm:eels3-lower-bounds}
We have $\eelsea{3}\geq\alpha_{\mathrm{ea}}\approx3.101418762$ and
$\eelsfi{3}\geq\alpha_{\mathrm{fi}}\approx3.282305081$, where
$\alpha_{\mathrm{ea}}\in(3,7/2)$ and $\alpha_{\mathrm{fi}}\in(3,4)$ are the
unique roots of $\alpha_{\mathrm{ea}}^3-2\alpha_{\mathrm{ea}}^2+4\alpha_{\mathrm{ea}}-23=0$
and $\alpha_{\mathrm{fi}}^3-\alpha_{\mathrm{fi}}^2+5\alpha_{\mathrm{fi}}-41=0$,
respectively.
\end{theorem}

The two bounds follow from the same adversarial argument, with different
choices of target positions and emergence times. We first consider targets
emerging at time $0$ and bound the positions of two distinct agents. We
then choose two further inputs for which neither agent can detect the
target during a specified time interval. The third agent cannot visit
both possible target positions before that interval ends, and the
remaining evacuation distance gives the contradiction.

Unlike the lower bound of Section~\ref{sec: lower bounds for 2 agents},
this argument does not assume origin-separated trajectories. As in
Section~\ref{sec:eels2-finite-lower-bound}, we obtain necessary constraints
by considering different inputs for the same algorithm. Here we choose
these inputs so that the constraints can be solved explicitly, without
enumerating event orders.

Fix a feasible algorithm $\mathcal A$ for \eels{3}. By
Section~\ref{sec:ProblemDefinition-results}, the pre-detection trajectory
$p_i$ of each agent $A_i$ is independent of the target position and
emergence time. Each target placement below defines a separate input
for this same algorithm. If the target in input $(x,T)$ has not been
detected before time $t\geq T$, agent $A_i$ is still at $p_i(t)$ and
must reach $x$, so $E(x,T)\geq t+|p_i(t)-x|$.

The first lemma gives the elementary lower bound $3$, valid for any
number of agents, and an inequality for one agent's trajectory. The
inequality, rather than the bound $3$ alone, is needed in the next lemma.

\begin{lemma}
\label{lem:eels3-lower-first-visit}
Suppose that $\mathcal A$ has competitive ratio at most $C$ under either
measure. Then $C\geq3$. Moreover, for every $r\geq1$, there are an agent
$A_i$ and a sign $\sigma\in\{-1,+1\}$ such that
$$
\sigma p_i(t)\geq t-(C-3)r
\text{ for every }0\leq t\leq r.
$$
\end{lemma}

\begin{proof}
Let $u$ be the first time any agent reaches $-r$ or $r$, and choose
$A_i$ and $\sigma$ so that $p_i(u)=\sigma r$. Feasibility and continuity
ensure that $u$ exists, and the speed bound gives $u\geq r$.
For $\epsilon>0$, consider input $(-\sigma(r+\epsilon),0)$. Its target
cannot be detected before $u$, and $A_i$ must travel another
$2r+\epsilon$ after time $u$ to reach it. Since both offline benchmarks
equal $r+\epsilon$,
$$
C\geq\frac{E(-\sigma(r+\epsilon),0)}{r+\epsilon}
\geq\frac{u+2r+\epsilon}{r+\epsilon}
\geq\frac{3r+\epsilon}{r+\epsilon}.
$$
Letting $\epsilon\to0$ gives $C\geq3$ and $u\leq(C-2)r$.
For every $0\leq t\leq r$, the speed bound then gives
$\sigma p_i(t)\geq r-(u-t)\geq t-(C-3)r$.
\end{proof}

We next obtain an estimate for a second agent by considering a target
on the opposite side. Here $C$ is an upper bound on the competitive
ratio, not necessarily its exact value. Thus the assumption $C>3$
does not exclude an algorithm whose competitive ratio is $3$.

\begin{lemma}
\label{lem:eels3-lower-two-agents}
Suppose that $\mathcal A$ has competitive ratio at most $C$ under either
measure, where $3<C<5$. Let $R,S\geq1$ and $\delta=(C-3)R$ satisfy
\begin{equation}
\label{eq:eels3-lower-radii}
\begin{aligned}
\delta&<S,\\
(5-C)R&>(C-1)S.
\end{aligned}
\end{equation}
Then there are two distinct agents $A_i,A_j$ and a sign
$\sigma\in\{-1,+1\}$ such that
$$
\begin{aligned}
\sigma p_i(t)&\geq t-\delta &&, 0\leq t\leq R,\\
\sigma p_j(t)&\leq-t+\delta' &&, 0\leq t\leq S,
\end{aligned}
$$
where $\delta'=((C-3)S+\delta)/2<\delta$.
\end{lemma}

\begin{proof}
Since $C>3$, the second inequality in~\eqref{eq:eels3-lower-radii}
implies $R>S$. Lemma~\ref{lem:eels3-lower-first-visit}, with $r=R$,
gives an agent $A_i$ and a sign $\sigma$ satisfying
$\sigma p_i(t)\geq t-\delta$ for $0\leq t\leq R$.

Consider input $(-\sigma S,0)$ and let $d$ be its detection time.
Suppose that $d\geq R$. Then $A_i$ follows $p_i$ through time $R$
and must subsequently reach $-\sigma S$. Hence
$$
\begin{aligned}
E(-\sigma S,0)&\geq R+\sigma p_i(R)+S\\
&\geq2R-\delta+S\\
&=(5-C)R+S>CS,
\end{aligned}
$$
contradicting the competitive-ratio bound. Therefore $S\leq d<R$.
At detection, we similarly obtain
$$
CS\geq E(-\sigma S,0)\geq d+\sigma p_i(d)+S\geq2d+S-\delta,
$$
so $d-S\leq\delta'$. Since $\sigma p_i(d)\geq S-\delta>0$,
an agent detecting $-\sigma S$ at time $d$ is different from $A_i$.
Let $A_j$ be such an agent. For every $0\leq t\leq S$, its speed bound
gives $\sigma p_j(t)\leq-S+(d-t)\leq-t+\delta'$.
Finally, $S<R$ and $\delta=(C-3)R$ imply $\delta'<\delta$.
\end{proof}

We next compare two inputs with a common emergence time. If two agents
cannot visit either target position during a given time interval, only
the remaining agent can detect either target during that time interval.
The following lemma is used with different target positions in the proofs
of the two competitive-ratio bounds.

\begin{lemma}
\label{lem:eels3-lower-delayed-pair}
Consider inputs $(\sigma a,T)$ and $(-\sigma b,T)$, where $a,b\geq1$,
$\sigma\in\{-1,+1\}$, and $0\leq T<H$. Suppose that $H-T\leq a+b$
and there are two distinct agents $A_i,A_j$ satisfying, for every
$t\in[T,H]$,
$$
\begin{aligned}
\sigma p_i(t)&>a,\\
\sigma p_j(t)&<-b.
\end{aligned}
$$
Then
$$
\max\{E(\sigma a,T),E(-\sigma b,T)\}
\geq H+\min\{a-\sigma p_j(H),b+\sigma p_i(H)\}.
$$
\end{lemma}

\begin{proof}
Suppose that the target is detected before $H$ on each of the two inputs,
at times $d_+,d_-\in[T,H)$, respectively. Neither $A_i$ nor $A_j$
can make either detection, so both detections must be made by the
remaining agent $A_k$. Since $A_k$ follows $p_k$ until detection on
each input, $p_k(d_+)=\sigma a$ and $p_k(d_-)=-\sigma b$.
The speed bound gives
$$
a+b=|p_k(d_+)-p_k(d_-)|\leq|d_+-d_-|<H-T\leq a+b,
$$
a contradiction. Thus the target is not detected before $H$ on at least
one of the two inputs. If that input is $(\sigma a,T)$, agent $A_j$
must still reach $\sigma a$, so
$E(\sigma a,T)\geq H+a-\sigma p_j(H)$.
Otherwise, agent $A_i$ must still reach $-\sigma b$, giving
$E(-\sigma b,T)\geq H+b+\sigma p_i(H)$. The claimed inequality follows.
\end{proof}

We are now ready to prove Theorem~\ref{thm:eels3-lower-bounds}.

\begin{proof}[Proof of Theorem~\ref{thm:eels3-lower-bounds}]
The polynomial $C^3-2C^2+4C-23$ has positive derivative for $C\geq3$,
and takes values $-2$ and $75/8$ at $3$ and $7/2$, respectively.
Similarly, $C^3-C^2+5C-41$ has positive derivative for $C\geq3$,
and takes values $-8$ and $27$ at $3$ and $4$, respectively.
Thus the two roots in the theorem exist and are unique in the stated
intervals. We prove the two bounds separately.

\emph{The emergence-aware lower bound.}
Suppose that $\mathrm{CR}_{\mathrm{ea}}(\mathcal A)<\alpha_{\mathrm{ea}}$,
and choose $\max\{3,\mathrm{CR}_{\mathrm{ea}}(\mathcal A)\}<C<\alpha_{\mathrm{ea}}$.
We will use target positions $1$ and $-1$ with emergence time just after
$1+\delta$. Set $R=\delta/(C-3)$ and $S=3+\delta$, where $\delta>0$
is to be chosen. Whenever Lemma~\ref{lem:eels3-lower-two-agents} applies,
it gives distinct agents $A_i,A_j$ and a sign $\sigma$ such that
$\sigma p_i(S)\geq3$ and $\sigma p_j(S)<-3$.
Lemma~\ref{lem:eels3-lower-delayed-pair} will therefore give evacuation
time at least $S+4$ for one of the two inputs. We choose $\delta$ to
satisfy $(5-C)R>(C-1)S$ and $C(2+\delta)<S+4$, or equivalently,
$$
\frac{3(C-3)(C-1)}{2+3C-C^2}
<\delta<
\frac{7-2C}{C-1}.
$$
Since $3<C<\alpha_{\mathrm{ea}}<7/2$, we have $C-1>0$ and
$2+3C-C^2>1/4$. Also $C^3-2C^2+4C-23<0$, because this polynomial
is strictly increasing and vanishes at $\alpha_{\mathrm{ea}}>C$.
Consequently,
$$
\frac{7-2C}{C-1}
-\frac{3(C-3)(C-1)}{2+3C-C^2}
=
-\frac{C^3-2C^2+4C-23}{(C-1)(2+3C-C^2)}>0,
$$
so the interval for $\delta$ is nonempty. These choices give $R>S>3$
and $\delta<S$, so Lemma~\ref{lem:eels3-lower-two-agents} applies.

Choose $0<\epsilon<2$ small enough that, for $T=1+\delta+\epsilon$,
we still have $C(T+1)<S+4$. For every $t\in[T,S]$, the agents and sign
supplied by Lemma~\ref{lem:eels3-lower-two-agents} satisfy
$$
\begin{aligned}
\sigma p_i(t)&\geq1+\epsilon>1,\\
\sigma p_j(t)&\leq-1-\epsilon-(\delta-\delta')<-1.
\end{aligned}
$$
Also $S-T=2-\epsilon<2$. Lemma~\ref{lem:eels3-lower-delayed-pair},
with $a=b=1$ and $H=S$, gives
$$
\max\{E(1,T),E(-1,T)\}\geq S+4>C(T+1).
$$
Both emergence-aware offline benchmarks equal $T+1$, contradicting the
choice of $C$.

\emph{The fully-informed lower bound.}
Suppose that $\mathrm{CR}_{\mathrm{fi}}(\mathcal A)<\alpha_{\mathrm{fi}}$,
and choose $\max\{3,\mathrm{CR}_{\mathrm{fi}}(\mathcal A)\}<C<\alpha_{\mathrm{fi}}$.
We now use target positions $\sigma A\lambda$ and $-\sigma B\lambda$
with common emergence time $\lambda$, where the sign $\sigma$ will be
supplied by Lemma~\ref{lem:eels3-lower-two-agents}. We choose $0<A,B<1$
and then $\lambda$ so that $A\lambda,B\lambda\geq1$, as required by the
problem definition. Neither target can be detected before it emerges at
time $\lambda$, and both fully-informed offline benchmarks equal $\lambda$.

We take $R=D\lambda/(C-3)$, $S=X\lambda$, and $\delta=D\lambda$, so that
$\delta'$ in Lemma~\ref{lem:eels3-lower-two-agents} equals
$\Gamma\lambda$, where $\Gamma=((C-3)X+D)/2$.
We choose $X=1+A+B$, making $S-\lambda$ equal to the distance between
the two target positions. To apply Lemma~\ref{lem:eels3-lower-delayed-pair}
with $H=S$ and obtain a contradiction, we seek constants satisfying
\begin{equation}
\label{eq:eels3-lower-fi-parameters}
\begin{aligned}
A+D&<1,\\
B+\Gamma&<1,\\
2X+A-\Gamma&>C,\\
2X+B-D&>C.
\end{aligned}
\end{equation}
The inequalities $\sigma p_i(t)\geq t-D\lambda$ and
$\sigma p_j(t)\leq-t+\Gamma\lambda$ from
Lemma~\ref{lem:eels3-lower-two-agents}, together with the first two
conditions, keep both agents from visiting either target position
throughout $[\lambda,S]$. The last two conditions make the evacuation
lower bound in Lemma~\ref{lem:eels3-lower-delayed-pair} exceed $C\lambda$.

To select the constants, we first solve the corresponding equalities
with $D=D_0$ and $\Gamma=\Gamma_0$:
$$
\begin{aligned}
A+D_0&=1,\\
B+\Gamma_0&=1,\\
2\Gamma_0&=(C-3)X+D_0,\\
2X+A-\Gamma_0&=C,\\
2X+B-D_0&=C,\\
X&=1+A+B.
\end{aligned}
$$
Solving gives
$$
\begin{aligned}
A&=\frac{C^2+C-11}{9},\\
B&=\frac{-C^2+2C+8}{9},\\
X&=\frac{C+2}{3},\\
D_0&=\frac{-C^2-C+20}{9},\\
\Gamma_0&=\frac{(C-1)^2}{9}.
\end{aligned}
$$
For $3<C<4$, these values satisfy $0<A,B,D_0<1<X$.
Keeping $A,B,X$ fixed and decreasing $D$ below $D_0$ also decreases $\Gamma$ below $\Gamma_0$, making all inequalities in~\eqref{eq:eels3-lower-fi-parameters} strict.

It remains to choose $D$ so that Lemma~\ref{lem:eels3-lower-two-agents} applies. The inequality $(5-C)R>(C-1)S$ is equivalent to $D>D_{\min}$, where $D_{\min}=(C-3)(C-1)X/(5-C)>0$.
Since $C<\alpha_{\mathrm{fi}}<4$, we have $5-C>0$ and
$C^3-C^2+5C-41<0$. Hence
$$
D_0-D_{\min}
=-\frac{2(C^3-C^2+5C-41)}{9(5-C)}>0.
$$
Choose $D_{\min}<D<D_0$ and $\lambda\geq\max\{1/A,1/B\}$.
Then $R>S>1$ and $\delta<\lambda<S$, so all hypotheses of Lemma~\ref{lem:eels3-lower-two-agents} hold. Let $A_i,A_j$ and $\sigma$ be supplied by that lemma. 
For every $t\in[\lambda,S]$,
$$
\begin{aligned}
\sigma p_i(t)&\geq t-D\lambda\geq(1-D)\lambda>A\lambda,\\
\sigma p_j(t)&\leq-t+\Gamma\lambda\leq(-1+\Gamma)\lambda<-B\lambda.
\end{aligned}
$$
Since $S-\lambda=(A+B)\lambda$, Lemma~\ref{lem:eels3-lower-delayed-pair},
with $a=A\lambda$, $b=B\lambda$, and $H=S$, gives
$$
\begin{aligned}
\max\{E(\sigma A\lambda,\lambda),E(-\sigma B\lambda,\lambda)\}
&\geq\lambda\min\{2X+A-\Gamma,2X+B-D\}\\
&>C\lambda.
\end{aligned}
$$
This contradicts the fully-informed guarantee. 
Together with the emergence-aware argument, this proves
$
\eelsea{3}\geq\alpha_{\mathrm{ea}}
$ and 
$\eelsfi{3}\geq\alpha_{\mathrm{fi}}.
$
\end{proof}

For fixed $C$, the parameter conditions used in both arguments are linear.
In the emergence-aware case the only free parameter is $\delta$, while in
the fully-informed case we choose $A,B,X,D,\Gamma$. 
Thus each proof may also be viewed as a small linear feasibility problem whose variables describe the adversarial target placements and emergence times. 
A feasible choice of these parameters produces inputs contradicting the assumed competitive ratio. This is different from Section~\ref{sec:eels2-finite-lower-bound}. 
There, the LP variables describe possible visitation times and agent positions, and the lower bound follows only after all possible cases are shown to be infeasible. 
Here we can solve the parameter conditions explicitly, and the boundary of feasibility yields the two cubic equations in Theorem~\ref{thm:eels3-lower-bounds}.

A computer-assisted extension could retain further target placements and the positions of all three agents at selected visits, then consider the possible visiting agents and event orders. 
The two-agent formulation would need to be extended. When the middle of three agents detects the target, the remaining evacuation distance need not equal the distance between the two extreme agents. 
In particular, inequality~\eqref{eq:eels2-detection-spread} cannot be used unchanged.
No such enumeration is needed for the bounds proved here. Further target placements may provide stronger constraints, but an improvement would require a new argument or a complete computer-assisted verification.

\section{Conclusions}

Eventually Emerging Linear Search behaves differently from classical
wireless evacuation already for very small numbers of agents. For one
agent, the additional uncertainty created by the emergence time does not
change the optimal competitive ratio, which remains the classical value
$9$. For two agents the picture changes: even with instantaneous wireless
communication, eventual emergence makes the problem strictly harder than
classical two-agent evacuation on the line. A third agent then provably
improves the optimal performance again, under both competitive measures.
Thus, for both the emergence-aware and fully-informed benchmarks, our
results separate the one-, two-, and three-agent problems. They also
indicate that the main difficulty lies in coordinating exploration with
carefully timed revisitation, rather than in communicating the target
location after detection.

Several structural questions remain open. We conjecture that our
two-agent algorithm is in fact optimal without the origin-separated
restriction, namely
$
\eelsea{2}=\eelsfi{2}=3+\sqrt6.
$
It is also open whether the emergence-aware and fully-informed competitive
ratios are ever different at the optimum. Our three-agent algorithms give
different guarantees for the two measures, but this does not establish a
separation between \eelsea{n} and \eelsfi{n} for any $n$. 
More generally, both optimal ratios are nonincreasing in the number of
agents, since an additional agent may follow the trajectory of an existing
one. Our results show that the decrease is strict from one to two agents
and from two to three agents. We conjecture that it remains strict for
every $n\geq3$:
$
\eelsea{n+1}<\eelsea{n}
$
and 
$
\eelsfi{n+1}<\eelsfi{n}.
$
Determining the asymptotic behavior of these quantities as the number of
agents grows is equally natural.

Finally, the success of the three-peak refinement suggests that further
local refinements of the trajectories may continue to improve the
three-agent upper bounds, and more generally may be needed for larger
teams. In this paper our objective was instead to establish the first
separations and develop tools for analyzing such trajectories and proving
lower bounds. Other natural directions include randomized search and the
face-to-face communication model. The latter appears particularly
interesting, since uncertainty about when the target emerges would then
interact with the additional difficulty of communicating its discovery.

\bibliographystyle{splncs04}
\bibliography{eels-bib-arxiv}

\end{document}